\documentclass[graybox]{svmult}

\usepackage{type1cm}          
\usepackage{makeidx}             
\usepackage{graphicx}           
\usepackage{multicol}              
\usepackage[bottom]{footmisc}
\usepackage{newtxtext} 
\usepackage{newtxmath}   
\usepackage{amsmath}
\usepackage{tcolorbox}
\usepackage{hyperref}
\hypersetup{
    colorlinks=true,
    linkcolor=cyan,
    filecolor=green,      
    urlcolor=magenta,
    pdftitle={Marie Farge: Turbulence and continuous wavelets},
    pdfpagemode=FullScreen,
    }

\makeindex          

\begin{document}

\title*{The evolution of turbulence theories \\ and the need for continuous wavelets}
\author{Marie Farge, \\ {\it 17 December 2022}}
\institute{Marie Farge \at LMD-CNRS, Ecole Normale Sup\'erieure, 24 rue Lhomond 75231 Paris Cedex 5, France \\
\email{marie.farge@ens.fr}}

\maketitle

\abstract{In the first part of this article, I summarise two centuries of research on turbulence. I also critically discuss some of the interpretations which are still in use, since turbulence remains an inherently non-linear problem that is still unsolved to this day. In the second part, I tell the story of how Alex Grossmann introduced me to the continuous wavelet representation in 1983, and how he instantly convinced me that this is the tool I was looking for to study turbulence. In the third part, I present a selection of results I obtained in collaboration with several students and colleagues to represent, analyse and filter different turbulent flows using the continuous wavelet transform. I have chosen to present both theories and results without the use of equations, in the hope that reading this article will be more enjoyable.}


\section{Turbulence}
\label{sec:1}

\subsection{Some definitions}
\label{subsec:1.1}

A fluid flow is called turbulent when it exhibits an unsteady, unstable, chaotic and mixing behaviour. By fluid I mean a continuous movable and deformable medium, so that liquids, gases and plasmas are considered to be fluids as long as the scale of the observer is much larger than the mean free path of the motions of their constituent particles. It is important to stress that turbulence is a characteristic of the flow and not of the fluid. In this article I will only consider incompressible, viscous, isotropic and Newtonian fluids ({\it i.e.}, such that the viscous stress tensor is proportional to the deformation rate tensor) and assume that the flow is governed by the Navier--Stokes equation. This fundamental equation of fluid mechanics expresses the conservation of momentum and predicts the evolution of two fields: the velocity and the pressure of the flow, as a function of two parameters: the density and the kinematic viscosity of the fluid, together with the geometry of the domain and the corresponding boundary conditions. 

\bigskip

The turbulence level is quantified by the Reynolds number, which corresponds to the ratio between the norm of the convective term and the norm of the dissipative term of Navier--Stokes' equation. Its convective term is non-linear and generates flow instabilities, while its dissipative term is linear and generates viscous damping by converting kinetic energy into thermal energy. In experimental fluid mechanics and in engineering, the Reynolds number is empirically estimated as the product between a characteristic velocity of the flow and a characteristic length of the solid boundaries, divided by the kinematic viscosity ({\it i.e.}, viscosity divided by the fluid density). One distinguishes three flow regimes which are sorted by increasing values of the Reynolds number: 

\begin{itemize}
\item
 the laminar regime at low Reynolds number (typically between $0$ and $10^2$), where the flow is quasi-steady, stable, non-chaotic, so that its behaviour can be deterministically predicted,
\item
 the weak turbulence regime at moderate Reynolds number (typically between $10^2$ and $10^5$), where the flow is unsteady, unstable, chaotic, therefore its behaviour can no longer be deterministically predicted, nor statistically predicted because it does not mixing sufficiently to allow well-converged statistics,
\item
 the strong turbulence regime, often called `fully-developed turbulence', at high Reynolds number (typically above $10^5$), where the flow is unsteady, unstable, chaotic and mixes enough to get well-converged statistics, therefore its behaviour can only be predicted statistically, but not deterministically.
 \end{itemize}
 
 \bigskip

 \noindent
 Let us consider some typical applications:
 
\begin{itemize}
\item
 in hydraulics ({\it i.e.}, pipes, pumps), internal geophysics  ({\it i.e.}, magma) or naval engineering ({\it i.e.}, sailing boats, tankers) the Reynolds number varies in the range $10^2$ to $10^6$,
\item
 in aeronautics ({\it i.e.}, cars, airplanes, rockets, shuttles) the Reynolds number varies in the range $10^6$ to $10^9$,
\item
 in external geophysics ({\it i.e.}, oceans, atmosphere) the Reynolds number varies in the range $10^9$ to $10^{12}$,
 \item
 in astrophysics the Reynolds numbers are larger than $10^{12}$.
 \end{itemize}

\bigskip

As turbulent flows are highly unstable, they transport and mix various quantities ({\it e.g.}, momentum, scalar tracers, particles) much more efficiently than laminar flows. The transport of momentum by turbulence leads, on a microscopic scale, to an equipartition of kinetic energy due to the fluid viscosity, which the observer perceives, on a macroscopic scale, as a conversion of kinetic energy into thermal energy. Since the quantity of kinetic energy that has reached equipartition can no longer provide work, it is said to be lost through "turbulent dissipation". Similarly, if a turbulent flow carries particles, these are diffused much more efficiently than in a laminar flow, which is known as "turbulent diffusion". We can easily observe this in everyday life: when you add sugar to your coffee and stir it with a spoon, thanks to turbulent diffusion you obtain a sweet coffee in a few seconds, whereas it would have taken about a day to achieve the same result without stirring, leaving only viscous dissipation to act.
 
 \bigskip

\noindent
In this article we will focus on strong turbulence, which corresponds to large Reynolds numbers and therefore to either high velocities of the flow (strong convection), large scales of the solid interfaces (large containers or obstacles) or small viscosities of the fluid (weak viscous dissipation).

\bigskip
 
\subsection{Brief history}
\label{subsec:1.2}

In the $16^{th}$ century Leonardo da Vinci chose the term {\it `turbolenza'} to characterise the regime when a fluid flow becomes non-linearly unstable and generates vortices. Understanding the mechanisms responsible for this behaviour remains to this day an open problem of interest to physicists and mathematicians alike. The etymology of the word 'turbulence' comes from two Latin words: {\it `turbo, inis'}, which means `vortices', and {\it `turba,ae'}, `crowd'; therefore a turbulent flow can be seen as a crowd of vortices in non-linear interaction.
\bigskip

In the $18^{th}$ century Leonhard Euler worked for King Frederick II of Prussia, for whom he wrote a treatise on 'New principles of gunnery' published in 1742. However, Euler was not satisfied with it, which led him to suggest to the Berlin Academy of Sciences for its Mathematical Prize of 1750 the problem of the resistance exerted by a fluid on a moving body. A few years earlier Jean Le Rond d'Alembert had already obtained this prize by solving the problem of the trade winds, for which he had introduced partial derivatives. To study this new problem d'Alembert introduced a new partial differential equation to predict the evolution of an incompressible inviscid fluid in motion. Euler refused to award him the prize and d'Alembert, vexed, published his paper in 1752 in Paris, under the title {\it `Essai d'une nouvelle th\'eorie sur la r\'esistance des fluides'} \cite{D'Alembert 1752}. Surprisingly the solution of d'Alembert's equation led to the proof that a fluid does not exert resistance on a moving body, which is contrary to observations and gave rise to d'Alembert's paradox. Euler modified d'Alembert's equation by adding a pressure term, and it became Euler's equations, but this did not solve d'Alembert's paradox. Euler wrote his article in French in 1755 under the title {\it `Principes g\'en\'eraux du mouvement des fluides'} and published it in 1757 \cite{Euler 1757}.
 
\bigskip

In the $19^{th}$ century Jean-Claude Adh\'emar Barr\'e de Saint-Venant, George Stokes and Claude-Louis Navier solved d'Alembert's paradox and suggested the role of viscosity to explain the resistance that a fluid flow exerts on a solid body. This led to the Navier--Stokes equation that Navier published in 1822,  which is the fundamental equation of fluid mechanics \cite{Navier 1822}.  It describes the flow evolution for both the laminar and the turbulent regimes; indeed, a flow becomes turbulent when the non-linear transport term of the Navier--Stokes equation, due to the motion of the flow, dominates the linear dissipation term, due to the viscosity of the fluid. In the $20^{th}$ century Jean Leray conjectured in 1934 that the turbulent regime is characterised by the loss of uniqueness of the solutions of the Navier--Stokes equation \cite{Leray 1934}. Today proving that the solutions of the Navier--Stokes equation for an incompressible fluid always remain regular is one of the seven Millennium Prize unsolved problems \cite{Fefferman 2006}, for which in year 2000 the Clay Mathematics Institute offered a one million dollar reward for the solution of each problem.

 \bigskip
 
\subsection{Few remarks}
\label{subsec:1.3}

When studying turbulent flows, we are overwhelmed by their apparent complexity, which leads us to describe them as `disordered' or `random'. But, as already stated four centuries ago by Spinoza, it is important to understand that the notion of `order' is subjective and that we call `disordered' a system whose behaviour appears to us too complicated to be described in detail: {\it `Because those who do not understand the nature of things but only imagine them, affirm nothing concerning things, and take the imagination for the intellect, they firmly believe, in their ignorance of things and of their own nature, that there is an order in things. For when things are so disposed that, when they are presented to us through the senses, we can easily imagine them, and so can easily remember them, we say that they are well-ordered; but if the opposite is true, we say that they are badly ordered, or confused. And since those things we can easily imagine are especially pleasing to us, men prefer order to confusion, as if order were anything in nature more than a relation to our imagination'} \cite{Spinoza 1677}. 

\bigskip

Maxwell, in the article on {\it `Diffusion'} that he wrote for the 5th edition of the {\it Encyclop{\ae}dia Britannica} published in 1877, also emphasised the fact that the terms `order', `disorder' and `dissipation' are subjective. He explained that:  {\it `Dissipated energy is energy which cannot lay hold of and direct at pleasure such as the energy of the confused agitation of molecules which we call heat. Now, confusion, like the correlative term order, is not a property of material things in themselves, but only in relation to the mind which perceives them. A book does not, provided it is neatly written, appear confused to an literate person, or to the owner who understands it thoroughly, but to any other person unable to read it appears inextricably confused. Similarly the notion of dissipated energy could not occur to a being who could trace the motion of every molecule and seize it at the right moment. It is only to a being in the intermediate stage, who can hold of some forms of energy while others elude his grasp, that energy appears to be passing inevitably from the available to the dissipated state' } \cite{Maxwell 1877}. This is why to study turbulence we should take into account the observer and its scale of observation, together with the problem we want to solve. For instance, if we need to adapt an engine to an airplane we have to estimate the airplane's drag, but to decide the safety distance required between two airplanes requires to know the length of the trailing vortices which might destabilise the following airplane. For the first problem we need to only compute low-order averages which are sensitive to frequent events (they correspond to the centre of the probability distribution functions), while for the second problem we need to compute high-order averages which are sensitive to rare events (they correspond to the tails of the probability distribution functions). 

\bigskip

One objective of the theory of turbulence is to define two kinds of observable quantities: those whose evolution we will deterministically predict, and those we will only statistically model without tracking their detailed dynamics. This program was already stated a century ago by Richardson in an article entitled {\it `Diffusion regarded as a compensation for smoothing}' where he wrote that: {\it `By an arbitrary choice we try to divide motions into two classes: (a) those which we treat in detail, (b) those which we smooth away by some process of averaging. Unfortunately these two classes are not always mutually exclusive. [...] Diffusion is a compensation  for neglect of detail. [...] The form of the law of diffusion depends entirely upon the arbitrarily chosen method of averaging, which is always implied when diffusion or viscosity are mentioned. This calls attention to the desirability of making much more explicit statements about smoothing operations than has hitherto been the  custom'} /Richardson et al. 1930/. Richardson emphasises a key question that is crucial for turbulence research: how to define the averaged quantities we wish to measure and predict in order to describe turbulent flows? The best discussion I know for defining appropriate averages in turbulence is in an article written in 1956 by Kamp\'e de F\'eriet entitled {\it `The notion of average in turbulence theory'} \cite{Kampe de Feriet 1956}.

\bigskip

We should be aware that our definition of turbulence depends on the theoretical tools we have at our disposal to describe and model it. If one relies on the theory of dynamical systems, turbulent flows are seen as a collection of vortices with chaotic dynamics in physical space, and characterised by a strange attractor in phase space. If one relies on stochastic theory, one emphasises the randomness of turbulent flows, which can be characterised by the probability distribution function of a large ensemble of different realisations of the same flow. Our definition of turbulence also depends on the technical tools we have at hand. For instance, the development of hot wire anemometry in the 1950s allowed experimentalists to obtain point-wise measurements of many flow realisations and therefore to calculate reliable statistics, such as energy spectra and probability distribution functions, but this technique does not give meaningful information (for instance the instantaneous spatial distribution of velocity, pressure or temperature fields) about an individual flow realisation. It is the generalised availability of computers beginning in the early 1980s, both for laboratory (data acquisition and processing) and numerical experiments (direct numerical simulation and large eddy simulation), that changed our views on turbulent flows by producing multi-point measurements (generally in the form of grid point sampling) and by giving access to the detailed space-time structure of the various fields of interest: velocity, vorticity (the curl of velocity), pressure, temperature, concentration of various scalar quantities, etc. The change in viewpoint allowed by computers has been advocated by experimentalists, such as Ahlers who testifies that: {\it `I believe that the most important experimental development of the 1970's was the advent of the computer in the laboratory [...]. Data acquisition and processing did not only provide us a new tool but they also gave us completely new perspectives on what types of experiments to do'} \cite{Aubin 1997}.

\bigskip
\bigskip

  
\section{Turbulence theories}
\label{sec:2}

I will distinguish three different approaches to studying and modelling turbulent flows. The kinetic statistical approach is inspired by the kinetic theory of gases, developed by Maxwell,  Boltzmann and Einstein among others, and separates flows into mean and fluctuating motions, supposing a spectral gap between them. This leads to several turbulent viscosity models suggested by Boussinesq and Reynolds, and to the mixing length model introduced by Prandtl. The second approach is probabilistic and relies on the theory of stochastic processes developed by Wiener, Khinchin and Kolmogorov, among others and involves random functions and probability measures; it predicts a scaling law for the energy spectrum. The third approach is deterministic and focuses on the vorticity field of individual flow realisations. It analyses the formation and interaction of coherent vortices which emerge out of turbulent flows. One looks for a low-dimensional discrete dynamical system that might exhibit the same chaotic behaviour as the infinite dimensional continuous turbulent flow.

\subsection{Kinetic statistical theories}
\label{subsec:2.1}

In order to try to master the complexity of turbulent flows, the first statistical approach decomposed the velocity field into its mean value and  fluctuations, by analogy with the kinetic theory of gases which distinguishes the mean motion from the fluctuating (thermal) motion of molecules. This statistical kinetic approach was introduced by Saint--Venant and Boussinesq assuming the existence of `fluid molecules' \cite{Boussinesq 1877}, and later by Reynolds in 1894 \cite{Reynolds 1894} and by Lorentz in 1896 \cite{Lorentz 1896}.  After decomposing the velocity field into a mean contribution plus fluctuations, one rewrites the Navier--Stokes equation to predict the evolution of the mean velocity as a function of  fluctuations; this procedure yields the Reynolds equations. However, one encounters  difficulties due to the non-linear advection term of the Navier--Stokes equation : the second-order moment of the velocity fluctuations, called the `Reynolds stress tensor', depends on the third-order moment, which in turn depends on the fourth-order moment, and so on {\it ad infinitum}. At each order one finds that there are more unknowns than equations and faces a closure problem. In order to close this hierarchy of equations, the usual strategy is to add  another equation, or system of equations, chosen from some {\it ad hoc} phenomenological hypotheses. In particular, one must assume that there exists a scale separation, namely that fluctuating motions are sufficiently decoupled from the mean motions to guarantee that the average of the product (coming from the non-linear term of Navier--Stokes equation) is equal to the product of averages (the $5^{th}$ Reynolds postulate \cite{Monin et al. 1965}).

\bigskip

To close the hierarchy of Reynolds equations, Prandtl introduced in 1925 a scale, called the `mixing length', a characteristic of the velocity fluctuations. Following the hypothesis suggested by Boussinesq \cite{Boussinesq 1877} and by analogy with molecular diffusion which regularises velocity gradients for scales smaller than the molecular mean free path, Prandtl assumed  that there exists a turbulent diffusion, {\it i.e.} an enhanced diffusion due to the flow non-linear instability, which smoothes the velocity fields on scales smaller than the mixing length; he could thus rewrite the Reynolds stress tensor as a turbulent diffusion term.  As soon as one considers turbulent flows having large Reynolds number, the mixing length hypothesis fails because the analogy with the kinetic theory of gases no longer holds. Indeed, if molecular motions can be modelled by a diffusion equation (Laplacian operator applied to a mean velocity field with the kinematic viscosity as transport coefficient), it is because there is a wide spectral gap between the scales as seen by the observer and the scales dominated by molecular motions. Such decoupling no longer exists for the strongly turbulent regime since the non-linear interactions couple all scales of motion and there is no spectral gap. This is a major obstacle in any attempt to model turbulence using moment equations, and therefore the closure problem remains open. An important direction of research is to find a new representation of turbulent flows for which such separation could exist. It would no longer be based on a decoupling in scales, but on a decoupling between out of statistical equilibrium motions and well-thermalized motions in statistical equilibrium. The energy of the latter being equipartitioned between all degrees of freedom, it would be possible to model their effect on the former by a dissipative term, which we will call 'turbulent dissipation'.

\bigskip

Modelling of turbulence should rely on a careful statistical analysis of turbulent flows, observed in the laboratory or in numerical experiments. Traditionally, one studies the long term velocity correlation, assuming the turbulent flow has reached a statistically steady state, such that time covariance remains steady. Covariance and cross-correlation functions was first introduced by Einstein, in an article written in French and published in 1914 in Switzerland, to study the statistics of long time series of fluctuating quantities \cite{Einstein 1914}, \cite{Yaglom 1986}. In 1935 Taylor \cite{Taylor 1935} formulated, in addition to statistical stationarity, the hypothesis of statistical isotropy (and consequently homogeneity), namely he assumed that statistical observables are invariant under both translation and rotation of the fields. Using the statistical tools introduced by Wiener (inspired in turn by the earlier work of Taylor on Brownian motion  \cite{Taylor 1921}), Taylor suggested studying isotropic turbulence by measuring its energy spectrum, defined by the modulus of the Fourier transform of the two-point correlation of the velocity increments  \cite{Taylor 1935}. Taylor also assumed that the spatial velocity distribution changes slowly as it is carried past the point at which the frequency spectrum (the Fourier transform of the time correlation) is measured, which in this case yields the same energy spectrum for both time and space correlations. This is known as `Taylor's hypothesis' and is commonly used in most laboratory experiments to interpret time correlations as space correlations. These new statistical tools introduced by Taylor to analyse turbulent flows brought about a change in research on turbulence and are still in use today. Although in his work on turbulent diffusion Taylor \cite{Taylor 1921} was the first to introduce stochastic tools in turbulence, he always adopted a dynamic point of view. This explains why he never showed a strong interest in Kolmogorov's theory (as mentioned by Batchelor in his biography of Taylor \cite{Batchelor 1996}). In his famous article, entitled {\it `The Spectrum of Turbulence'} \cite{Taylor 1938}, Taylor expressed his intuition that energy dissipation is not densely distributed in space and that turbulent flows are intermittent, {\it i.e.}, the sparser their fluctuations, the  stronger they are. He stated the hypothesis that intermittency is related to the spottiness of the spatial distribution of vorticity: {\it `The fact that small quantities of very high frequency disturbances appear, and increase as the speed increases, seems to confirm the view frequently put forward by the author (himself) that the dissipation of energy is due chiefly to the formation of very small regions where the vorticity is very high'} \cite{Taylor 1938}. Understanding intermittency is still a very important open problem, possibly the essential one, to be solved before we can build a satisfactory theory of turbulence.

 \bigskip
 
\subsection{Probabilistic statistical theories}
\label{subsec:2.2}

In order to overcome the closure problem, due to the fact that there is no scale separation to decouple large scale from small scale motions in turbulent flows, another statistical approach has been put forward. It replaces the observation of an individual realisation of the flow with the calculation of the correlation between different measurements made on many realisations of this flow, whose number must be large enough to ensure statistical convergence. The ensemble averages are constructed on the assumption that the probability law of the process governing turbulent flows is known, and thus all its moments, which is unfortunately not yet the case. This probabilistic approach was initiated by Gebelein in 1935 \cite{Gebelein 1935} and developed by many scientists, often independently, among them Kamp\'e de F\'eriet \cite{Kampe de Feriet 1939}, Millionshchikov \cite{Millionshchikov 1939}, Kolmogorov \cite{Kolmogorov 1941}, Obukhov \cite{Obukhov 1941}, Onsager \cite{Onsager 1945}, Heisenberg and Von Weizs\"acker \cite{Heisenberg et al. 1948}. Since Gibbs, such a probabilistic approach has become standard in statistical physics, but the difficulty in applying it to turbulence arises from the fact that turbulent flows are open thermodynamic systems due to the injection of energy by external forces and its dissipation by viscous frictional forces. To resolve this difficulty, in 1941 Kolmogorov proposed the existence of an `energy cascade' in spectral space, based on the hypothesis that external forces which inject energy into the flow act only at the lower wavenumbers, while frictional forces which dissipate energy act only at the highest wavenumbers. In the limit of very large Reynolds numbers, Kolmogorov assumed that there exists an  intermediate range of wavenumbers, called the inertial range, for which energy is conserved and only transferred from low to high wavenumbers at a constant rate $\epsilon$. Kolmogorov also supposed that the flow statistics are homogeneous (invariant by translation) and isotropic (invariant by rotation), and  that the skewness of the velocity increment probability distribution is non-zero which implies that turbulent flows are non-Gaussian. All these hypotheses led him to the prediction that the modulus of the energy spectrum scales according to the power-law $\epsilon ^{2/3} k^{-5/3}$, $k$ being the modulus of the wavenumber and $\epsilon$ the rate of energy transfer \cite{Kolmogorov 1941}. 

\bigskip

Kolmogorov's statistical theory of homogeneous isotropic three-dimensional turbulence is not based on the Navier--Stokes equation because he only used the fact that Euler equation, the fundamental equation describing inviscid fluid flow, conserves energy. In 1953, Batchelor published a textbook which is still a reference about homogeneous isotropic turbulence \cite{Batchelor 1953}. In 1959, Kraichnan, who was one of the last assistants to Albert Einstein at the Institute of Advanced Studies in Princeton before moving to the Courant Institute in New York, published an important article entitled {\it The structure of isotropic turbulence at very high Reynolds numbers}. Based on the same hypotheses as Kolmogorov he proved that in spectral space the non-linear interactions are local and he proposed the `Direct Interaction Approximation' which is still an important method to model three-dimensional turbulence \cite{Kraichnan 1959}. In the same vein as Kolmogorov's theory for three-dimensional turbulence, Kraichnan proposed in 1967 the statistical theory of homogeneous isotropic two-dimensional turbulence \cite{Kraichnan 1967}. For this, he relied on the fact that in two dimensions the Euler equation preserves not only the energy but also the enstrophy (the norm of the square of the vorticity). From there, he proved that there is an `inverse energy cascade' in spectral space, as non-linear interactions within two-dimensional turbulent flows no longer transfer energy but enstrophy to high wavenumbers, thus forcing energy to be transferred to low wavenumbers \cite{Kraichnan 1967}.

\bigskip

In fact, the hypothesis of a turbulent cascade and the resulting scaling of the energy spectrum can only be valid for ensemble averages. If one considers each flow realisation, the energy ($L^2$-norm of velocity) and the enstrophy ($L^2$-norm of vorticity) are generated very locally in physical space, at locations where there are boundaries or internal shear layers, and therefore they are generated very non-locally in wavenumber space (resulting from the definition of the Fourier transform and the related uncertainty principle). Likewise, the spatial support of dissipation is highly spotty for a given flow realisation and hence also non-local in wavenumber space. These  observations are in contradiction with the hypothesis of a low-wavenumber injection and a high-wavenumber dissipation of energy necessary  to maintain an inertial range. Such observations already been made more than 400 years ago by Leonardo da Vinci when he wrote: {\it `Where turbulence of water is generated, where turbulence of water maintains for long, where turbulence of water comes to rest'} \cite{Frisch 1995}. Obviously da Vinci was describing the evolution of one turbulent flow realisation in physical space and not an ensemble average in wavenumber space. His remark has therefore nothing to do with the notion of turbulent cascade which is a concept formulated for ensemble averages and which tells nothing about an individual realisation. Vinci was observing the formation of vortices in boundary layers and internal shear layers, their advection by the velocity field they collectively generate, and their dissipation resulting from the non-linear interactions between them. The frequent confusion between observations made in physical space and the concept of turbulent cascade developed in spectral space was denounced by Kraichnan as early as 1974 in an article entitled {\it `On Kolmogorov's inertial-range theories'} \cite{Kraichnan 1974}. Kraichnan noted: {\it `The terms ``scale of motion'' or ``eddy of size $l$'' appear repeatedly in the treatment of the inertial range. One gets an impression of little, randomly distributed whirls in the fluid, with the fission of the whirls into smaller ones, after the fashion of Richardson's poem. This picture seems to be drastically in conflict with what can be inferred about the qualitative structures of high Reynolds numbers turbulence, from laboratory visualisation techniques and from plausible application of the Kelvin's circulation theorem'} \cite{Kraichnan 1974}. Incidently we should notice that Richardson's poem mentioned by Kraichnan describes the formation of smaller and smaller structures at the interface of clouds, but not in the bulk of a turbulent flow \cite{Richardson 1922}. Indeed, the break-up of vortices into smaller and smaller vortices is not mechanically possible in an incompressible fluid, as it is a continuous divergence-free medium; only non-linear instabilities caused by deformation (such as the Kelvin-Helmholtz instability) can generate smaller vortices in fluid flows. Unfortunately this image of a breaking vortex is still widely used to explain the non-linear cascade of energy characteristic of turbulent flows and confuses students, as it did for me over forty years ago.
 
\bigskip

Due to evidence of small-scale intermittency observed by Townsend in 1951 \cite{Townsend 1951} and following a remark of Landau who pointed out that the dissipation rate $\epsilon$ should fluctuate in space, Kolmogorov corrected his theory published in 1941 and presented his new theory during the `1st International Colloquium on Turbulence' held in Marseille in 1961 \cite{TCM 1962a}. For this he added an `intermittency correction' to the energy spectrum exponent \cite{Kolmogorov 1962}, which opened a debate that is still strong today. In 1974 Kraichnan wrote: {\it `The 1941 theory is by no means logically disqualified merely because the dissipation rate fluctuates. On the contrary, we find that at the level of crude dimensional analysis and eddy-mitosis picture the 1941 theory is as sound a candidate as the 1962 theory. This does not imply that we espouse the 1941 theory. On the contrary, the theory is made implausible by the basic physics of vortex stretching. The point is that this question cannot be decided a priori; some kind of non-trivial use must be made of the Navier--Stokes equation'}  \cite{Kraichnan 1974}. Kraichnan, although a master of the statistical approach, claims that one needs to first understand the generic dynamics of the Navier--Stokes equation before being able to construct a statistical theory that includes intermittency: {\it `If the Kolmogorov law $E(k) \propto k^{-5/3-\mu}$ is asymptotically valid, it is argued that the value $\mu$ depends on the details of the non-linear interaction embodied in the Navier--Stokes equation and cannot be deduced from overall symmetries, invariances and dimensionality'} \cite{Kraichnan 1974}.

\bigskip

Pauli liked to say that there exist true theories, false theories and theories which are neither true nor false, namely, following Popper's terminology, theories which are `unfalsifiable'. In order to fit observations, unfalsifiable theories add new parameters, rather than change their hypotheses. Consequently these theories can be very accurate and optimal for describing existing data, but they are not able to predict new facts and they lack predictability. A typical example is the geocentric theory of the solar system which since Ptolemy was able to describe the motions of all known planets with a small number of epicycles. When new observations became available new epicycles were added in order to preserve the corpus of the theory. The theory was very useful in practice, but unable to predict the existence of an unknown planet, as Leverrier did for Neptune. Unfalsifiable theories are poor from an epistemological point of view, but they produce a large number of publications because each adjustment, necessary to match new observations, requires new parameters to amend the model. This may be what Kraichnan had in mind when he wrote: {\it `Once the 1941 theory is abandoned, a Pandora's box of possibilities is open. The 1962 theory of Kolmogorov seems arbitrary, from an {\it a priori} viewpoint [...]. We make the point that even in the general framework of some kind of self-similar cascade, and of intermittency which increases with the number of cascade steps, the 1962 theory is only one of many possibilities'} \cite{Kraichnan 1974}. One can also add a new class to Pauli's picture: `too true' theories, namely theories one cannot falsify even in situations where their hypotheses are not valid, which is illustrated by Kraichnan's comment on Kolmogorov 41's theory: {\it `Kolmogorov's 1941 theory has achieved an embarrassing success. The $-5/3$ spectrum has been found not only where it reasonably could be expected, but also at Reynolds numbers too small for a distinct inertial range to exist as in boundary layers and shear flows where there are substantial departures from isotropy, and such strong effects from the mean shearing motion that the stepwise cascade appealed to by Kolmogorov is dubious'} \cite{Kraichnan 1974}. The robustness of the energy spectrum is not entirely surprising because it is the Fourier transform of the two-point correlation of the velocity increments and therefore it is insensitive to rare events, such as coherent vortices which are produced in shear layers and in boundary layers. But this robustness of Kolmogorov's theory is lost as soon as one considers higher-order statistics which, unlike lower-order statistics, are sensitive to rare events, as we shall discuss in the following paragraph.

 \bigskip

\subsection{Deterministic theories}
\label{subsec:2.3}

In parallel with the statistical probabilistic approach based on ensemble averages, there has been a tendency to analyse each turbulent flow realisation separately, in order to study its dynamics and deterministically predict its evolution by identifying the active components. In 1951, Townsend \cite{Townsend 1951, Townsend 1956} was the first to suggest, from laboratory experiments, the existence of coherent structures that emerge out of turbulent flows through non-linear instability; he observed that they appear to be responsible for the chaotic and intermittent behaviour of turbulent flows and play an essential role in their transport properties. He explained that: {\it `The most natural hypothesis is that, under the action of distortion, vorticity distributions that are initially diffuse are concentrated into sheets and lines of vorticity. [...] The resultant turbulence pattern will be `spiky'. [...] There results an intermittent distribution of highly concentrated [...] vortex lines and sheets, in qualitative agreement with observations}' \cite{Townsend 1951}. In 1955, Theodorsen in the article, {\it The Structures of Turbulence} \cite{Theodorsen 1955}, formulated the hypothesis that vortex tubes, such as horseshoe or hairpin vortices, are the coherent structures which drive the flow and are responsible for the eddy motions and mixing properties characteristic of the strong turbulent regime. It was probably under Onsager's influence that Theodorsen assumed that vortex tubes play an essential dynamic role, as they were colleagues at the University of Trondheim in Norway and had certainly discussed such a puzzling subject as turbulence together.  Indeed, Onsager had published in 1949 an article entitled {\it Statistical Hydrodynamics} \cite{Onsager 1949} where he proposed a dynamic model made of singular vortex tubes, from which he built a statistical kinetic theory of turbulence. Here is how he explained his model: {\it `The formation of large, isolated vortices is extremely common, yet spectacular phenomenon in unsteady flow. Its ubiquity suggests an explanation on statistical grounds. To that end, we consider $n$ parallel vortices of intensities (circulations) $k_1, ..., k_n$ in an incompressible, frictionless fluid, [...] vortices of opposite sign will tend to approach each other, [...] vortices of the same sign will tend to cluster [...]. It stands to reason that the large compound vortices formed in this manner will remain as the only conspicuous features of the motion; because the weaker vortices, free to roam practically at random, will yield rather erratic and disorganised contributions to the flow.'} \cite{Onsager 1949}. Taylor and Onsager were thus the pioneers of the dynamic deterministic approaches to understand turbulence, although they both were conscious that in practice one can only predict and measure statistically defined observables due to the complexity of turbulent flows. Since their theoretical predictions the presence of coherent structures in strongly turbulent flows has been observed in both laboratory and numerical experiments.
 
\medskip

\subsubsection{Laboratory experiments}
\label{subsubsec:2.3.1}

In 1971, at Stanford University Kline designed a new technique for visualizing and measuring turbulent flows by injecting hydrogen bubbles into the flow, which led him to discover elongated coherent structures in a turbulent boundary layer ({\it i.e.}, a fluid flow in contact with a solid wall) \cite {Kline et al. 1971}. In 1974, at Caltech another crucial experiment was performed by Brown and Roshko, who confirmed the existence of coherent structures in the strong turbulence regime \cite{Brown and Roshko 1974}. They studied a plane mixing layer ({\it i.e.}, two fluid flows with a strong velocity gradient at their interface) and showed that it exhibits coherent structures even at very large Reynolds number. This was very surprising, because those years turbulence was characterised by the breakdown of any organised structure and this was thought to be the cause of the randomness of the flow. They concluded their article by stating: {\it `It seems remarkable, at first, that a flow with such organised structure [...] could have the attributes usually associated with a turbulent flow : randomness, broad energy spectrum, etc. [...] One mechanism, of course, is the well-known `cascade to higher wavenumbers', which fills out the high-wavenumber end of the spectrum of eddy sizes. We visualize this mechanism as connected with internal instabilities (internal to the larger eddy), rather than breakdown of the eddy into smaller `pieces' (quite the opposite process is at work). The sort of thing we have in mind is illustrated in some pictures obtained by Pierce \cite{Pierce 1961} of small-scale instabilities on vortex layers which have rolled up into a large structure. Three-dimensional vortex stretching effects would also fall into this category of internal instabilities. [...] On the important problem of the properties and role of the large coherent structures, there is much still to be done. The recent work of Winant and Browand \cite{Winant and Browand 1974} describes the crucial process of amalgation of the structures and its relation to the growth of the layer. More understanding of the larger-scale interactions resulting from those events, and their relation to the turbulent character of the flow, is needed.} \cite{Brown and Roshko 1974}. I find it shocking that even today many professors continue to explain the turbulent cascade by the fracturing of vortices into smaller and smaller ones, according to Richardson's 1922 parody of Swift's poem about fleas \cite{Richardson 1922}, probably as a joke he was making with a fine British humour. Indeed, as early as 1974, eminent experimentalists, like Kline, Brown, Roshko and Dimotakis, as well as an outstanding theorist, Kraichnan \cite{Kraichnan 1974}, had provided clear arguments against `vortex breaking', but they were unfortunately preaching in the wilderness... This interpretation has always seemed absurd to me from a mechanical point of view. Indeed, a fluid vortex cannot break up like a solid stone, but rather deforms when subjected to stress; nor can the turbulent cascade be considered as a gear system, which needs a gap between the rotating parts to function. I consider the marketing and branding of ideas by a few over-ambitious researchers to be as misleading, and therefore dangerous, as those to which we are subjected under commercial pressure...

\medskip

The results of Brown and Roshko \cite{Brown and Roshko 1974} were confirmed two years later by Dimotakis and Brown for a mixing layer at Reynolds $3 \times 10^6$ \cite{Dimotakis and Brown 1976}. Ten years later, Roshko discovered that the coherent structures he had observed in the turbulent planar mixing layer destabilise themselves into secondary vortices organised in the direction of flow, and thus contribute to increasing the mixing properties of such a turbulent flow \cite{Bernal and Roshko 1986}. In 1975 an excellent review on `Coherent structures in turbulence' was published by Davies, who stated that: {\it `Recently, however, coherent structures have been clearly observed in flows that otherwise display all the characteristics of fully developed turbulence [...]. In the light of current knowledge it seems reasonable, therefore, to regard turbulence as an assembly of repetitive ordered structures which interact strongly with each other as they travel downstream. Point velocity or other measurements, normally made at fixed points, consist of a time history record obtained from a succession of many such structures passing the measuring stations. Since motions within those structures which are relatively remote from each other become statistically independent, such time history records display the statistical characteristics of a random process.[...] It should be clearly understood that the organized structure of such concentrated regions of vorticity differs in many respects from that of the turbulent eddy or the wave models of statistical theories of turbulence'}. The same year, in 1975, Laufer published another review, entitled {\it `New Trends in Experimental Turbulence Research' }\cite{Laufer 1975}, where he confirmed the essential role that the laboratory experiments of Kline and Roshko have played to change our view about turbulence: {\it `In the past ten years two important observations were reported that had significant impact on subsequent turbulence research (Kline \& Runstadler 1959 \cite{Kline and Runstadler 1959}, Brown \& Roshko 1971 \cite{Brown and Roshko 1971}). Ironically, these were made not by sophisticated electronics instrumentation, but visually with rather simple optical techniques. The essence of these observations was the discovery that turbulent flows of simple geometry are not as chaotic as has been previously assumed: There is some order in the motion with an observable chain of events reoccurring randomly with a statistically definable mean period. This surprising result encouraged researchers to reexamine the line of inquiry for designing their experiment, and they began seriously questioning the relevance of some of the statistical quantities they had been measuring. It was soon realized, for instance, that retaining some phase information in the statistics and obtaining more detailed spatial information are essential for a quantitative explanation of the visual observations. This of course became possible only with the rapidly developing computer techniques of today. Considererable progress has been made, especially in the utilization of digital techniques, which are proving to be most useful in the study of the quasi-ordered motion.'} \cite{Laufer 1975}. 

\medskip

Actually it is possible to topologically characterise coherent structures via the stress (velocity gradient) tensor, whose symmetric part corresponds to the irrotational strain, while its antisymmetric part corresponds to the rotation undergone by a fluid element. Its eigenvalues allow the flow to be separated into different regions where Lagrangian dynamics are different. In two dimensions, there are two types of regions which are either elliptic or hyperbolic \cite{Weiss 1991}. Indeed, imaginary eigenvalues correspond to elliptic regions where rotation dominates strain and for which fluid trajectories stay close, which is characteristic of coherent vortices. Real eigenvalues correspond to hyperbolic regions where strain dominates rotation and for which fluid trajectories separate exponentially, characterising the hyperbolic stagnation points of the background flow. In three-dimensional flows, where vortex stretching plays a key role, there are more types of topological regions. Unfortunately, the classical theory of turbulence is blind to the presence of coherent structures, because they are advected by the flow in a homogeneous and isotropic random fashion. They are also highly unstable  and their time and space support may be very small. Consequently, their presence only affects high-order velocity structure functions ({\it i.e.}, moments of order $p$ of the velocity difference evaluated at two points in space separated by a constant distance). They have been measured only in the 1990s and the results contradict  Kolmogorov's theory, which predicts a linear dependence, with a slope $1/3$, between the scaling exponent of the velocity structure functions and their order. Moreover, it has been found that there are two distinct non-linear dependences for odd and for even order structure functions \cite{Van den Water 1993}. Actually Kolmogorov's prediction is only observed for second-order structure functions but deviates for $p>2$. Indeed, one should not forget that the `turbulent cascade' is only a hypothesis of Kolmogorov's theory and that the observed broadband energy spectrum of turbulent flows can be explained without this hypothesis. 

\medskip

\subsubsection{Dynamic interpretations}
\label{subsubsec:2.3.2}

Another deterministic point of view, which we will call `dynamic', is based on the assumption that the non-linear dynamics of turbulent flows generate quasi-singularities in physical space, tending towards point vortices and vorticity filaments in two-dimensional flows, or towards vortex tubes and vorticity sheets in three-dimensional flows. This dynamic view was advocated by Kraichnan in an article written in 1974 \cite{Kraichnan 1974} where he explained: {\it `The stretching mechanism has led a number of authors to conjecture that the small-scale structure should consist typically of extensive thin sheets or ribbons of vorticity, drawn out by the stirring action of their own shear field. In this picture, the randomness lies in the distribution of thickness and extension of the thin sheets and ribbons, and in the way they are folded and tangled through the fluid. A typical small-scale structure is thought to be small in one or two dimensions only, not in the third'} \cite{Kraichnan 1974}. His last remark implies that if a turbulent flow produces such {\it `thin sheets or ribbons of vorticity'}, the enstrophy spectrum and the energy spectrum of one realisation averaged over all angles is broadband. Therefore, if it is the case for one flow realisation this is {\it a fortiori} also true for an ensemble of such realisations. 

\medskip

The dynamic interpretation of the energy spectrum was developed in the 1970s by Saffman for two-dimensional turbulence \cite{Saffman 1971}, and in the 1980s by Lundgren for three-dimensional turbulence \cite{Lundgren 1982}. In 1971 Saffman explained that the non-linear dynamics of two-dimensional turbulent flows tend to form vortex patches with velocity jumps with along lines, that give an energy spectrum scaling in $k^{-4}$. In 1982 Lundgren assumed that the non-linear dynamics of three-dimensional turbulent flows tend to form vortex tubes rolling up into spirals, in such a way that the energy spectrum scales as $k^{-5/3}$. These ideas were further developed by Gilbert \cite{Gilbert 1988} and Moffatt \cite{Moffatt 1993}, who suggested that the scaling behaviour of the energy spectrum for two-dimensional turbulence can be explained by the wind-up of a weak vortex by a neighbouring strong vortex, which produces spiral structures of vorticity filaments and leads to a power-law of the energy spectrum with an exponent between $-3$ and $-4$. Therefore, Saffman's deterministic prediction of a $k^{-4}$ scaling \cite{Saffman 1971} and Kraichnan's statistical prediction of a $k^{-3}$ scaling \cite{Kraichnan 1967} could be reconciled as two limiting cases. In fact, most of direct numerical simulations of two-dimensional turbulence, computed by integrating the two-dimensional Navier-Stokes equations without any {\it ad hoc} turbulence model, predicted a power-law energy spectrum whose exponent which is close to $-4$ or even steeper, thus larger than the $-3$ exponent predicted by Kraichnan's statistical theory of two-dimensional turbulence {Kraichnan 1967}. 

\medskip

In 1991, Farge and Holschneider \cite{Farge et al. 1991} suggested that spiral vorticity filaments generate cusp-shaped vortices, because the viscosity of the fluid smoothes the filaments and glues them together. Moreover, viscosity regularises their cores and vorticity thus remains bounded, in contrast to singular vortices, {\it e.g.}, point vortices therefore they called them `quasi-singular vortices'. They also proved that if vorticity locally scales as $r^{-1/2}$, with $r$ the distance to the vortex core, the energy spectrum scales in $k^{-4}$, as predicted by Saffman \cite{Saffman 1971} for two-dimensional homogeneous isotropic turbulence. Both predictions were deterministic and gave the same exponent of the energy spectrum, but they differed in the non-linear dynamics producing quasi-singular structures in physical space: Saffman \cite{Saffman 1971} assumed velocity jumps along lines, while Farge and Holschneider \cite{Farge et al. 1991} proposed cusp-shaped vortices. Farge and Holschneider \cite{Farge et al. 1991} also conjectured that such vortices emerge out of an initial random vorticity distribution by an inviscid non-linear instability, such as the Kelvin-Helmholtz instability, which accretes vorticity onto the strongest singularities of the initial random distribution (corresponding to the tails of the vorticity probability distribution function); the same mechanism could also explain the formation of vortices at the wall. In 1992 they have shown that these cusp-like quasi-singularities remain stable under the Navier--Stokes dynamics \cite{Farge et al. 1992a}, and that the strain they impose in their vicinity controls the flow and inhibits further instability that otherwise would develop there \cite{Kevlahan et al. 1997}. If we extrapolate previous results to very large or infinite Reynolds numbers, we might guess that finite time singularities exist as solutions of Navier--Stokes equation, but this is still an open question, which is one of the seven `Millennium Prize Problems' \cite{Fefferman 2006}. In 1982 Caffarelli, Kohn and Nirenberg \cite{Caffarelli et al. 1982} proved that, if finite time singularities would exist as solutions of the three-dimensional Navier--Stokes equation, their space-time support should be of measure one. Therefore, if singularities or quasi-singularities exist, they can only be rare events (because if their time support tends to one, their space support tends to zero), which can hardly be detected using standard statistical methods, {\it e.g.}, two-point correlations, since low-order statistics are insensitive to rare events.

\medskip

\subsubsection{Numerical experiments}
\label{subsubsec:2.3.3}

The advent of the first computer Z1, built by Zuse in Germany in 1938, allowed the development of numerical experiments, {\it i.e.}, the experimental study of the solutions of equations which cannot been found by analytical methods, as it is the case for the Navier--Stokes equation as soon as the flow becomes turbulent. In an article published in 1986 \cite{Farge 1986} (English version \cite{Farge 2007}), I explained why numerical experimentation is for me complementary to theory and laboratory experimentation, as a new way of doing well thought experiments. The first researchers who have foreseen such a role that computers would  play in mathematics and physics were von Neumann and Ulam. The latter recounts in his autobiography \cite{Ulam 1976} that: `{\it Almost immediately after the war Johny and I also began to discuss the possibilities of using computers heuristically to try to obtain insights into questions of pure mathematics. By producing examples and by observing the properties of special mathematical objects one could hope to obtain clues as to the behavior of general statements which have been tested on examples. [...] In the following years in a number of published papers, I have suggested and in some cases solved a variety of problems in pure mathematics by such experimenting or even merely observing'} \cite{Ulam 1976}. In 1955, Fermi, Pasta and Ulam \cite{Fermi 1965} carried out an illuminating numerical experiment by studying a weakly non-linear interacting system of particles. They were surprised to discover that its evolution did not lead to equipartition of energy and maximum entropy, as they expected, but exhibited a quasi-periodic behaviour, which was in contradiction with the ergodic hypothesis. In 1950, the first numerical simulation of a turbulent flow was made by Charney, Fj\''ortoft, Smagorinsky and von Neumann, who designed a model to predict the evolution of the atmosphere of the whole Earth during one day. Their model was based on the two-dimensional barotropic equation, which assumes the atmosphere is stably stratified and in quasi-geostrophic equilibrium. With the help of von Neumann's wife, they computed it on the ENIAC, a computer built in 1946 by Eckert and Mauchly at the Ballistic Research Laboratories of Aberdeen in Maryland, and compared the velocity and pressure fields they had computed for 5, 30, 31 January and 13 February 1949 with those actually observed on those days. They found large forecast errors and concluded that these were due to the overly restrictive hypotheses of their model and the coarse resolution of their computational grid of size 736 km $\times$ 736 km. In 1961, they published their results in an article \cite{Charney et al. 1961} where they noted that: {\it`It may be of interest to remark that the computation time for a 24-h forecast was about 24h, that is, we were just able to keep pace with the weather. However, much of this time was consumed by manual and I.B.M. operations, namely by reading, printing, reproducing, sorting, and interfiling of punch cards. In the course of the four 24-h forecasts about 100,000 standard I.B.M. punch cards were produced and 1,000,000 multiplications and divisions were performed'} \cite{Charney et al. 1961}. 

\medskip

In 1974, Orszag and Israeli published an article entitled {\it `Numerical simulation of viscous incompressible flows'}, where they stated that: {\it  At this time, numerical simulation remains an art. As such, much depends on the skill of the artist. Nevertheless, there has been progress toward making simulation a science [...] There are at least three classes of numerical artists. First, there are those who simulate exceedingly complicated flows that push the state-of-the-art and the machines to their capacity (or perhaps beyond). In support of such simulations, there is often urgency in the gathering of data in order to meet some pressing practical problem. In the second group, there are those who continually test and compare numerical schemes, always looking for the "perfect" scheme but hardly ever getting results of genuine fluid-dynamical interest. [...] The third group includes those who use accurate flow simulation codes to produce computer results that are both reliable and hard to achieve in the laboratory. These worthy few use the computer to extend basic fluid-dynamical knowledge. [...] We believe that the third level of numerical experimentation should be the goal of computational fluid dynamics. The principal role of computers in fluid dynamics should be to give physical insight into dynamics'} \cite{Orszag and Israeli 1974}. In 1976, Reynolds published a remarkable review article entitled {\it `Computation of Turbulent Flows'} \cite{Reynolds 1976}, where he explained that: {\it `The computation of turbulent flows has been a problem of major concern since the time of Osborne Reynolds. Until the advent of the high-speed computers, the range of turbulent-flow problems that could be handled was very limited. The advances during this period were made primarily in the laboratory, where basic insights into the general nature of turbulent flows were developed, and where the behaviors of selected families of turbulent flows were studied systematically. For the engineer there were only a limited number of useful tools such as boundary-layer prediction methods based on the momentum-integral equation with a high empirical content. Features such as sudden changes in boundary conditions, separation, or recirculation could not be predicted by these early methods with any degree of reliability. Very specific empirical work remained an essential ingredient of any engineer's analysis. Midway through this century computers began to have a major impact. First it became possible to handle more difficult boundary layers by complex integral analyses involving several first-order ordinary differential equations. By the mid-1960s there were several workers actively developing turbulent-flow computation schemes based on the governing partial differential equations. The first such methods used only the equations for the mean motions, but second-generation methods began to incorporate turbulence partial differential equations'} \cite{Reynolds 1976}. In an article published in 1974 Leonard had proposed a new method to compute turbulent flows, that he called `Large Eddy Simulation (LES)', which in Reynolds' review was presented as very promising. Leonard justified the motivation of his new method in the following way: {\it `Numerical simulation of all the scales of a turbulent flow, even at modest Reynolds numbers, is generally not practical; however, most information of interest can be obtained by simulating the motion of the large-scale, energy containing eddies'} \cite{Leonard 1975}. The principle of the LES numerical simulation is to filter the Navier--Stokes equations in order to deterministically calculate the scales of each realisation of the turbulent flow up to a predefined cut-off scale, which corresponds to the size of the computational grid, and to statistically model the effect of the discarded subgrid scales on the resolved scales. Leonard explained that: {\it `The large-scale fluctuations satisfy the filtered or averaged momentum and continuity equations. Averaging the nonlinear advection term yields two terms; one is the Reynolds stress contribution from the subgrid-scale turbulence, and the other is the filtered advection term for the large scales. In some models, the energy cascade is viewed solely as an energy loss of the large-scales because of an artificial viscosity arising from subgrid-scale motions. However, in most cases of interest, motions on the order of the dissipation length scale cannot be treated explicitly, and modifications of the Navier-Stokes equations must be introduced to simulate properly the energy cascade. Noting that the large-scale motions vary in a non negligible way over an averaging volume'} \cite{Leonard 1975}. 

\medskip

In the 1970s the arrival of the first vector supercomputer, the Cray 1 delivering up to $100$ MFLOPs ($10^8$ floating point operations per second), the first multiprocessor supercomputer, the Cray X-MP, and the first network-available parallel supercomputer, the ILLIAC-IV, allowed the rapid development of LES simulations. The first numerical experimentation of a three-dimensional turbulent channel flow were made at Stanford University in California by Kim and Moin, using the Large Eddy Simulation (LES) method and the ILLIAC-IV of the nearby NASA-Ames Research Center in Moffett Field. In their article published in 1982 they explained that: {\it `Fully developed turbulent channel flow has been simulated numerically at Reynolds number $13 800$, based on centre-line velocity and channel half-width. The large-scale flow field has been obtained by directly integrating the filtered, three-dimensional, time-dependent Navier--Stokes equations. The small-scale field motions were simulated through an eddy-viscosity model. The calculations were carried out on the ILLIAC I V computer with up to $516 096$ grid points. The computed flow field was used to study the statistical properties of the flow as well as its time-dependent features. The agreement of the computed mean-velocity profile, turbulence statistics, and detailed flow structures with experimental data is good. The solvable portion of the statistical correlations appearing in the Reynolds-stress equations are calculated. Particular attention is given to the examination of the flow structure in the vicinity of the wall'} \cite{Moin and Kim 1982, Moin and Kim 1985}. The visualisations of their results have had a decisive impact in proving that coherent structures generically appear in the solutions of the Navier-Stokes equations in the strong turbulence regime and that they play an essential dynamic role. In 1984, Rogallo and Moin published a review article on the `Numerical simulation of turbulence' \cite{Rogallo and Moin 1984}, where they explained why the LES approach was useful: {\it `When the scale range exceeds that allowed by computer capacity, some scales must be discarded, and the influence of these discarded scales upon the retained scales must be modeled. We shall distinguish between completely resolved and partly resolved simulations by referring to them as "direct" and "large-eddy" (LES), respectively [...] The attraction of direct simulation is that it eliminates the need for ad hoc models, and the justification often advanced is that the statistics of the large scales vary little with Reynolds number and can be found at the low Reynolds numbers required for complete numerical resolution. This approach has been successful for unbounded  flows where viscosity serves mainly to set the scale of the dissipative eddies, but it has not been successful for wall-bounded  flows, such as the channel  flow, where computational capacity has so far not allowed a Reynolds number at which turbulence can be maintained. This is typical of many  flows of engineering interest and forces the development of the LES approach}' \cite{Rogallo and Moin 1984}. 

\medskip

As is rightly pointed out by Rogallo and Moin \cite{Rogallo and Moin 1984}, in the 1980s, Direct Numerical Simulation (DNS) of strong turbulence was out of reach for three-dimensional flows, even with the fastest supercomputers available in those years. In contrast, it was possible for two-dimensional flows for which much higher Reynolds numbers were accessible, because in two dimensions the ratio between the largest and smallest resolved scales is higher than in three dimensions for the same computing power. Thus the numerical experiment using DNS became a very useful tool for studying atmospheric dynamics, because at mid-latitudes the atmospheric flow is mostly two-dimensional by the thinness of the atmosphere compared to its extent, and by the rotation of the Earth. In addition, the Earth's curvature can be neglected, allowing the atmospheric flow to be calculated in a tangent plane without walls. The atmospheric dynamics is then simulated by solving the two-dimensional Navier-Stokes equations with periodic boundary conditions using a pseudo-spectral numerical method, as proposed by Orszag in 1969 \cite{Orszag 1969, Orszag 1970a, Orszag 1970b}, which allowed to study the evolution of cyclones and anticyclones. In 1984, McWilliams published an article, entitled {\it `The emergence of isolated coherent vortices in turbulent flow'} \cite{McWilliams 1984}, where he showed how ubiquitous are the coherent vortices which emerge out of an initially random two-dimensional turbulent flow computed by DNS in a periodic domain. He explained that: {\it `A variety of examples have been presented of the emergence and persistence of isolated concentrations of vorticity in turbulence flows. These constitute a primafacie case for the considerable commonness of the occurrence'} \cite{McWilliams 1984}. In the 1990s, other impressive numerical experiments using high-resolution DNS have been made by Moser and Rogers at NASA-Ames Research Center, who computed the evolution of a plane mixing layer separating two uniform streams of differing speed \cite{Rogers and Moser 1992, Moser and Rogers 1993}. They showed how two primary spanwise rollers, produced by the Kelvin-Helmholtz instability, pair together and enhance the mixing property of such an inhomogeneous turbulent flow. Since then, numerous numerical experiments carried out by DNS have shown that in turbulent flows, both two-dimensional and three-dimensional, vorticity tends to condense into coherent vortices that form during the flow evolution and seem to have their own dynamics and interaction laws, almost independently of the background flow in which they evolve. Although there is still no agreed definition of coherent vortices, they can be defined as local concentrations of vorticity, where rotation is stronger than deformation. They either emerge from random initial conditions, or form in boundary and shear layers by non-linear instability. Moreover, they have their own dynamics, namely the merging of vortices of the same sign, the linking of vortices of opposite sign and the emission of vorticity filaments when strongly deformed. In two dimensions, due to the orthogonality between the vorticity and velocity gradient vectors, there is no vortex stretching, whereas in three dimensions vortices can be stretched by velocity gradients, thus producing more vorticity.

\bigskip
\bigskip


\section{Inadequacies of current theories}
\label{sec:3}

After more than a century of research on turbulence \cite{Reynolds 1883}, no single convincing theoretical explanation has given rise to a consensus among engineers, physicists and mathematicians. There exists a large number of {\it ad hoc} models, called `phenomenological', widely used by engineers to solve industrial applications where turbulence plays a role, but many parameters of those turbulence models cannot be derived from first principles and must be determined by performing experiments in wind tunnels or water tanks. In fact, it is still not known whether strong turbulence has the assumed universal behaviour (independent of initial and boundary conditions) in the limit of infinitely large Reynolds numbers and infinitely small scales. Our understanding of turbulence is impaired by the fact that we have not yet identified the `proper questions' to ask, nor the `appropriate objects', namely the structures and elementary interactions from which it would be possible to construct a satisfactory statistical mechanics, or kinetic theory, of strong turbulence. Ignorance of the elementary physical mechanisms at work in turbulent flows arises in part from the fact that:

\begin{itemize}
\item
 we do not take into account coherent vortice, because we use two-point correlation functions, 
\item
we think in terms of Fourier modes that are delocalised,
\item
 that we consider $L^2$-norms, such as energy and enstrophy, instead of higher-order norms. 
\end{itemize}

\noindent
Our present lack of understanding of the dynamics of coherent vortices arises from several reasons:

\begin{itemize}
\item
We focus on the velocity field, which depends on the inertial frame we choose, and not on the vorticity field, which is frame-independent and therefore preserves the Galilean invariance. 
\item
We study the flow evolution in a Eulerian frame of reference, instead of a Lagrangian frame attached to each fluid particle. It would be more appropriate to follow the evolution of vorticity or of circulation, because vortex tubes are advected by the flow and motion is conserved in the absence of dissipation (Helmholtz laws and Kelvin's circulation theorem).
\item
Traditionally we perform one-point measurements and two-point correlations which are insensitive to intermittent and rare features such as coherent vortices.
\item
The infinite time limit, which is usual in statistical physics is not appropriate for many applications of turbulence, where we are interested in the transient evolution and look for real-time methods to reduce or enhance turbulence, {\it e.g.}, to accelerate mixing. In  particular, the infinite-time limit is irrelevant in many meteorological situations where one cannot guarantee the statistical stationarity nor the time decorrelation (Markov process hypothesis) of the external forcing. 
\item
`Real life' problems are bounded in space, time and scale. Therefore, we should rather focus on statistically unsteady and inhomogeneous turbulent flows than on statistically steady and homogeneous turbulent flows, because the latter are so ideal that they do not exist in reality. Unfortunately, theoretical physicists and mathematicians rarely take into account solid boundaries, or initial conditions with the flow at rest, or scale cutoffs, because these conditions do not match the homogeneity, stationarity and self-similarity conditions they use to simplify the problems they wish to solve.
\item
Due to Kolmogorov's theory we only focus on the modulus of the energy spectrum, but not on its phase, and so we lose track of the spatial coherence which is coded in the phase of all Fourier modes but not in their modulus.
\item
When we develop numerical simulations, as a combination of a Navier--Stokes solver and a turbulence model, to compute fully-turbulent flows, we do not address the non-linear problem {\it per se}. Only direct numerical simulations do so because they solve the Navier--Stokes equation without any turbulence model, but reaching the strongly turbulent regime for three-dimensional flows requires tremendous computational means and only a few flow realisations can thus be obtained. Nor do we neither address the fact that we do not have a statistical equilibrium since we only compute one realisation of the flow. On the contrary, one assumes either a linear behaviour of the sub-grid scale motions (in the case of direct numerical simulation), or the existence of a scale separation, which assumes a statistical equilibrium for the sub-grid scale motions (in the case of large eddy simulation). This last point was already noticed in 1974 by Kraichnan when he wrote: {\it `Our basic point is that the inertial-range cascade represents strong statistical disequilibrium. This carries two implications. First, that analogies with equilibrium and near-equilibrium phenomena are unjustified. Second, that the structure of the inertial range depends on the actual magnitude of the coefficients coupling the degrees of freedom and not just on their overall symmetry and invariance properties. This is because cascade is a transport process and the coefficient magnitudes affect the rate of transport'} \cite{Kraichnan 1974}.
\end{itemize}

\bigskip

An open question to address is: what is the importance of coherent vortices for mixing and turbulence? Do they play an essential role which should be taken into account in our models, or can we neglect them? We should eliminate the current misconception which relates coherent vortices to small wavenumbers (misleadingly named `large eddies') of turbulent flows. This erroneous view arises from the fact that one tries to recover some dynamic picture from averaged quantities which have already lost track of the spatial and temporal coherence which characterises coherent vortices. In particular the energy spectrum in the inertial range is dominated by the background and not by the coherent vortices, because their space and time support is too small to have a sufficient weight in the integral when one computes second-order structure functions. This is not true for high-order structure functions and actually the presence of coherent vortices may explain the departure from Kolmogorov's prediction observed for high-order structure functions. On the contrary, when one considers the probability distribution function of vorticity, one finds that its non-Gaussian shape arises from the coherent vortices, which are responsible for its heavy tails. This is due to the fact that, although coherent vortices are quite rare in space and time, they are present in any realisation of a turbulent flow. The formation of coherent vortices is probably a far-reaching consequence of the incompressible Navier--Stokes dynamics, we should clarify. To answer the previous question concerning the role of coherent vortices in turbulent flows, we would need a clear definition, still lacking, of what they are and an appropriate method to extract them. 

\bigskip

The statistical  probabilistic approach may be too abstract and its link to experimental observations are difficult to ascertain in most cases. In order to compare its predictions with laboratory or numerical experiments, one has to check that ensemble averages converge to time or space averages, and therefore satisfy the ergodic hypothesis. One assumes that there is only one attractor which satisfies Sinai-Bowen-Ruelle conditions (for almost all initial conditions under which the time average exists and is unique) and that the observed turbulent flow has visited all possible phase-space configurations compatible with this attractor. Therefore, due to our limited understanding of turbulence, it has become increasingly important to perform many well-controlled experiments to gain better insight and suggest new models to describe the behaviour of high Reynolds number turbulent flows. There are two kinds of experimental approaches, each with its own limitations. Firstly, in laboratory experiments it is easy to measure one-, two- or many-point correlations and to accumulate long time statistics, but one cannot today directly measure the instantaneous spatial distribution of velocity and vorticity; for this experimentalists use indirect methods based on many particles ({\it e.g.}, with Particle Image Velocimetry) or a dye transported by the flow. Secondly, in numerical experiments, it is easy to measure the temporal evolution and spatial distribution of the velocity and vorticity fields for a given realisation of the flow, but it is still out of reach to compute ensemble averages from many evolutions of different realisations of the same turbulent flow, as it is necessary to ensure statistical convergence. The two approaches are in fact  complementary: laboratory experiments allow us to perform statistical analysis, while numerical experiments allow us to perform dynamic analysis. Unfortunately, it is therefore difficult to compare them. For instance, the statistical analysis deals with averages, describes turbulent flows in terms of fluctuations (often called  `turbulent eddies') and uses random functions and probability measures, while the dynamic analysis considers each flow realisation {\it per se}, describes the flow in terms of interacting coherent vortices, and deals with non-random functions or distributions. Much confusion in our understanding of turbulence is due to the fact that we try to retrieve dynamic insight from statistical averages, and statistical information from a single flow realisation. Moreover, the statistical analysis relies on the Fourier spectral representation, while the dynamic analysis relies on the spatial representation. We must be aware that we cannot reconcile these two representations, unless we use basis functions which are localised in both physical and wavenumber space, such as wavelets or wavelet-packets.

\bigskip

Kolmogorov's statistical theory of homogeneous isotropic turbulence is the simplest possible universal theory (simple in the sense of `Occam's razor' or Aristotle's logical simplicity principle). It is verified for second-order moments, two-point correlation and $L^2$-norm, but it fails to correctly predict  higher-order moments, $n$-point correlations and $L^p$-norms with order $p > 2$. We think that coherent vortices may explain this discrepancy and that their role is essential for understanding turbulence. Therefore we need to find another theoretical setting in which coherent vortices can be considered as building blocks of turbulent flows. Therefore we would like to construct a statistical mechanics of turbulent flows based on coherent vortices, but we still do not know what should be the appropriate invariant measure for this purpose. In any case, Kolmogorov's prediction for the two-point velocity correlation, second-order moment and energy will always be satisfied in the limit of infinite Reynolds number, because the weight of coherent vortices in those integrals becomes negligible in this limit. However, this is no longer true as one measures $n$-point correlations, higher-order moments and $L^p$-norms, because the contribution of coherent vortices in those integrals becomes increasingly significant as the order $p$ increases. In this picture dissipation results from the non-linear interactions between coherent vortices which produces incoherent enstrophy due to strong mixing, and therefore irreversibility and high entropy. The larger the Reynolds number, the more local in physical space (and therefore more non-local in spectral space) dissipation will be. Note that the Kolmogorov dissipative wavenumber is only an averaged quantity; we have conjectured that its variance in space is large and depends on the flow intermittency \cite{Farge et al. 1990b}, \cite{Farge 1992}. According to this picture, universality seems to be lost, because the density of coherent vortices depends on the initial conditions and on the forcing. But there may be a universal way of describing turbulent flows as the superposition of a coherent flow, made of vortices with a quantified amount of enstrophy which is dynamically active, and of a background incoherent flow, which is passive and can be seen as a thermal bath affecting only the coupling between coherent vortices. The prediction of Kolmogorov's theory may be verified only for the incoherent background flow which is homogeneous, Gaussian and well-mixed, while the coherent flow is not.

\bigskip

Today, we believe that the theory of strong turbulence is still in a pre-scientific phase, as we do not yet have an equation, nor a set of equations, that could be used to efficiently compute fully-developed turbulent flows from first principles. The Navier--Stokes equation is appropriate to study laminar flows and the transition to turbulence, but it is not useful to compute strongly turbulent flows, because its computational complexity becomes intractable when the Reynolds number is too large. Indeed, the number of degrees of freedom necessary to compute a turbulent flow by direct numerical simulation is proportional to $Re^{9/4}$ (where $Re$ is the Reynolds number) \cite{Liepmann 1979, Okamoto et al. 2007}. Thus, to compute the aerodynamic properties of an aircraft, say an Airbus 380 with a Reynolds number of the order of $10^9$, one has to solve of a linear system of equations of the order of $10^{20}$. For such strongly turbulent flows we hope to be able to define new averaged quantities, which would become the appropriate  observables to describe them, and to find the corresponding  transport equation to compute the evolution of these averaged quantities. Indeed, just as the Navier--Stokes equation can be derived from the Boltzmann equation by considering appropriate limits (Knudsen and Mach numbers tending to zero), appropriate averaging procedures to define new coarse-grained variables (velocity and pressure) and the associated transport coefficients (viscosity and density), the equation governing strong turbulence should be derived  as a further step in this hierarchy of embedded approximations. Unfortunately, the appropriate parameters are easier to define when going from the Boltzmann equation to the Navier--Stokes equation than from the Navier--Stokes equation to the `strong turbulence equation' if any. In the first case, only a linear averaging procedure, namely coarse-graining based on some known statistical equilibrium distribution of velocities, is required. In the second case, we must find an appropriate non-linear procedure, namely some conditional averaging yet to be defined. To do this, it is necessary to identify the dynamically active coherent structures that drive the evolution of the strong turbulence, characterise them and describe their elementary interactions. We should then design a non-linear filter \cite{Farge et al. 1992b, Farge and Phillipovitch 1993, Farge et al. 2001} to extract those elementary structures out of strongly turbulent flows. Unfortunately we may not be able to use classical statistical methods, as turbulent flows are not in statistical equilibrium and their statistics are not stationary. On the other hand, the residual background flow is sufficiently mixed to guarantee the ergodicity, stationarity and homogeneity required by Kolmogorov's theory, which can then be used to model the incoherent background flow. To conclude, I believe that the future of turbulence research will be a combination of both deterministic and statistical approaches. The deterministic approach will be needed to compute the evolution of the low-dimensional dynamical system which corresponds to the elementary structures out of statistical equilibrium, namely the coherent vortices. The statistical approach will also be needed to model the incoherent background flow, by finding an equivalent stochastic process which has the same statistical behaviour.



\begin{figure}[ht!]
\centering
\vspace{3cm}
\includegraphics[width=1.0\linewidth]{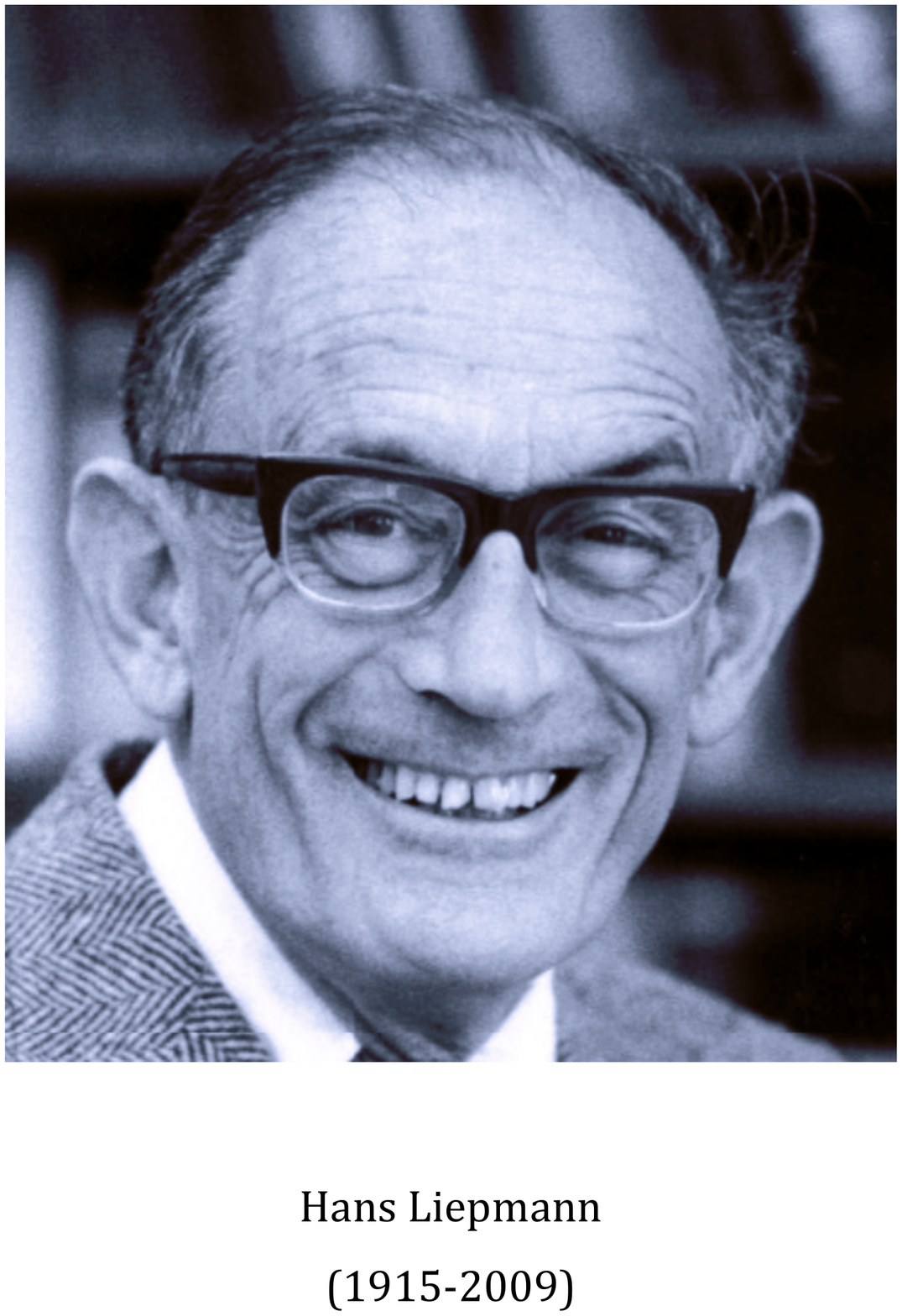}\\
  \nocaption
  \vspace{-0.5cm}
  \centerline{Hans Liepmann in his office at GALCIT, Caltech, Pasadena.}
  \label{Figure_Hans}
\end{figure}

\bigskip

Hans Liepmann, former Von Karman professor of aeronautics at Caltech, was fond of pointing out that in turbulence research the {\it `theory has been accused of acting like the well-known drunk who looks for his lost keys under the street light rather than in the dark alley where he lost them'} \cite{Liepmann 1979}. Indeed, researchers who study turbulence know well that it is related to the production of vortices in boundary layers or shear layers and their interactions, but as they do not have a sufficient theoretical grasp of their structure and dynamics they prefer to use the probabilistic statistical formalism of Kolmogorov. This forces them to accept several hypotheses (ergodicity of the flow, stationarity, homogeneity and isotropy of the statistics, energy production at low-wavenumber and dissipation at high-wavenumber, existence of an inertial range, {\it i.e.}, an intermediate wavenumber range where energy is conserved) and provides them powerful technics to predict the spectral distribution of the energy modulus in the inertial range for an ensemble of flow realisations. But Hans Liepmann considered that this is not the key needed to understand turbulence, because it does not capture the elementary dynamic features as observed in physical space. Moreover, it does not enable us to compute from first principles the evolution of a given turbulent flow realisation, nor to understand the near-wall dynamics. In the proceedings of the `1st International Conference on Turbulence' held in 1961 in Marseille Hans Liepmann made the following remark: {\it `It is clear that the essence of turbulent motion is vortex interactions. In the  particular case of homogeneous isotropic turbulence this fact is largely masked, since the vorticity fluctuations appear as simple derivatives of the velocity fluctuations. In general this is not the case, and a Fourier representation is probably not the ultimate answer. The proposed detailed models of an eddy structure represent, I believe, a groping for an eventual representation of a stochastic rotational field, but none of the models suggested so far has proven useful except in the description of a single process'} \cite{Liepmann 1962}. Twenty-five years later, during an international workshop on `Dynamical Systems and Statistical Mechanics Methods for Coherent Structures in Turbulent Flows' I organised at UC Santa Barbara in February 1997, Hans Liepmann gave us the following guidelines: {\it `Kolmogorov's theory has been counter-productive. It is OK for light or sound scattering by turbulent flows, but it is not useful for the main lines of turbulence. [...] In turbulence you have long range forces, and it is difficult to extrapolate from  Boltzmann's gas, which has short range forces. Therefore I am uneasy about Reynolds equations. [...] As long as we are not able to predict the drag on a sphere or the pressure  drop in a pipe from continuous, incompressible and Newtonian assumptions, without any other complications (namely from first principles), we will not have made it!'}. In 1999 he developed the same view in a letter to George Batchelor where he wrote:  {\it `To me the crux of the turbulence problem are the fluxes of momentum, energy and mass. To put it more drastically: as long as we cannot predict pipe flow or the drag of a sphere for all Reynolds numbers we have no real theory'} \cite{Narasimha et al. 2013}. I recommend that you read Hans Liepmann's article `The Rise and Fall of Ideas in Turbulence', published in 1979 in the `American Scientist' \cite{Liepmann 1979}: it is the best review I know of to describe the state of research on turbulence and to clearly formulate the questions that need to be answered; moreover, all of his criticisms and recommendations are still relevant today. For those who read French I also highly recommend the article that Patrick Chassaing published in 2019 in the `Comptes-Rendus de l'Acad\'emie des Sciences de Paris', entitled {\it La notion de moyenne en turbulence} \cite{Chassaing 2019}.

\bigskip
 
In 1961, the first international colloquium devoted to turbulence (TCM-1961) was organised in Marseille by Favre, on the occasion of the inauguration of his `Institut de M\'ecanique Statistique de la Turbulence'. It was an exceptional international event, the proceedings of which were published by the CNRS in a book \cite{TCM 1962a}, and some of the debates were even reported in the general press \cite{TCM 1962b}. Indeed, Favre managed to bring together in Marseille the main representatives of both probabilistic statistical theories, with Kolmogorov, Obukhov, Millionshchikov, Yaglom, Batchelor, Kamp\'e de F\'eriet, Kraichnan and Lumley, and the main representatives of the deterministic theories, with Liepmann, Roshko, Laufer, Von Karman, Taylor, Saffman and Moffatt.  This `1st Turbulence Colloquium Marseille 1961' had a strong impact on turbulence theory; for example, it was on this occasion that Kolmogorov corrected his theory of isotropic homogeneous turbulence by modifying the exponent of the energy spectrum he predicted in 1941 \cite{Kolmogorov 1941} to take into account intermittency \cite{Kolmogorov 1962} as Landau suggested him during a seminar in Kazan. To celebrate the fiftieth anniversary of the first international colloquium on turbulence (TCM-1961) which was held in Marseille in 1961, I organised with Keith Moffatt and Kai Schneider the `Turbulence Colloquium Marseille 2011'  (TCM-2011)\cite{TCM 2011a} at the CIRM (`Centre International de Rencontres Math\'ematiques'), a beautiful place above the famous Calanques of Marseille; incidently, this is where in the early 1980s we  had our first meetings to work on continuous wavelets with Alex \cite{Farge et al. 2012}. We limited TCM-2011 to eighty participants by invitation, due to the maximum accommodation capacity of the CIRM. We structured it as the 1961 colloquium with the same objectives: to minimise the number of invited lectures, given by senior researchers selected for the importance of their contribution to the study of turbulence, and to maximise the open discussions between junior and senior researchers. There were eight half-day sessions on: turbulence before TCM-1961 (given by Eckert who was writing a book on `The Turbulence Problem' \cite{Eckert 2019}, mathematics for turbulence, homogeneous turbulence, turbulent shear layers and wakes, pipe and channel flow turbulence, turbulent boundary layer, atmospheric turbulence and magneto-hydrodynamic turbulence. Each session began by one hour review lecture given by a senior scientist who summarised how the topics have evolved since TCM-1961, it was followed by one hour coffee-break, where posters on recent research results were presented by their authors, and it ended with a one-hour open discussion, where conclusions were collectively drawn and written down by a junior scientist acting as a session secretary, as was done in 1961. The book of the proceedings have been published \cite{TCM 2011b} and many documents are available from {\it http://turbulence.ens.fr}. At the end of the colloquium we organised a surprise exam \cite{TCM 2011c}, where each participant was proposed to answer nine questions about turbulence and two prizes were awarded: one for the junior winner (Nguyen van yen) and one for the senior winner (Dubrulle), who  both received the book on the history of turbulence in the 20th century, which Cambridge University Press had just published \cite{Davidson et al. 2011}). The questions, designed by a jury of three mathematicians (Bardos, Doering and Titi), were difficult and subtle as they dealt with problems which are still open:
 
\begin{itemize}
\item
1. What is your understanding of the claim: Turbulent Flows have finite degrees of freedom? What are these degrees of freedom?
\item
2. What is your understanding of an average?
\item
3. Do you think that any notion of solution of Euler or Navier-Stokes should conserve the energy, or satisfy an energy balance? What if a solution does not conserve energy?
\item
4. What is your understanding of dissipation anomaly? Can you formulate the problem as precisely as possible?
\item
5. Do you think that Navier-Stokes or Euler solutions develop a finite time singularity? Is this consistent with the physical observations? If a singularity is proven to be formed, what is the mechanism? How should we modify the models?
\item
6. Do you think the project of searching for singularity computationally makes sense? In which geometry? Do you think that a smooth physical boundary could be the real source of singularity formation? Or should we stick to periodic boundary conditions?
\item
7. Is atmospheric turbulence 3d or 2d? How should we test this matter? What are the right mathematical quantifiers in this case?
\item
8. Is the range of applications of the incompressible Navier-Stokes equations restricted to incompressible fluids?
\item
9. Does triad wavenumber interaction have much to do with fluid motion? What is the equivalent picture in the physical domain?
\end{itemize}


\clearpage


\begin{figure}[ht!]
\centering
\vspace{5cm}
\includegraphics[width=0.7\linewidth]{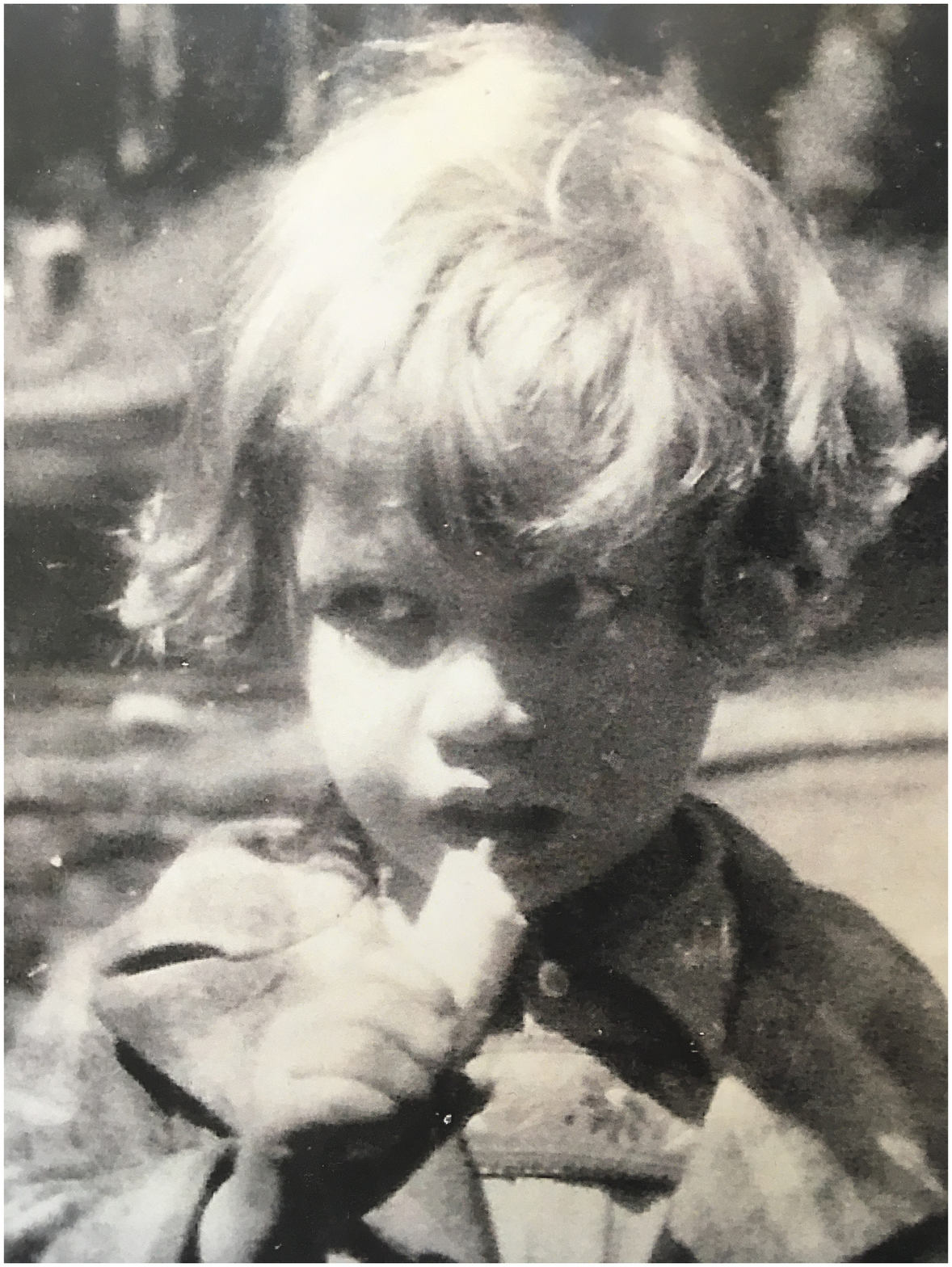}\\
  \nocaption
  \vspace{+1cm}
  \centerline{Alex Grossmann as a child in Zagreb.}
  \label{Figure_Alex}
\end{figure}

\begin{figure}[ht!]
\centering
\vspace{3cm}
\includegraphics[width=1.0\linewidth]{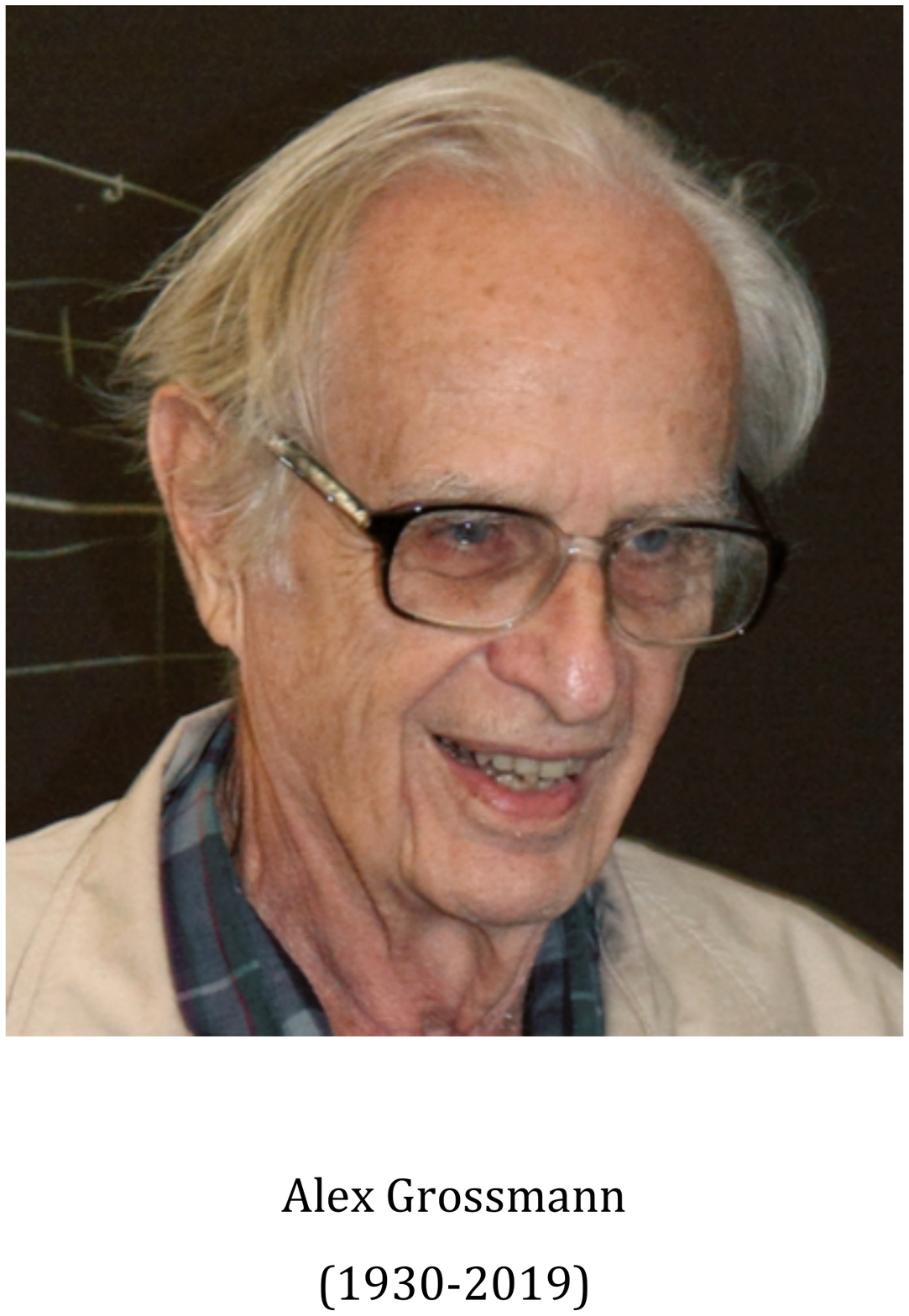}\\
  \nocaption
  \vspace{-2cm}
  \centerline{Alex Grossmann giving a talk in 2013 at CIRM in Marseille.}
  \label{Figure_Alex}
\end{figure}

\clearpage

\section{The need for continuous wavelets}
\label{sec:4}

\subsection{How I met Alex}
\label{subsec:4.1}

In 1980, while I was a postdoc at Harvard University, Max Krook, a theoretical physicist of South-African origin who discovered the 'Krooked solution' of the Boltzmann equation, suggested that I meet Alex and Robert Coquereaux, who were visitors in the Physics Department. On November 21, 1980, Max Krook invited me to dine with them at `Legal Sea Foods' on Boston harbour, and that was where I first met Alex and his wife Dickie. When I returned to Paris in 1981 I joined the CNRS and became a member of the `Laboratoire de M\'et\'eorologie Dynamique' of `Ecole Normale Sup\'erieure', where I am still based today. This new affiliation led me to focus on two-dimensional turbulence, because at the scale of the Earth the dynamics of the atmosphere can be viewed as two-dimensional, due to the thinness of the atmosphere and the constraint imposed by the rotation of the Earth. To study atmospheric turbulence I relied on numerical experiments, which was a path of research complementary to observation, laboratory experiment and theory \cite {Farge 1986}.

\bigskip

\begin{figure}[ht!]
\hspace{-3cm}
\includegraphics[width=1.4\linewidth]{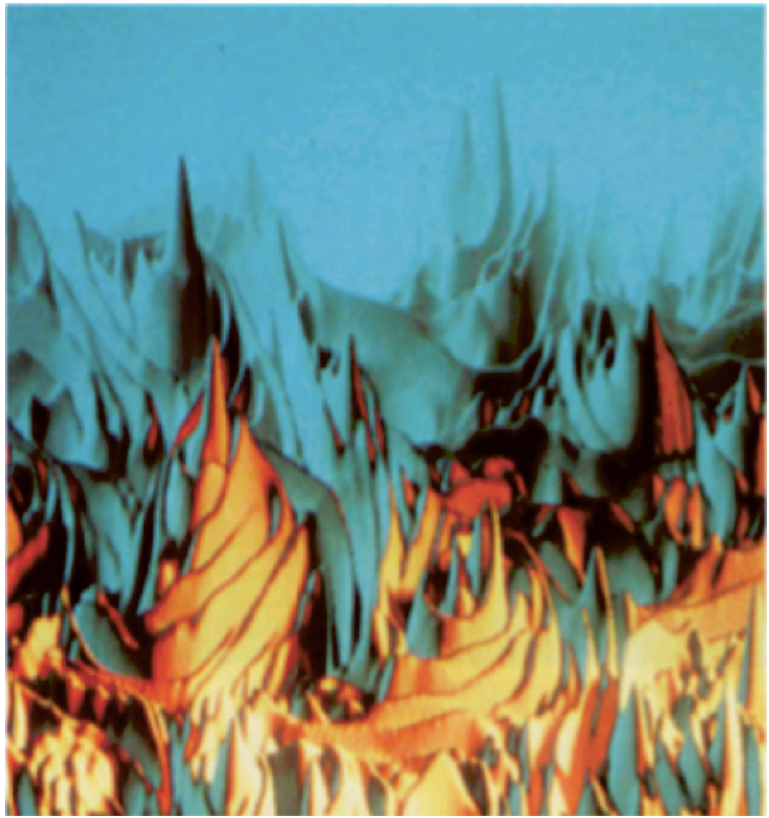}\\
  \caption{Vorticity field of a two-dimensional turbulent flow, computed by direct numerical simulation of Navier--Stokes equation with periodic boundary conditions, using a pseudo-spectral method with resolution $N=512^2$. It is visualised with a cavalier representation \cite{Farge 1987}. We observe the cusp-like vortices which drive the flow evolution.}
  \label{Figure1}
\end{figure}

\begin{figure}[ht!]
\centering
\vspace{5cm}
\includegraphics[width=1.0\linewidth]{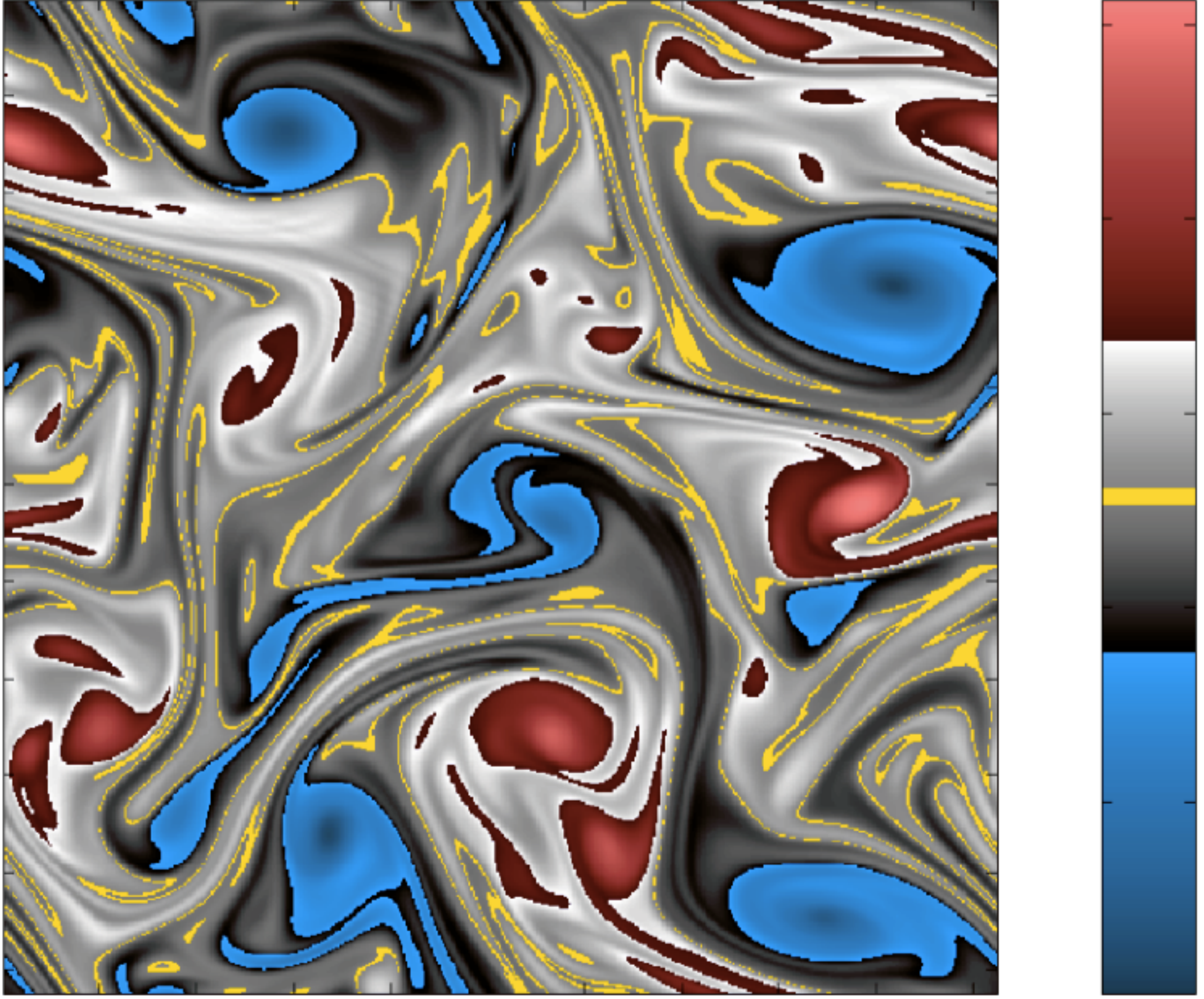}\\
  \caption{Vorticity field of a two-dimensional turbulent flow, computed by direct numerical simulation of Navier--Stokes equation with periodic boundary contitions, using a pseudo-spectral method with resolution $N=512^2$. It is visualised with a cartographic representation \cite{Farge 1987} where strong positive vorticity is red, strong negative vorticity is blue, weak vorticity is grey and zero vorticity is yellow. We observe the merging of same-sign vortices and the binding of opposite sign vortices to form vortex dipoles.}
  \label{Figure2}
\end{figure}

\begin{figure}[ht!]
\hspace{-3cm}
\includegraphics[width=1.4\linewidth]{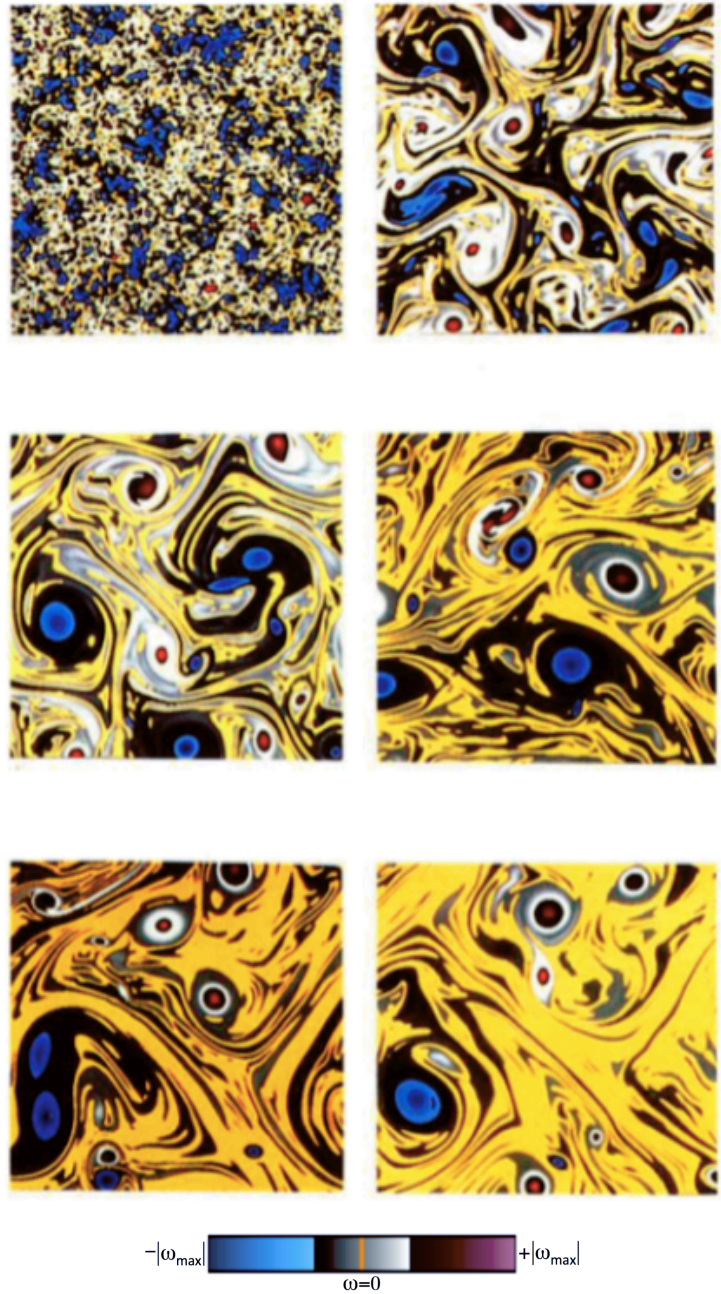}\\
  \caption{Evolution of the vorticity field from an initial random distribution, computed by direct numerical simulation of the two-dimensional Navier--Stokes equation with periodic boundary contitions, using a pseudo-spectral method with resolution $N=512^2$. It is visualised with a cartographic representation \cite{Farge 1987}, with the same colour-scale as [Figure \ref{Figure2}]. We observe the formation of coherent vortices emerging out of the initial random flow, followed by the merging of same sign vortices which eject vorticity filaments into the background flow, together with the binding of opposite sign vortices which form vortex dipoles.}
  \label{Figure3}
\end{figure}

\begin{figure}[ht!]
\centering
\includegraphics[width=0.6\linewidth]{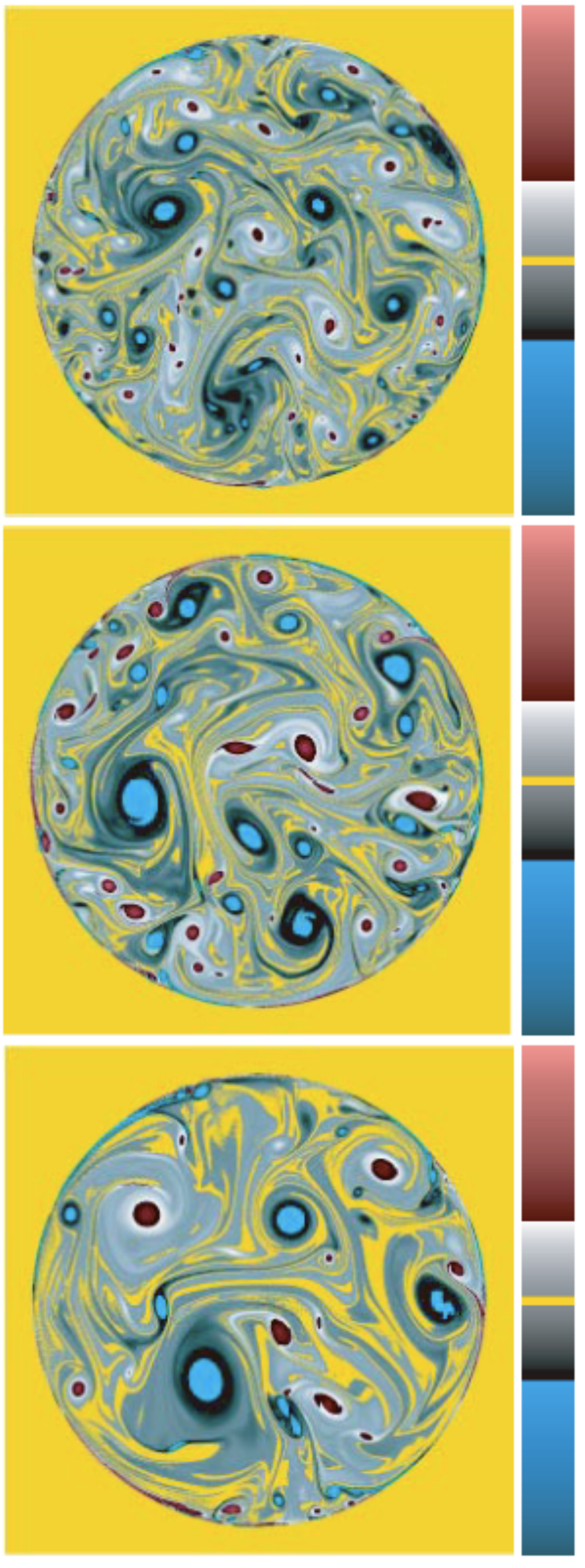}\\
  \caption{Evolution of the vorticity field from an initial random distribution, computed by direct numerical simulation of the two-dimensional Navier--Stokes equation in a circular container with no-slip boundary conditions, using a pseudo-spectral method and volume penalisation to take into account the solid wall, with resolution $N=1024^2$. It is visualised with a cartographic representation \cite{Farge 1987} with the same colour-scale as [Figure \ref{Figure2}]. We observe the formation of coherent vortices at the solid wall, due to Kelvin-Helmholtz instability, which spread into the bulk flow and exhibit same sign vortex merging and opposite sign vortex binding, as we have previously noticed.}
  \label{Figure4}
\end{figure}

\begin{figure}[ht!]
\centering
\vspace{5cm}
\includegraphics[width=1.0\linewidth]{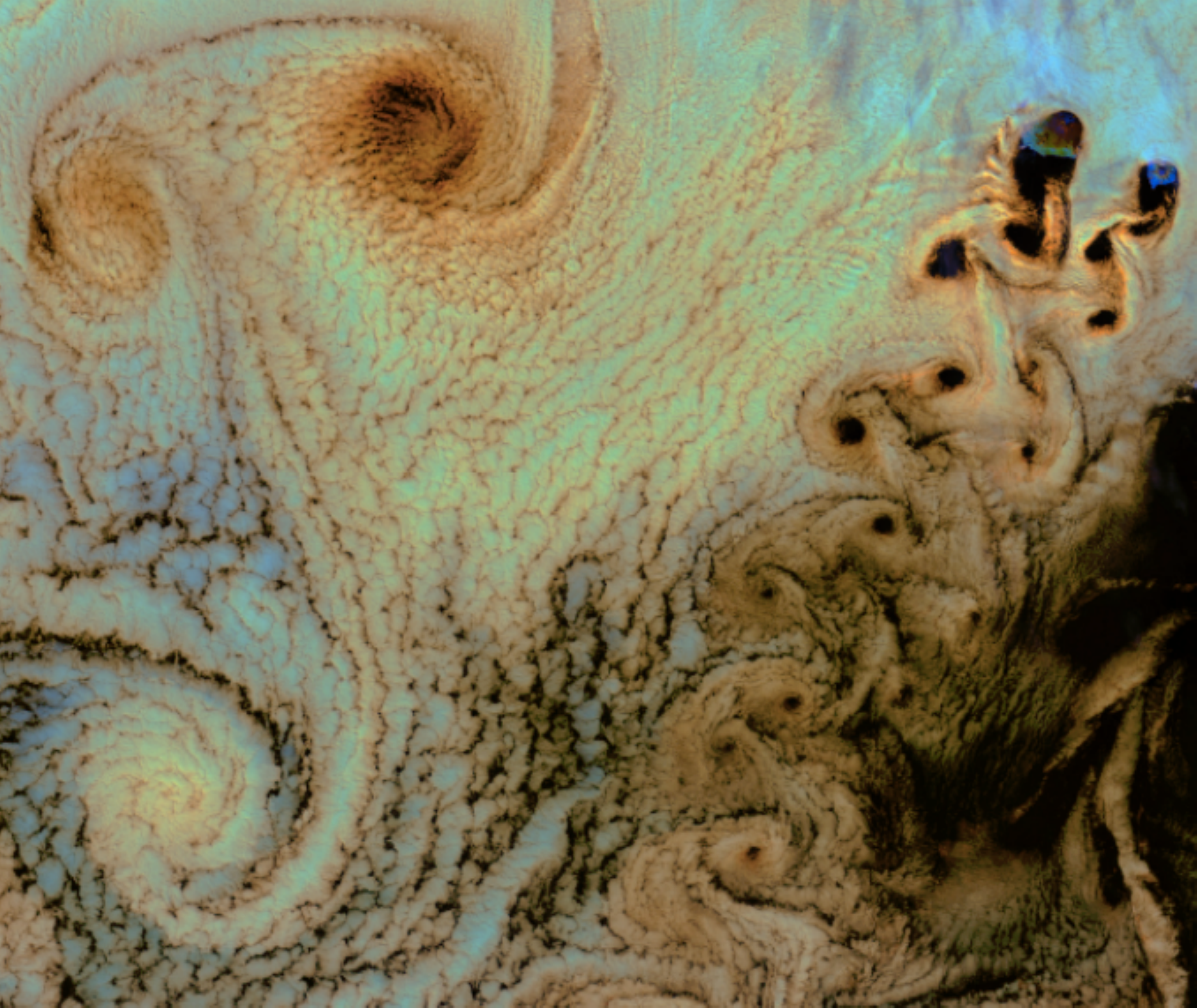}\\
  \caption{Satellite photograph of the Pacific Ocean from NASA. The clouds reveal the presence of two-dimensional vortices, because at large scale the Earth's rotation constraints the atmospheric turbulent flow to be two-dimensional.  We observe a large vortex dipole and many smaller ones generated by the von Karman vortex street in the wake of the Guadalupe Islands (Baja California, Mexico).}
  \label{Figure5}
\end{figure}

In 1981, I developed a numerical code that integrates the two-dimensional Navier--Stokes equation using a pseudo-spectral method, where the spatial derivatives are computed in spectral space and the products in physical space, which requires transforming the velocity and vorticity fields back and forth using a fast Fourier transform. Note that in two dimensions vorticity is a scalar field while velocity is a vector field. Since at each time step I thus had both the spatial and spectral representation of these fields, I sought to visualise the turbulent flows on both sides of the Fourier transform. I first focused on the vorticity and studied its evolution in the physical space using different types of visualisation: {\bf [Figure \ref{Figure1}]} and 
{\bf [Movie \href{http://wavelets.ens.fr/TURBULENCE/3_MOVIES/Movie_1_Vorticity_evolution_of_a_2D_turbulent_flow_with_periodic_boundary_conditions_computed_by_DNS_and_visualised_with_a_cavalier_view.mpg}{1}]}
show the vorticity field visualised with a cavalier representation \cite{Farge 1987}, while in {\bf [Figure \ref{Figure2}]} it is visualised with a cartographic representation \cite{Farge 1987}. These two different visual representations of turbulent fields are complementary ways of studying the emergence and interactions of coherent vortices, which are responsible for the mixing character of turbulent flows and their unpredictability. I also wanted to understand the struggle between the non-linear terms and the linear terms of Navier--Stokes equation: the former produce vortices, due to the instability of the flow, while the latter destroy them, due to the viscosity of the fluid. {\bf [Figure \ref{Figure3}]} and 
{\bf [Movie \href{http://wavelets.ens.fr/TURBULENCE/3_MOVIES/Movie_2_Vorticity_evolution_of_a_2D_turbulent_flow_with_periodic_boundary_conditions_computed_by_DNS_and_visualised_with_a_cartographic_view.mpg}{2}]} show the evolution of a two-dimensional turbulent flow in a periodic domain from initial random conditions, which simulates the atmospheric dynamics at global scale; it was computed by direct numerical simulation of the two-dimensional Navier--Stokes equation, using a pseudo-spectral method with resolution $N=512^2$. {\bf [Figure \ref{Figure4}]} and 
{\bf [Movie \href{http://wavelets.ens.fr/TURBULENCE/3_MOVIES/Movie_3_Vorticity_evolution_of_a_2D_turbulent_flow_in_a_circular_domain_with_no_slip_boundary_conditions_computed_by_DNS_and_visualised_with_a_cartographic_view.mpg}{3}]} 
show the evolution of a two-dimensional turbulent flow in a closed domain to study the interaction between the atmospheric flow and obstacles, such as mountains or islands (see {\bf [Figure \ref{Figure5}]}); it was computed by direct numerical simulation of the two-dimensional Navier--Stokes equation in a circular container with no-slip boundary conditions, using a pseudo-spectral method and volume penalisation to take into account the solid wall, with resolution $N=1024^2$. For both numerical experiments we see the formation of vortices that either emerge from the initial random flow, or are produced in contact to the solid wall. Later, as the flow evolves, we observe that if one vortex encounters another of the same sign, they rotate around each other and their mutual shearing produces vorticity filaments that are ejected far away, and then they usually merge. On the other hand, if one vortex encounters another of opposite sign, they bind and form a dipole that rapidly self-propels through the flow. {\bf [Figure \ref{Figure5}]}, a satellite photograph of the Pacific Ocean from NASA, confirms that the presence of similar vortex dipoles that the atmospheric flow produces in the wake of the Guadalupe Islands (Baja California, Mexico).

\clearpage

 
\smallskip

\noindent

\clearpage

 
\smallskip

\noindent

\clearpage

 
\smallskip

\noindent

\clearpage

  
\clearpage

  
\clearpage

In section 2.2, entitled `Probabilistic statistical theories', I have briefly presented Kolmogorov's theory of homogeneous isotropic turbulence. I would like to remind you of it and go into a bit more detail to help you understand the problem I was trying to solve before Alex told me about wavelets. As we have seen, when the non-linear terms of the Navier--Stokes equation strongly dominate its linear terms, the `fully-developed' or strong turbulence regime is reached, for which Kolmogorov developed in 1941 the theory of homogeneous isotropic turbulence. He assumed that the kinetic energy is injected into the flow at a low-wavenumber (called the `integral wavenumber') and dissipated at a high-wavenumber (called the `Kolmogorov wavenumber'), beyond which the viscosity of the fluid damps the non-linear instabilities and converts kinetic energy into thermal energy. In the spectral band between these two characteristic wavenumbers (called the `inertial range'), Kolmogorov supposed that the kinetic energy is conserved and only transferred from low- to high-wavenumbers at a constant rate (called `turbulent cascade').  Based on these hypotheses and few others, he predicted that in the inertial range the energy spectrum behaves as a power law, such that  the logarithm of the modulus of energy varies like the modulus of the wavenumber to the power $-5/3$. Throughout the inertial range the non-linear transport of the flow conserves energy but, beyond the Kolmogorov wavenumber, the viscosity of the fluid prevents instabilities from developing and dissipates energy, leading to an exponential energy decay at large wavenumbers. 

\bigskip

In 1982, inspired by Kolmogorov's theory I began to analyse in both spectral space and physical space the vorticity fields resulting from my numerical experiments. For this I used a low-pass filter to extract its contribution to the conservative inertial range, and a high-pass filter to get its contribution to the dissipative viscous range. I then reconstructed both contributions in physical space and visualised their evolutions separately. My goal in proceeding so was to observe in physical space the spatial support of the viscous dissipation that Kolmogorov had defined in spectral space. The result was quite puzzling, and therefore challenging, because the more turbulent the flow was the more the filtered vorticity field was full of wiggles and narrower the high-pass filter worse it was. So I realised that I was confronted with Heisenberg's uncertainty principle (that I prefer to call `Fourier uncertainty principle') and that it was impossible to overcome this obstruction... 

\bigskip

From time to time I was going to Marseille to see my dentist, Marius Mayoute, who was living at the `Roy d'Espagne', a southern district of Marseille. 
Every time I went there, I took the opportunity to visit Dickie and Alex because they also lived at the `Roy d'Espagne'. It was during one of these visits, on February 9, 1983, that Alex told me for the first time about the wavelet transform he was developing with Jean Morlet. Jean had introduced few years before a new transform, which he called the `cycle-octave transform', that decomposes a signal varying in time into both instants and time scales. The cycle-octave transform was developed for discrete signals only and his inverse transform did not reconstruct them perfectly. Roger Balian advised Jean Morlet to ask Alex Grossmann for help. Alex accepted with pleasure and he designed the mathematical setting of the continuous wavelet transform and found the admissibility condition which insures its invertibility. In my first discussion with Alex about wavelets, he explained that their goal was to represent phenomena on both sides of the Fourier transform in a way that was optimal with respect to the uncertainty principle. This discussion acted on me like a scientific thunderbolt. Indeed, I felt that there should be a way to overcome the obstruction I was facing when trying to visualise turbulent flows in both physical and spectral space. 

\bigskip

I asked Alex to show me a typical wavelet and he chose the Mexican hat function, which is the second derivative of the Gaussian function. This was a second thunderbolt for me, because a few months earlier I had received a letter from David Ruelle, who was at that time in Australia, to inform me that he had found a new solution of the Navier--Stokes equation in two dimensions where the spatial distribution of vorticity is precisely a Mexican hat. I quote his letter dated February 5th 1983: {\it 'I spent some time here reading Kraichnan and Montgomery's paper \cite{Kraichnan Montgomery 1980} (more readable than I thought) and thinking about two-dimensional turbulence. Without much illumination, by the way. However, I have come to wonder whether vortices, which one `sees' in a two-dimensional fluid, can be identified with quasi-punctual concentrations of vorticity, as is usually done. In fact there is a Navier--Stokes solution (*) that could also be considered as a vortex. This solution has a non-zero angular momentum, but the vorticity integral is zero. This solution seems quite stable by perturbations, it spreads slowly by diffusion. If the viscosity tends to zero and the product between it and time remains close to one, the diffusion is very slow. (*) This solution is characterised by a core where the vorticity has one sign, surrounded by a layer of vorticity of opposite sign. I wonder if there is a way to show this experimentally or numerically'}. After receiving David Ruelle's letter I confirmed to him that in my numerical simulations of two-dimensional turbulence I observed similar Mexican hat vortices. They appear when two vortices come sufficiently close together and develop strong interactions that lead them to merge and form a single vortex, which then relaxes into a Mexican hat vorticity distribution, that we called a `shielded vortex' (see {\bf [Figure \ref{Figure2}]} and http://wavelets.ens.fr/TURBULENCE). 

\bigskip

Thanks to the discussion with Alex and to David's letter, I saw the direction in which I wanted to go to represent, analyse and simulate turbulent flows: no longer as a superposition of waves delocalised in physical space, as with the Fourier representation, but as a superposition of vortices localised in both space and wavenumber, as with the wavelet representation. Of course, I was still far from my goal, because Alex had not yet found the exact reconstruction formula, but I hoped to reach it one day. If I speak of `scientific thunderbolt', it is because during this discussion with Alex in 1983 I had almost instantly the vision of the research project on which I am still working today, and which is far from being completed... 

\bigskip
 
\subsection{Research programme}
\label{subsec:4.2}

During my studies I was never satisfied with the lectures I attended on turbulence because I did not understand how one could explain the energy transfers (from small to large wavenumbers), called 'turbulent cascade', produced by the fracture of large vortices into smaller and smaller fragments until they reach such a small size that the viscosity of the fluid is able to destroy them by viscous dissipation. This {\it doxa} was suggested by Richardson in 1922, inspired by the famous poem of Swift in Gulliver's Travels, which reads : {\it 'So, naturalists observe a flea hath smaller fleas that on him prey, and these have smaller yet to bite them, and so proceed ad infinitum.'}  \cite{Richardson 1922}. This has always been nonsense for me, because I do not understand how an incompressible deformable fluid can break into pieces, because the fluid is a continuous medium, while breaking pebbles creates a vacuum and there is discontinuity between the fragments. It is very likely that this image arose at the origin of Kolmogorov's theory of homogeneous isotropic turbulence in 1941. Indeed, Obukhov measured experimentally the energy spectrum of various turbulent flows and asked Kolmogorov to explain why this spectrum has a power law behaviour. Kolmogorov was working at that time on applied problems for the mining industry and he predicted that, if one breaks rocks, the distribution of their size follows a power law. As I explained in the first part, the nonsense I am talking about does not arise from Kolmogorov's theory, which predicts the behaviour of ensemble averages but not that of each realisation. Thus many scientists use Kolmogorov's theory outside its scope without understanding it, which as a consequence leads to the misinterpretation I noticed when I was student and denounce today.

\bigskip

In 1983, the same year as Alex introduced me to wavelets, I was invited to celebrate the 60th anniversary of Ren\'e Thom in Cerisy (France) and to explain what numerical simulation means. I tried to convince the mathematicians present that it is a new way to do research by experimenting with the fundamental equations and observing their solutions using numerical methods and computers  \cite{Farge 1986}, \cite{Farge 2007}. Since most non-linear equations are not solvable with analytical integration techniques, we use numerical experimentation to explore their approximate solutions, knowing that: {\it 'Although this may seem a paradox, all exact science is dominated by the idea of approximation'} \cite{Russell 1931}. Indeed, numerical experimentation is a third path for studying phenomena, which is complementary to theory and laboratory experimentation. This branch of research should not be a surprise since, as with 'differentiation of the species', the distinctions between mathematicians and physicists, then between theoreticians and experimentalists, emerged progressively over the last centuries. Numerical simulation is the tool I have chosen to do research and to experiment with different turbulent flows, in two and three space dimensions and for various geometries. I would like to study strong turbulence from a fundamental point of view without using {\it ad hoc} turbulence models, as engineers do because they must resort to wind tunnel tests to identify the parameters of their models. On the contrary, my goal is, either to compute turbulent flows from first principles only, or to find turbulence models with the fewest possible adjustable parameters. Due to the intrinsically chaotic nature of turbulent flows one can predict their evolution in a deterministic way only on very short time scales, as is the case for example in meteorology. To predict over long times, which is necessary, for example, in climatology, one then needs to take a statistical viewpoint. Thus, the art of strong turbulence consists in elaborating statistical observables able to capture the non-linear dynamics and out of equilibrium behaviour of the flow, without relying on a statistical equilibrium nor on a spectral gap, we do not have for them, in contrast to classical statistical physics. 

\bigskip

Using the wavelet representation I would like to revisit the program suggested by Reynolds in 1883 \cite{Reynolds 1883}, which decomposes each flow realisation into mean and fluctuations to simplify the Navier--Stokes equation. This led him to propose the Reynolds equations which predict the evolution of the mean fields only, and model the effect of the fluctuations on the mean by adding an extra term, the so-called `Reynolds stress tensor'. From there an industry of `turbulence models' has developed to solve a large variety of industrial problems, in aeronautics, combustion, meteorology, and many other domains of application. The art of turbulence modelling consists in replacing the Reynolds stress tensor by an {\it ad hoc} model whose parameters must be estimated from data gathered from field , laboratory or wind tunnel experiments. The simplest and widely used model is to replace the Reynolds stress tensor by a dissipation operator with a strong turbulent viscosity, whose parameter must be estimated or measured. In 1938 Tollmien and Prandtl suggested that {\it `turbulent fluctuations might consist of two components, a diffusive and a non-diffusive. Their ideas that fluctuations include both random and non random elements are correct, but as yet there is no known procedure for separating them.'} \cite{Dryden 1948}. Inspired by this quotation I found in an article published in 1948 by Hugh Dryden (who founded NACA which later became NASA), I suggested to split each flow realisation into mean, non random fluctuations, due to the non-linear transport of the flow, and random fluctuations, due to the linear dissipation due to the viscosity of the fluid.  

\bigskip

In the first part of this article we have seen that turbulence, as its Latin etymology suggests, is the state of an incompressible fluid flow when it produces a crowd ({\it turba, ae}) of vortices ({\it turbo, inis}) interacting strongly with each other. Indeed, turbulent flows are driven by non-linear instabilities which generate vortices, whose motion can no longer be damped by the fluid viscosity, and whose evolution is no longer predictable over long times due to their chaotic and mixing behaviour. My research is focused on the regime of  strong turbulence which corresponds to very large Reynolds numbers. In contrast to the past, to study such a highly non-linear regime we can today benefit from new tools, in particular the direct numerical simulation of the Navier--Stokes equation computed on parallel computers, together with the high-resolution visualisations and measures from laboratory experiments. This provides insight into the dynamically essential features which are the coherent vortices emerging out of turbulent flows due to their instabilities. We seek an appropriate representation of turbulent flows that can decouple the out-of-statistical equilibrium motions to be computed, from the well thermalised motions to be eliminated as turbulent dissipation. The representation we are looking for should identify a gap to separate the dynamically active from the passive components of turbulent flows. The Fourier representation does not exhibit such a spectral gap since the energy spectrum behaves as a power law. In 1988 I suggested to use the wavelet representation to study the evolution of turbulent flows in both space and scale, because it provides the best compromise to deal with the uncertainty principle and allows us to observe the flow evolution in both space and scale at once. 

\bigskip

This research programme consists, as a first step, of analysing turbulent flows in order to highlight their dynamically active structures, {\it i.e.}, vortices, and to study how these form and deform, due to the instability of the flow and their mutual interactions. To achieve this, we need to find an appropriate representation as these structures become more and more intermittent as the Reynolds number increases. The next step is to develop conditional statistics that can capture the non-linear dynamics and account for its complexity. Our objective is to compute turbulent flows using the smallest number of degrees of freedom necessary to predict its evolution, deterministically over short time scales, and statistically over long time scales. I am looking for a dynamic system with a large number of degrees of freedom that would be projected onto an appropriate base, so that at each time step only the non-linearly active degrees of freedom would be taken into account. The difficulty with this method is that, in order to advance from one time step to the next, it is necessary to select and activate only those degrees of freedom that drive the evolution of the flow and to eliminate those that have become inactive. This requires the development of an adaptive method designed in an appropriate functional space.

\bigskip

\subsection{Why wavelets?}
\label{subsec:4.3}

\begin{figure}[ht!]
\centering
\includegraphics[angle=-90,width=1.0\linewidth]{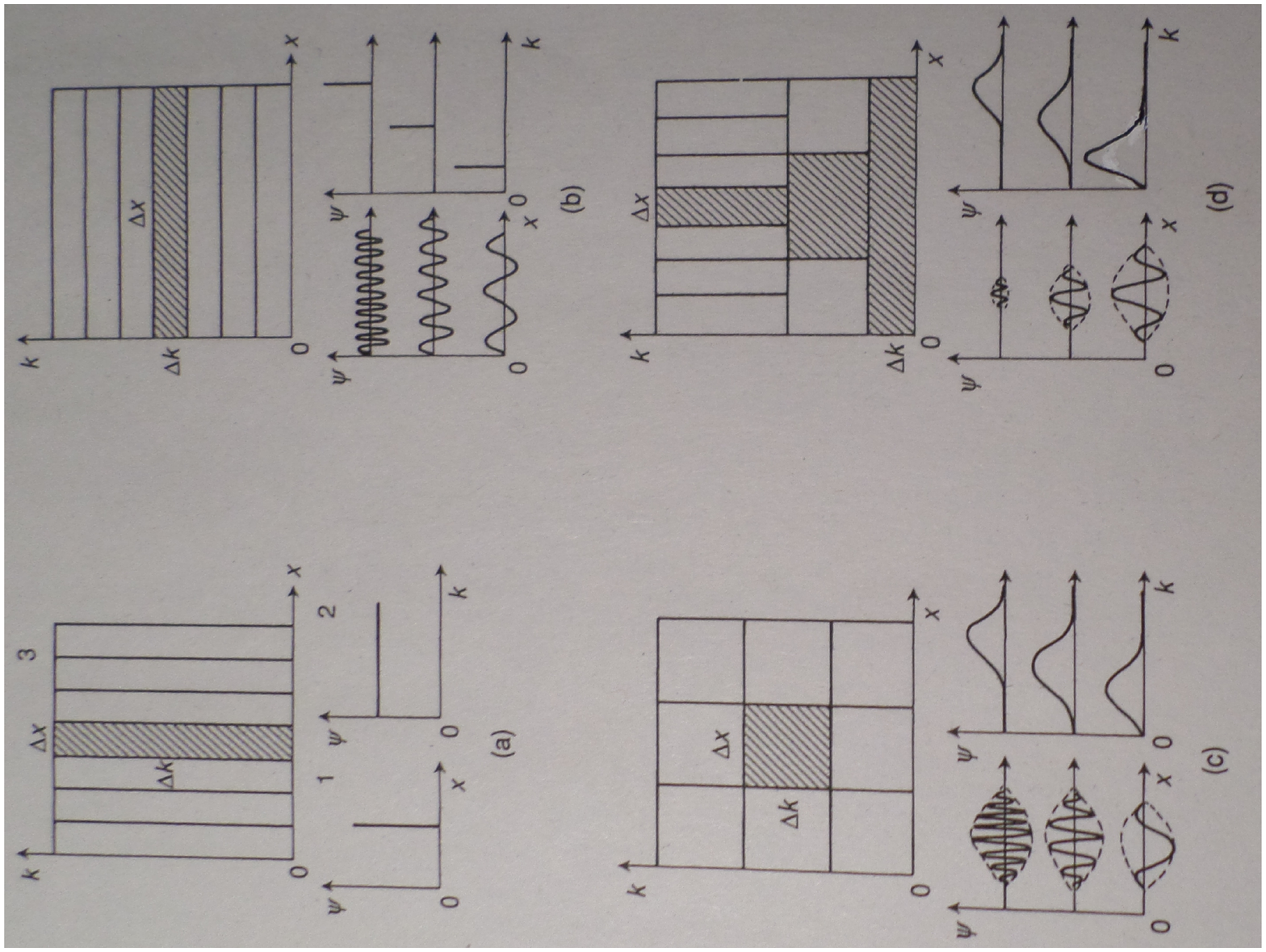}\\
  \caption{The information plane shows how different transforms adjust their spatial resolution $\Delta x$ and their spectral resolution $\Delta k$ to optimise against the uncertainty principle: $\Delta x . \Delta k \geq C$. The hashed regions correspond to the elementary information cells of constant area $C$ for: (a) Shannon transform, (b) Fourier transform, (c) Windowed-Fourier transform, (d) Wavelet transform \cite {Farge et al. 1988, Farge et al. 1993a}.}
  \label{Figure6}
\end{figure}

If, for instance, we would like to analyse a signal evolving in time, its representation in physical space does not provide information on its properties in spectral space, such its frequency content and their instant of emission. A more appropriate representation should combine both because they are complementary. This is especially needed to analyse strongly turbulent signals which are very intermittent, because they exhibit bursts or even quasi-singular features. Indeed, the `uncertainty principle' forbids us to perfectly represent a signal on both sides of the Fourier transform at the same time, because the product of the resolution in physical space $\Delta x$ and of the resolution in spectral space $\Delta k$ is equal or greater than a constant $C$. It is the minimal area $C$ of the elementary `information cell' one can get, which is shown as an hatched area on the `information plane' for different transforms (see {\bf [Figure \ref{Figure6}]}) \cite {Farge et al. 1988, Farge et al. 1993a}. Indeed, in choosing a representation there is always a compromise to be made: either a good resolution in physical space and a bad resolution in spectral space, as it is the case with the Shannon representation which samples a signal by convolving it with a Dirac comb (see {\bf [Figure \ref{Figure6}]}(a)), or a bad resolution in physical space and good resolution in spectral space, as it is the case with the Fourier representation (see {\bf [Figure \ref{Figure6}]}(b)). In order to get a better compromise Gabor proposed in 1946 \cite{Gabor 1946} the `windowed Fourier transform', which convolves the signal with a set of Fourier modes localised inside a Gaussian window of constant width (see {\bf [Figure \ref{Figure6}]}(c)). Unfortunately, as proved in 1981 by Balian \cite{Balian 1981}, the bases constructed with such windowed Fourier modes cannot be orthogonal. Three years later Grossmann and Morlet \cite{Grossmann Morlet 1984}, \cite{Grossmann Morlet 1985} devised a new transform, that they first called the `cycle-octave transform' and later the `wavelet transform'. It consists of convolving the signal with a `family' of functions generated by translating and dilating a chosen function, the `mother wavelet', which oscillates and has a zero mean. By adapting the resolution in physical space and in spectral space scale-by-scale  it realises the best compromise with respect to the uncertainty principle (see {\bf [Figure \ref{Figure6}]}(d)).

\bigskip

  
\clearpage

In 1985, after reading the article by Grossmann and Morlet \cite{Grossmann Morlet 1984}, Meyer tried to show that orthogonal wavelet bases cannot be constructed from smooth functions, as Balian had proved in 1981 \cite{Balian 1981} for the windowed Fourier transform. To his surprise, he discovered a basis of orthogonal wavelets constructed from trigonometric functions, belonging to the Schwartz class of infinitely differentiable functions of fast decay with and compact support in spectral space, that he presented at the Bourbaki seminar in February 1986 \cite{Lemarie Meyer 1986, Meyer 1987}. Nowadays, the basis constructed by Haar in 1909, for his thesis supervised by Hilbert, is recognised as the first orthogonal wavelet basis \cite{Haar 1910}, but the basis functions are piecewise constant and are not regular, even not continuous, which limits its applications. This is no longer the case for the wavelet bases proposed by Str\"omberg in 1981 because they are constructed from spline functions (piecewise polynomials) which are as smooth as one wishes \cite{Stromberg 1981}. In fact, when Meyer found the Meyer's wavelet, he was unaware of Str\"omberg's work and used a different method for it. In practice, orthogonal wavelet bases are constructed using functions whose regularity is chosen in order to obtain the best spectral decay according to the problem being addressed. 

\bigskip

In 1986, Daubechies, Grossmann and Meyer \cite{Daubechies et al. 1986} showed how to discretise the continuous wavelet representation to obtain wavelet frames. These are overcomplete representations consisting of functions (often called `atoms') which are not linearly independent as for orthogonal wavelets, but only quasi-orthogonal such that the energy remains bounded as long as the discretisation is sufficiently fine. In 1988, Daubechies \cite{Daubechies 1988} found several orthogonal wavelet bases consisting of compactly supported functions, which she defined using two discrete quadratic mirror filters so that the more regular the basis functions, the longer these filters. Note that, with the exception of Haar wavelets, previously introduced orthogonal wavelets were not compactly supported. In 1989, building on Daubechies' work Mallat \cite{Mallat 1989} designed a fast algorithm to calculate the orthogonal wavelet transform using wavelets defined by quadratic mirror filters and a method similar to the `algorithme \`a trous' developed by Holschneider, Kronland-Martinet, Morlet and Tchamitchian to compute in real-time the discrete wavelet transform of audio signals on $2^{10}$ octaves \cite{Holschneider et al. 1989}. This was a crucial step for the application of the wavelet transform, as the decomposition and reconstruction of a signal of $N$ samples are performed in $CN$ operations each, $C$ being the length of the quadratic mirror filter defining the chosen wavelet. Since then the fast wavelet transform algorithm has been widely used, particularly in designing more efficient numerical analysis techniques \cite{Beylkin et al. 1991}. In 1990, Malvar, Coifman and Meyer \cite{Malvar 1990}, \cite{Coifman Meyer 1991} found a new kind of window of variable width with which to construct orthogonal adaptative local cosine bases. The elementary functions of such bases are parametrised by their position, their scale (the width of the window) and their wavenumber (proportional to the number of oscillations inside each window). Then in 1991, Coifman, Meyer and  Wickerhauser \cite{Coifman et al. 1991}, \cite{Coifman Wickerhauser 1992}, introduced the so-called `wavelet-packets' which, similarly to compactly supported wavelets, are wave-packets of prescribed smoothness. Similar to wavelets, wavelet packets are defined by discrete quadratic mirror filters, from which many orthogonal bases are constructed and the best basis that optimally compresses the information is selected according to the maximum entropy principle. \cite{Wickerhauser 1994}.

\bigskip

Since Alex revealed the existence of the wavelet transform to me in 1983, I have, in collaboration with several students and colleagues, tested the following transforms to find those which might be useful for studying strong turbulence:
 
\begin{itemize}
\item
the continuous wavelet transform with complex-valued wavelets ({\it e.g.}, Morlet) to analyse the evolution of two-dimensional turbulent flows ({\it i.e.}, velocity and vorticity) computed by direct numerical simulation of the Navier--Stokes equation,
\item
the local cosine transform and the wavelet-packet transform with real-valued wavelets ({\it e.g.}, Coiflets) to extract coherent structures out of two-dimensional turbulent flow fields computed by direct numerical simulation of the Navier--Stokes equation,
\item
the orthogonal wavelet transform with real-valued wavelets ({\it e.g.}, Coiflets) to extract coherent structures and to denoise experimental turbulent plasma signals ({\it e.g.}, ion saturation current and density fluctuations) measured in various tokamaks, 
\item
the orthogonal wavelet transform with real-valued wavelets ({\it e.g.}, Coiflets) to extract coherent vortices out of two-dimensional and three-dimensional turbulent flows computed by direct numerical simulation of the Navier--Stokes equation,
\item
the orthogonal wavelet transform with real-valued wavelets ({\it e.g.}, Daubechies, Meyer and splines) and complex-valued wavelets ({\it e.g.}, Kingslets), to firstly compute the inviscid Burgers equation in one dimension and study how its solution forms shocks, and secondly to model the viscous Burgers equation by denoising its solution at each time step in order to check if we recover its exact analytical solution,
\item
the orthogonal wavelet transform with real-valued wavelets ({\it e.g.}, Coiflets) and complex-valued wavelets ({\it e.g.}, Kingslets), to compute the two-dimensional Euler equation in two dimensions and to model the two-dimensional Navier--Stokes equation by extracting coherent vortices and discarding the incoherent noise at each time step,
\item
the orthogonal wavelet transform, with real-valued wavelets ({\it e.g.}, Coiflets), to compute the three-dimensional Navier--Stokes equation and study the formation and interactions of vortices in the turbulent wake produced by the flapping wing flight of insects.
\end{itemize}

\bigskip

Alex Grossmann and Jean Morlet being the pioneers of the continuous wavelet transform, I will present here only a selection of the results I obtained with it between 1986 to 1996 in collaboration with several students and colleagues, using the complex-valued progressive `Morlet wavelet'. This selection is organised in three parts devoted to the representation, then the analysis and finally the filtering of two- and three-dimensional turbulent flows, in order to better understand and predict their non-linear behaviour.

\bigskip
\bigskip


\section{Continuous wavelets to represent turbulent flows}
\label{sec:5}

\subsection{Choice of the mother wavelet \\ and notation for the wavelet coefficients}
\label{subsec:5.1}

The `mother wavelet' is the function from which the `wavelet family' of translated and dilated wavelets is generated. It must be compatible with the admissibility condition required for the wavelet transform to be an isometry between physical space and wavelet space, which guaranties its invertibility without errors. For the purpose of analysis I recommend using a complex-valued progressive wavelet, such as the Morlet wavelet. Indeed, by visualising the modulus and the phase of complex-valued wavelet coefficients one eliminates the oscillations of real-valued wavelet coefficients, due to the quadrature between their real and imaginary parts. If the wavelet is real-valued, as is the case for orthogonal wavelets, the wavelet coefficients oscillate as much as the wavelet itself, at all scales and positions, and their visualisation and interpretation become tedious. Moreover, the redundancy of the continuous wavelet coefficients allows the information to be deployed in space and scale on the full grid, as opposed to the dyadic grid used for orthogonal wavelets which also impairs the readability of the wavelet coefficients. For data analysis I recommend complex-valued continuous wavelets, but for compression, orthogonal wavelets must be preferred, such as the Coifman 12 wavelet which offers a good compromise between spatial and spectral localisation. The main advantage of the orthogonal wavelet transform is its computational efficiency, since the computation of the orthogonal wavelet transform for a one-dimensional signal discretised on $N$ grid points requires only $2MN$ operations, where $M$ is the length of the quadratic mirror filter \cite{Farge 1992} associated with the chosen wavelet ({\it e.g.}, $M=12$ for the Coifman 12 wavelet). In contrast, the computation of the continuous wavelet transform of the same signal  of length $N$ requires $sN^2\log_2N$ operations, where $s$ is the number of voices per octave. The main drawback of the orthogonal wavelet transform is the loss of both the invariance under dilation and the invariance under translation, because the dyadic grid samples scale by octaves, and space by discrete steps whose size varies with scale.

\bigskip

In my first publication on wavelets in 1988 \cite{Farge et al. 1988}, I suggested the notation $\widetilde f$ for the wavelet coefficients of a function $f$, by analogy with the notation $\widehat f$ used for the Fourier coefficients. I emphasise the need for a standard notation in order to avoid the risks of misinterpretation if the same notation is used for different notions. Moreover, this notation has the advantage of preserving the information on the transformed variable, {\it e.g.}, $\widehat V$ for the velocity, or $\widehat P$ for the pressure. Indeed, it is important to be able to identify at a glance which variable is designated, and to know whether it is represented in physical space, Fourier space or wavelet space. Unfortunately, a few years after my suggestion of using the tilde to identify wavelet coefficients, it was also used to denote biorthogonal wavelets. Despite the risk of confusion resulting from the same symbol designating different concepts, I decided to stick to this notation for the reasons given above.

\bigskip

\subsection{Continuous wavelet representation in one dimension}
\label{subsec:5.2}

\begin{figure}[ht!]
\centering
\vspace{2cm}
\includegraphics[width=1.1\linewidth]{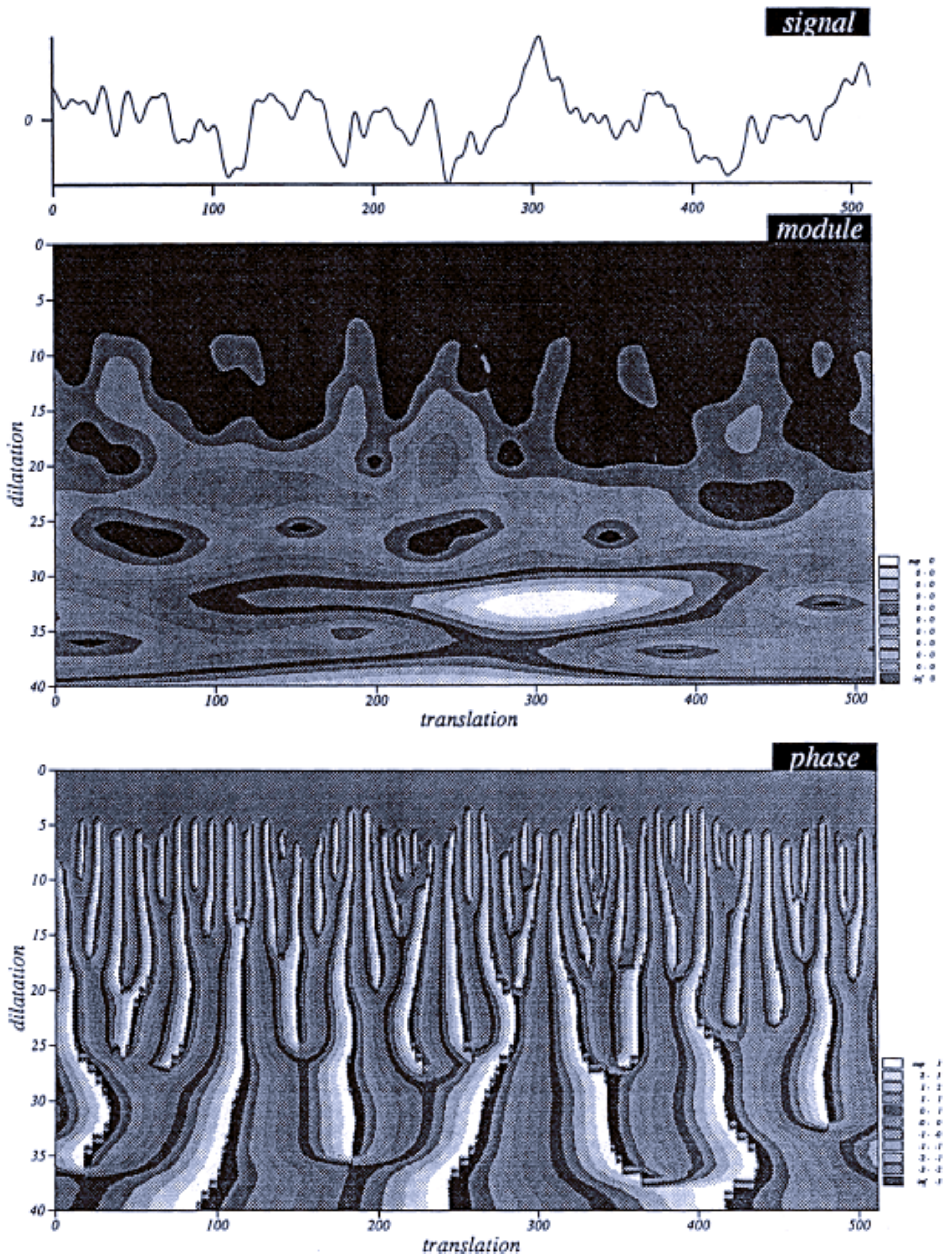}\\
  \caption{One-dimensional continuous wavelet representation, with a complex-valued Morlet wavelet, of a one-dimensional cut from the vorticity field of a two-dimensional turbulent flow, computed by a direct numerical simulation of the two-dimensional Navier--Stokes equation with periodic boundary conditions, using a pseudo-spectral method with resolution $N=512^2$. The complex-valued wavelet coefficients are plotted with a linear horizontal axis for the position, and a logarithmic vertical axis for the scale, with the largest scale at the bottom and the smallest scale at the top. Top: one-dimensional cut of the vorticity field. Middle: modulus of its complex-valued wavelet coefficients. Bottom: phase of its complex-valued wavelet coefficients \cite{Farge et al. 1988}.}
  \label{Figure7}
\end{figure}

In 1986 I studied the evolution of turbulent flows computed by direct numerical simulation of the Navier--Stokes equation in two dimensions and I represented in wavelet space one-dimensional cuts of the vorticity field taken at successive instants. I computed their one-dimensional continuous wavelet transform with a complex-valued Morlet wavelet, using the {\it Fortran} code developed by Ginette Saracco and Richard Kronland-Martinet. In collaboration with Gabriel Rabreau \cite{Farge et al. 1988} I found that initially the modulus and the phase of the wavelet coefficients of vorticity are spread out in space and in scale, but later when vortices interact within each others and merge, the modulus becomes increasingly concentrated in both space and scale, while the isolines of the phase point towards the centre of each vortex ({\bf [Figure \ref{Figure7}]}). The smallest scales of these turbulent flows tend to be concentrated in the centre of the vortices, because when two neighbouring vortices interact strongly they produce vorticity filaments and exhibit a cusp-like shape (as can be observed in {\bf [Figure \ref{Figure1}]}). In contrast, when vortices are far from each other they no longer produce vorticity filaments and, due to dissipation by the viscous fluid, they relax to a Gaussian shape \cite{Farge et al. 1988}, \cite{Farge 1989}. A few year later, in collaboration with Matthias Holschneider, we proved that such cusp-shaped vortices remain stable with respect to Navier--Stokes dynamics, due to the vorticity filaments they have produced, which protect them from destabilisation by other vortices \cite{Farge et al. 1992a}.

\bigskip


\clearpage

\begin{figure}[ht!]
\centering
\vspace{2cm}
\hspace{-2.38cm}
\includegraphics[width=1.2\linewidth]{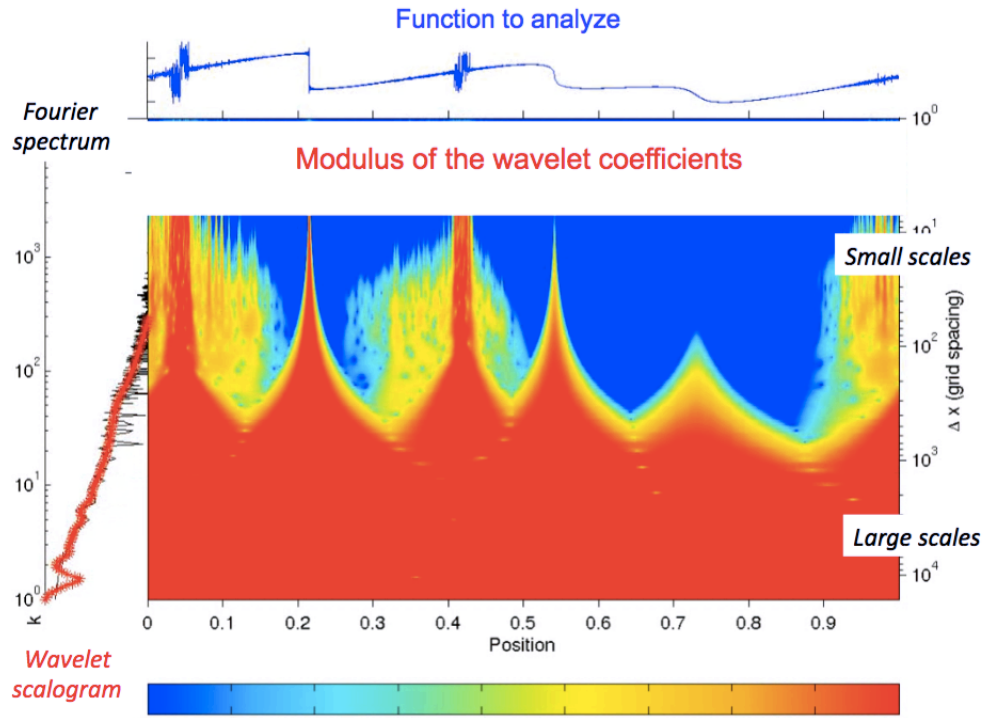}\\
  \caption{Continuous wavelet representation, with a complex-valued Morlet wavelet, of one instant of the solution of the one-dimensional Burgers equation with periodic boundary conditions, computed using a pseudo-spectral method with resolution $n=2^{14}=16384$. We show the solution in physical space (the one-dimensional plot above), the modulus of its wavelet coefficients (the two-dimensional plot below), and the Fourier spectrum in black and the scalogram in red  (the one-dimensional plot on the left). We observe the formation of shocks which produce resonances due to Galerkin truncation that induce numerical noise \cite{Pereira et al. 2013}.}
  \label{Figure8}
\end{figure}

A few years later, in collaboration with Jori Ruppert-Felsot who was doing his postdoc with me, we developed a {\it Matlab} code to compute the continuous complex-valued wavelet transform in one dimension. We applied it to represent with a Morlet wavelet the formation of shocks in the solution of the one-dimensional inviscid Burgers equation. Starting from a smooth initial condition we computed the evolution using a pseudo-spectral method with a finite number of Fourier modes at resolution $N=2^{14}= 16384$ \cite{Pereira et al. 2013}. In {\bf [Figure \ref{Figure8}]} one can see :
\begin{itemize}
\item
the solution of inviscid Burgers equation, as a function of the position $x=n_x \Delta x$, with $n_x=[1,N]$ and $\Delta x$ the grid size,
\item
the modulus of its complex-valued wavelet coefficients, as a function of the position $x$ and of the logarithm of the equivalent wavenumber $k=[1,\frac{N}{2}]$, which is the inverse of the scale divided by the centroid wavenumber of the chosen wavelet ({\it i.e.}, the barycentre of the wavelet support in Fourier space).
\item
the scalogram showing the enstrophy (square of the modulus of vorticity) as a function of the scale (the red line), 
\item
the spectrogram showing the enstrophy as a function of the wavenumber.
 \end{itemize}

\begin{figure}[ht!]
\centering
\includegraphics[width=1.3\linewidth]{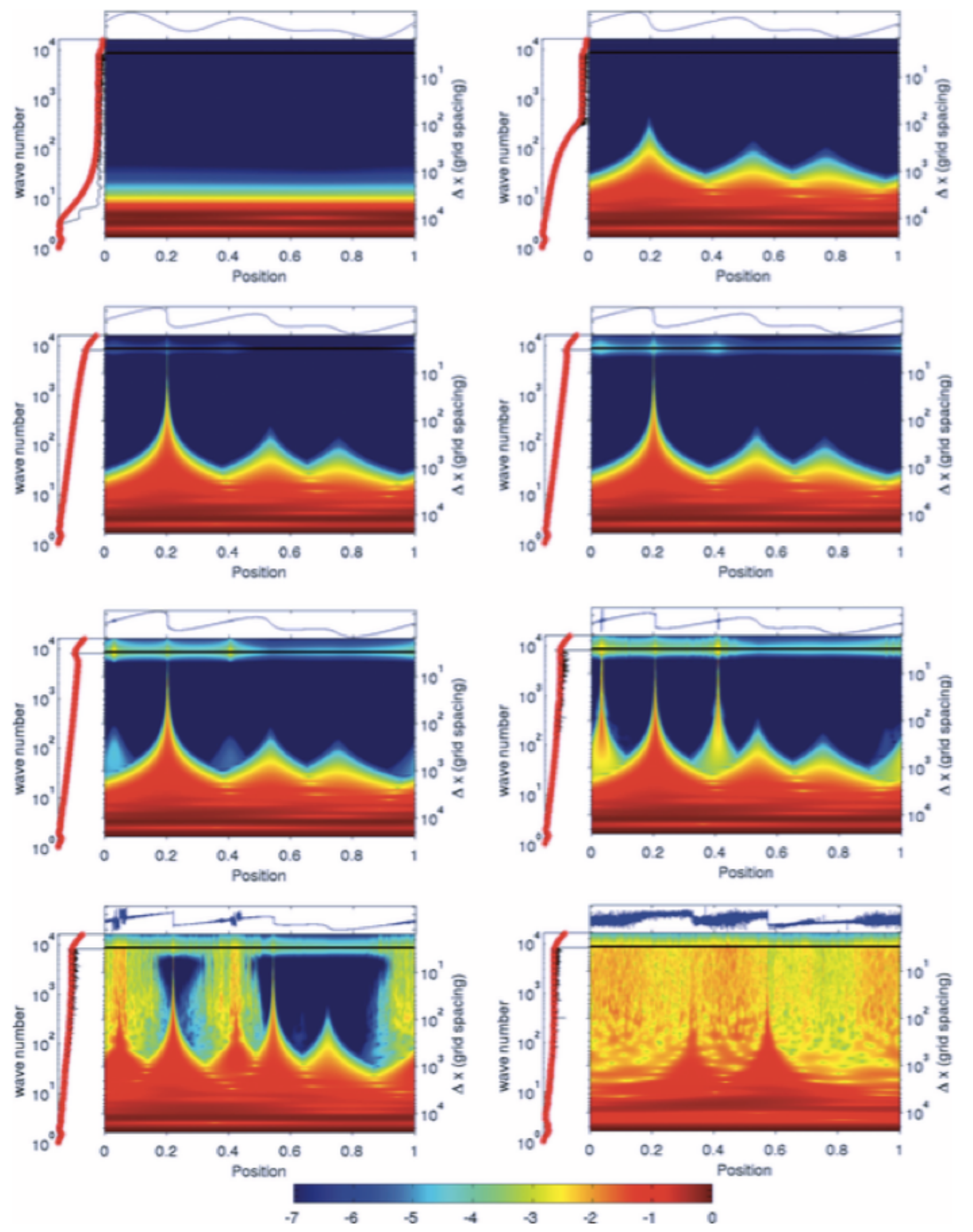}\\
  \caption{Continuous wavelet representation, with a complex-valued Morlet wavelet, of eight instants of the solution of the one-dimensional Burgers equation with periodic boundary conditions, computed using a pseudo-spectral method with resolution $n=2^{14}=16384$. For each instant we show the solution in physical space (the one-dimensional plot above), the modulus of its wavelet coefficients (the two-dimensional plot below), and the Fourier spectrum in black and the scalogram in red  (the one-dimensional plot on the left). One observes the formation of shocks which produce resonances and noise due to the Galerkin truncation \cite{Pereira et al. 2013}.}
  \label{Figure9}
\end{figure}

\noindent
The Burgers equation is interesting because it includes a quadratic non-linear term similar to that in the Navier--Stokes and Euler equations but, in contrast to them, it is integrable analytically. Therefore, we know its exact solution, which is very useful to test numerical Galerkin methods truncated at a finite number of modes. By representing the evolution of its solutions in both space and scale, using a continuous wavelet transform with a Morlet wavelet (see {\bf [Figure \ref{Figure9}]}), we observe that, as soon as shocks develop, truncation produces resonances which spread all over space and scale and become noise. This work has been published in \cite{Pereira et al. 2013}.

\clearpage


\clearpage


\clearpage

\subsection{Continuous wavelet representation in two dimensions}
\label{subsec:5.3}

\begin{figure}[ht!]
\centering
\hspace{-3.5cm}
\includegraphics[width=1.3\linewidth]{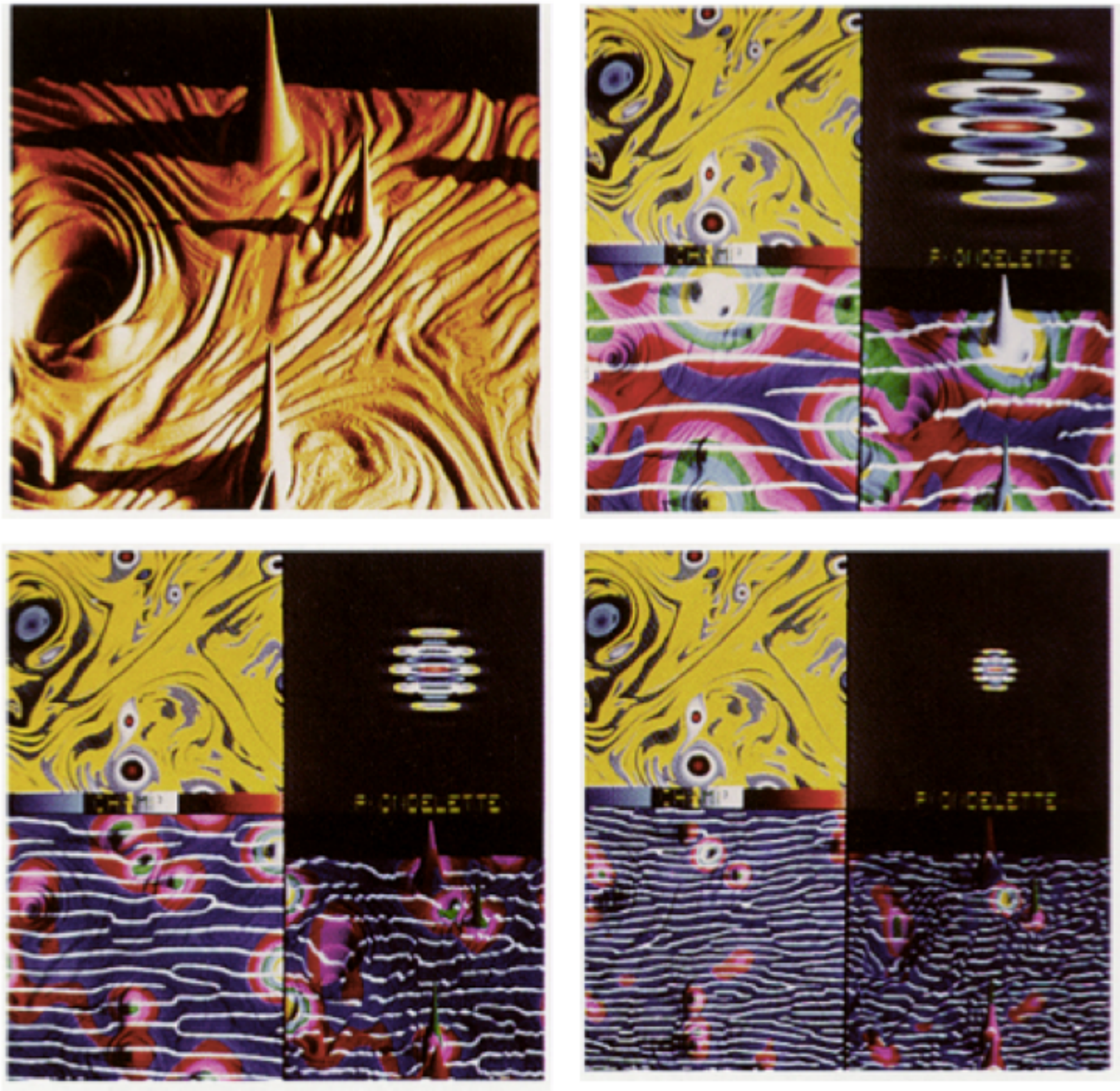}\\
 \caption{Cartographic representation \cite{Farge 1987} of the modulus and the phase of the continuous wavelet coefficients of a vorticity field, from a two-dimensional turbulent flow, computed by direct numerical simulation of the two-dimensional Navier--Stokes equation in a periodic domain, using a pseudo-spectral method at resolution $N=512^2$. Top, left: vorticity field. Top, right: its large-scale wavelet coefficients. Bottom, left: its medium-scale wavelet coefficients. Bottom, right: its small-scale wavelet coefficients \cite{Farge et al. 1990b}.}
  \label{Figure10}
\end{figure}

\begin{figure}[ht!]
\hspace{-3cm}
\includegraphics[width=1.4\linewidth]{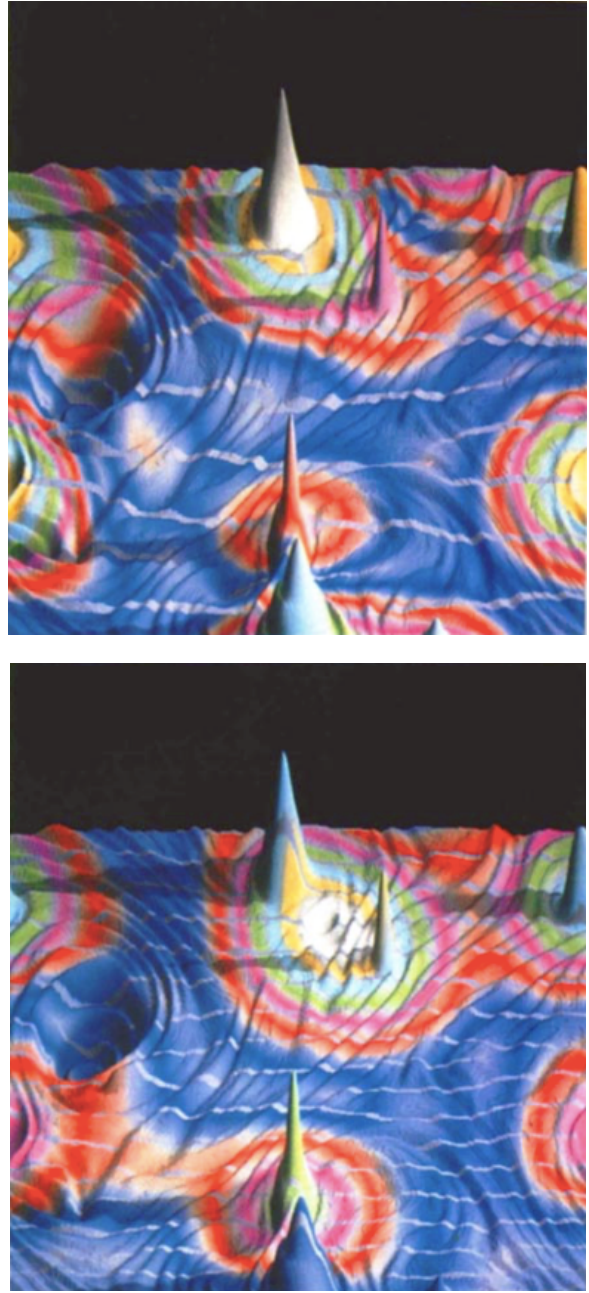}\\
 \caption{Cavalier representation \cite{Farge 1987} of the modulus and the phase of the continuous wavelet coefficients of a vorticity field, from a two-dimensional turbulent flow, computed by direct numerical simulation of the two-dimensional Navier--Stokes equation in a periodic domain, using a pseudo-spectral method at resolution $N=512^2$. Top: large-scale wavelet coefficients. Bottom: small-scale wavelet coefficients \cite{Farge et al. 1990b}.}
  \label{Figure11}
\end{figure}

\begin{figure}[ht!]
\centering
\includegraphics[width=1.6\linewidth]{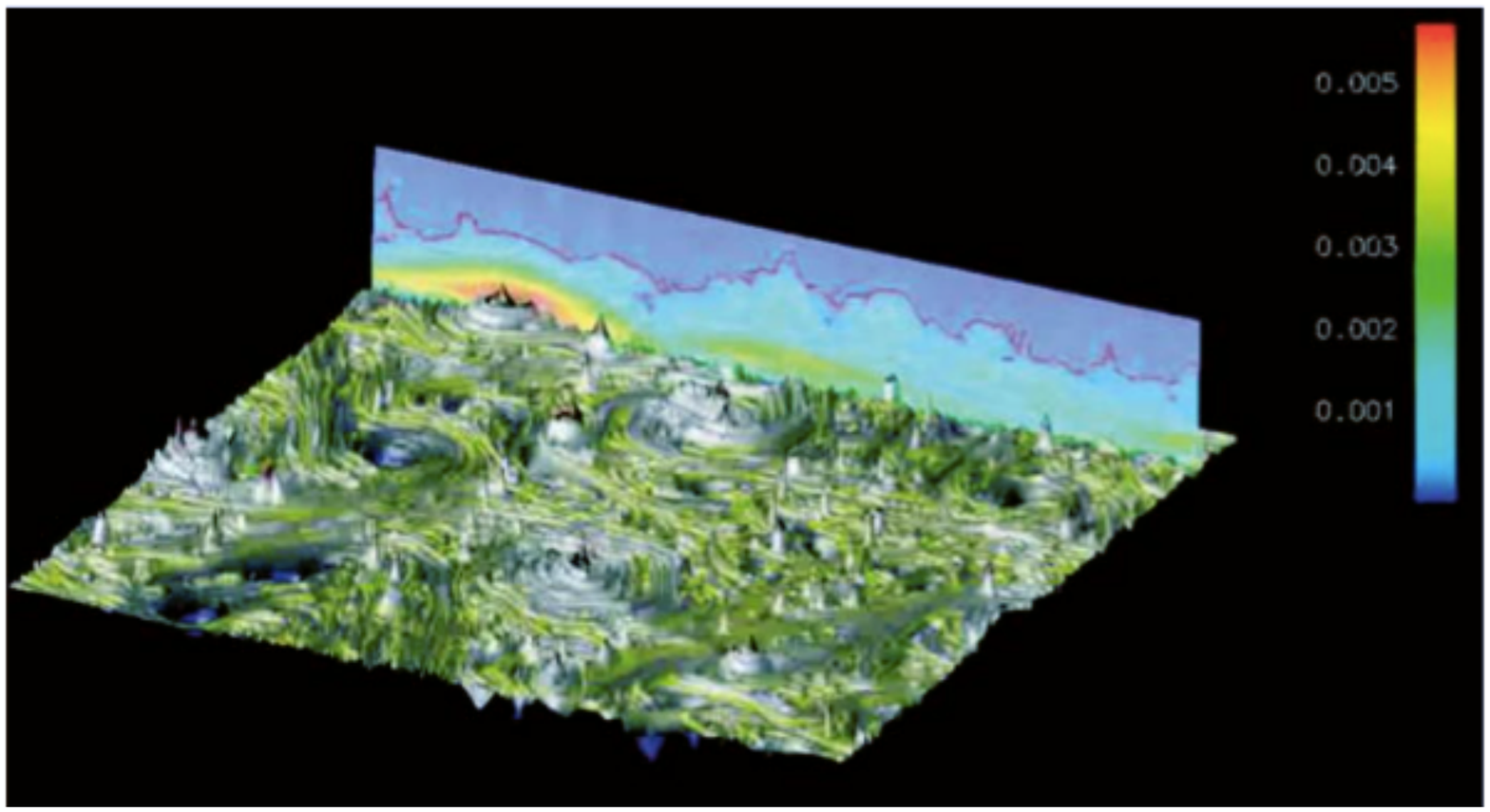}\\
  \caption{Continuous wavelet representation of a two-dimensional turbulent flow studied by numerical experiment. The horizontal plane is a cavalier representation \cite{Farge 1987} of a two-dimensional vorticity field computed by direct numerical simulation, which is an instantaneous solution of two-dimensional the Navier--Stokes equation. The vertical cut is a lin-log two-dimensional plot of the modulus of its continuous wavelet coefficients using a Morlet complex-valued wavelet, $|\widetilde{\omega}(\vec x, l)|$, represented with a rainbow colour scale, which shows the space and scale distribution of enstrophy ({\it i.e.}, vorticity squared). On this plane the red line indicates the value of the threshold $T$ which separates the enstrophy of the coherent flow due to the vortices, $|\widetilde{\omega}(\vec x, l)| > T$, from the incoherent background flow, $|\widetilde{\omega}(\vec x, l)| \leq T$ \cite{Schneider et al. 2006}.}
  \label{Figure12}
\end{figure}

\begin{figure}[ht!]
\centering
\includegraphics[width=1.4\linewidth]{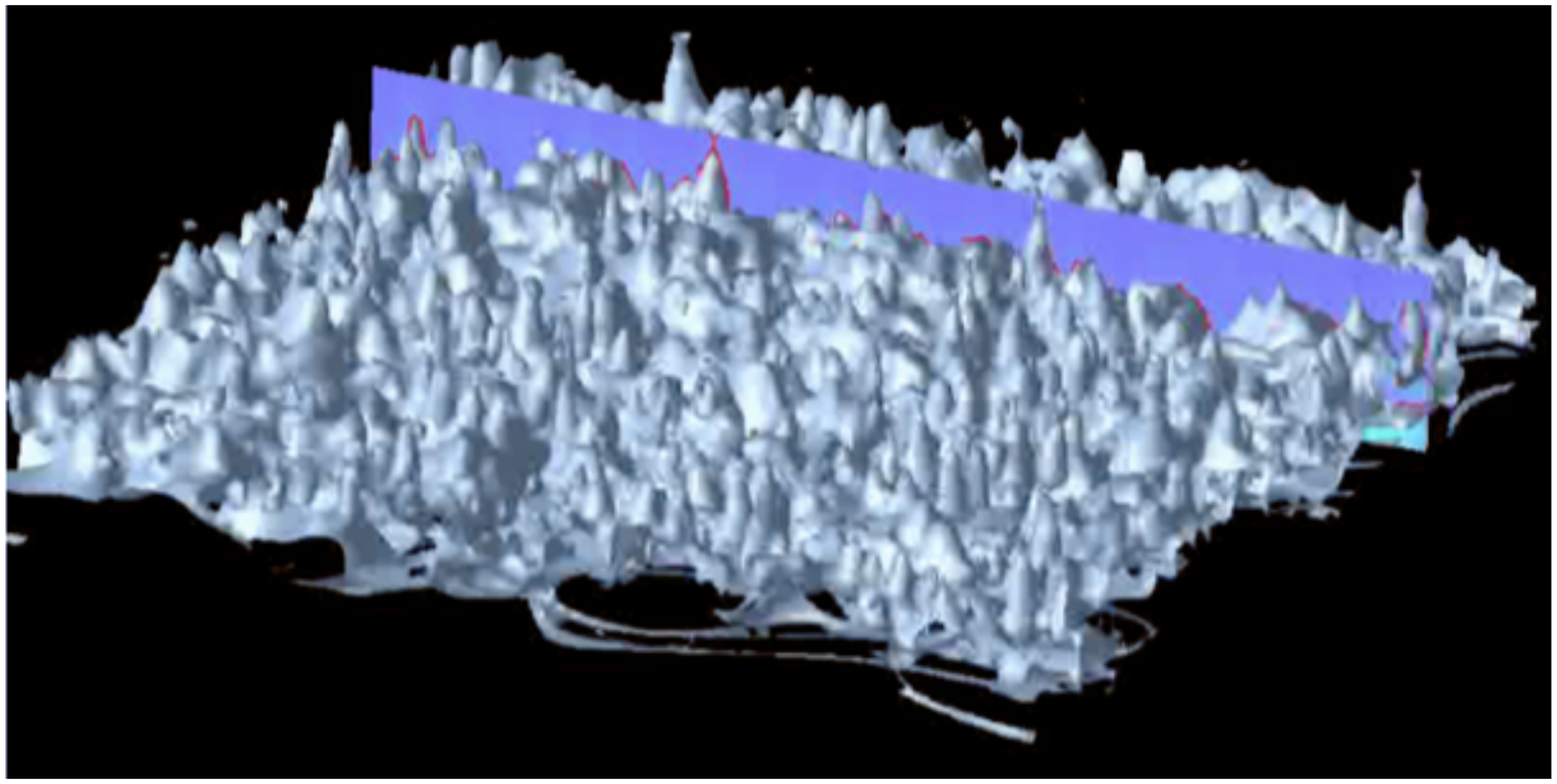}\\
  \caption{Continuous wavelet representation of a two-dimensional turbulent flow studied by numerical experiment. The continuous wavelet coefficients (using a Morlet complex-valued wavelet), $|\widetilde{\omega}(\vec x, l)|$, of a two-dimensional vorticity field computed by direct numerical simulation are shown on a three-dimensional lin-log plot, where the two linear horizontal axes correspond to space and the logarithmic vertical axis to scale. The grey isosurface $|\widetilde{\omega}(\vec x, l)| = T$ is represented in gray and separates the enstrophy of the coherent flow due to the vortices, where $|\widetilde{\omega}(\vec x, l)| > T$, from the incoherent background flow, where $|\widetilde{\omega}(\vec x, l)| \leq T$ \cite{Schneider et al. 2006}.}
  \label{Figure13}
\end{figure}

\begin{figure}[ht!]
\centering
\includegraphics[width=0.9\linewidth]{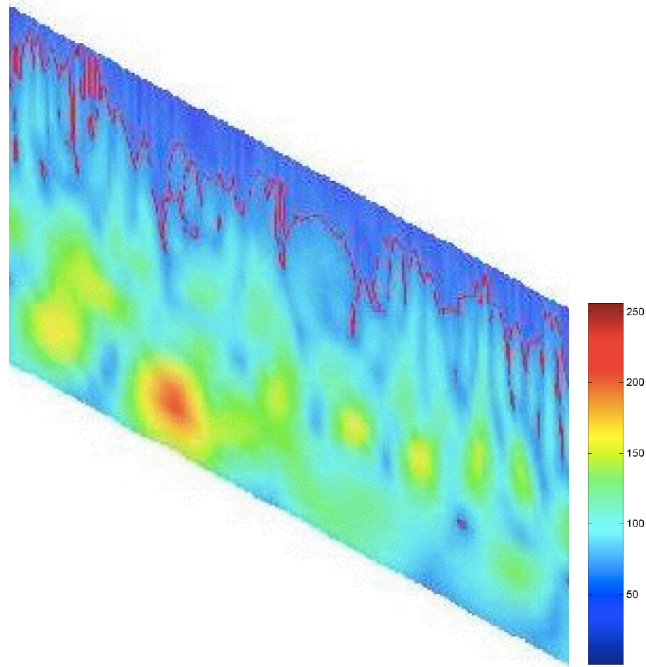}\\
  \caption{Continuous wavelet representation of a two-dimensional turbulent flow studied by laboratory experiment. The lin-log plot represents the modulus of the continuous wavelet coefficients (using a Morlet complex-valued wavelet), $|\widetilde{\omega}(\vec x, l)|$, of a two-dimensional vorticity field measured by particle image velocimetry (PIV) in a rotating tank. The red isoline $|\widetilde{\omega}(\vec x, l)| = T$ shows the separatrix between the threshold $T$ corresponds to a space-scale manifold which separates the enstrophy of the coherent flow due to the vortices, where $|\widetilde{\omega}(\vec x, l)| > T$, from the incoherent background flow, where $|\widetilde{\omega}(\vec x, l)| \leq T$.}
  \label{Figure14}
\end{figure}

In 1989 Romain Murenzi, while working on his PhD with Alex Grossmann, developed the continuous wavelet transform in higher dimensions \cite{Murenzi 1989}. For this they generalised the framework of the continuous wavelet transform which in one dimension is based on the affine group. They also generalised to higher dimensions the admissibility condition required for the choice of the analysing wavelet in order to guarantee an exact reconstruction, as proved by Alex Grossmann and Jean Morlet in 1984 \cite{Grossmann Morlet 1984}). They developed together the continuous wavelet transform in $n \geq 2$ dimensions by replacing the affine group by the Euclidean group with rotations. The same year I used their work to develop, in collaboration with Matthias Holschneider, a {\it Fortran} code to compute the continuous wavelet transform of a two-dimensional field $F(\vec x)$, with $\vec x \in R^2$. The resulting wavelet coefficients constitute a four-dimensional field $\widetilde F(\vec x, l, \theta)$, which depends on the two space variables $\vec x=(x,y)$ (scanned by the translation operator), plus one scale variable $l$ (scanned by the dilation operator), and one angular variable $\theta$ (scanned by the rotation operator). It is quite difficult to visualise such a four-dimensional complex-valued field. Therefore, in collaboration with Jean-Fran\c cois Colonna, we combined a cavalier view \cite{Farge 1987} of the field represented in physical space, with a cartographic view \cite{Farge 1987} of its wavelet representation, using a rainbow colour-scale for the modulus of the wavelet coefficients, and white isolines for the zeros of their phase to indicate the angular direction. We represented in wavelet space both a numerical experiment and a laboratory experiment: the former is a two-dimensional vorticity field computed by direct numerical simulation of the two-dimensional Navier--Stokes equation (see {\bf [Figure \ref{Figure10}, \ref{Figure11}, \ref{Figure12}, \ref{Figure13}, \ref{Figure14}]}), the latter is another two-dimensional vorticity field measured by particle image velocimetry in a rotating tank, where rotation forces the flow to remain two-dimensional. These results  \cite{Farge et al. 1990b} confirm those we had obtained with the one-dimensional wavelet representation \cite{Farge et al. 1988} (see {\bf [Figure \ref{Figure7}]}). Indeed, in both cases we observe that, during the flow evolution enstrophy (the square of the vorticity modulus, analogous to what energy is for velocity) tends to concentrate in the vortex cores.

\clearpage


\clearpage


\clearpage

\noindent
{\bf [Figure \ref{Figure12}]} shows one snapshot of 
{\bf [Movie \href{http://wavelets.ens.fr/TURBULENCE/3_MOVIES/Movie_4_Modulus_of_the_continuous_wavelet_coefficients_of_the_vorticity_of_a_2D_turbulent_flow_computed_by_DNS.mpg}{4}]}, where the modulus is visualised with a rainbow scale (with values increasing from blue to orange) and the red isoline corresponds to the threshold value $T$ separating the coherent flow and the incoherent flow.

\smallskip


\noindent

\clearpage

\noindent
{\bf [Figure \ref{Figure13}]} shows one snapshot of 
{\bf [Movie \href{http://wavelets.ens.fr/TURBULENCE/3_MOVIES/Movie_5_Interface_in_the_continuous_wavelet_space_of_vorticity_separating_the_coherent_vortices_and_the_background_flow.mpg}{5}]}, where the modulus is visualised with a cavalier view \cite{Farge 1987} of the iso-surface which corresponds to the value of the threshold $T$.

\smallskip


\bigskip

\noindent

\clearpage


\clearpage

\begin{figure}[ht!]
\centering
\vspace{3cm}
\includegraphics[width=1.2\linewidth]{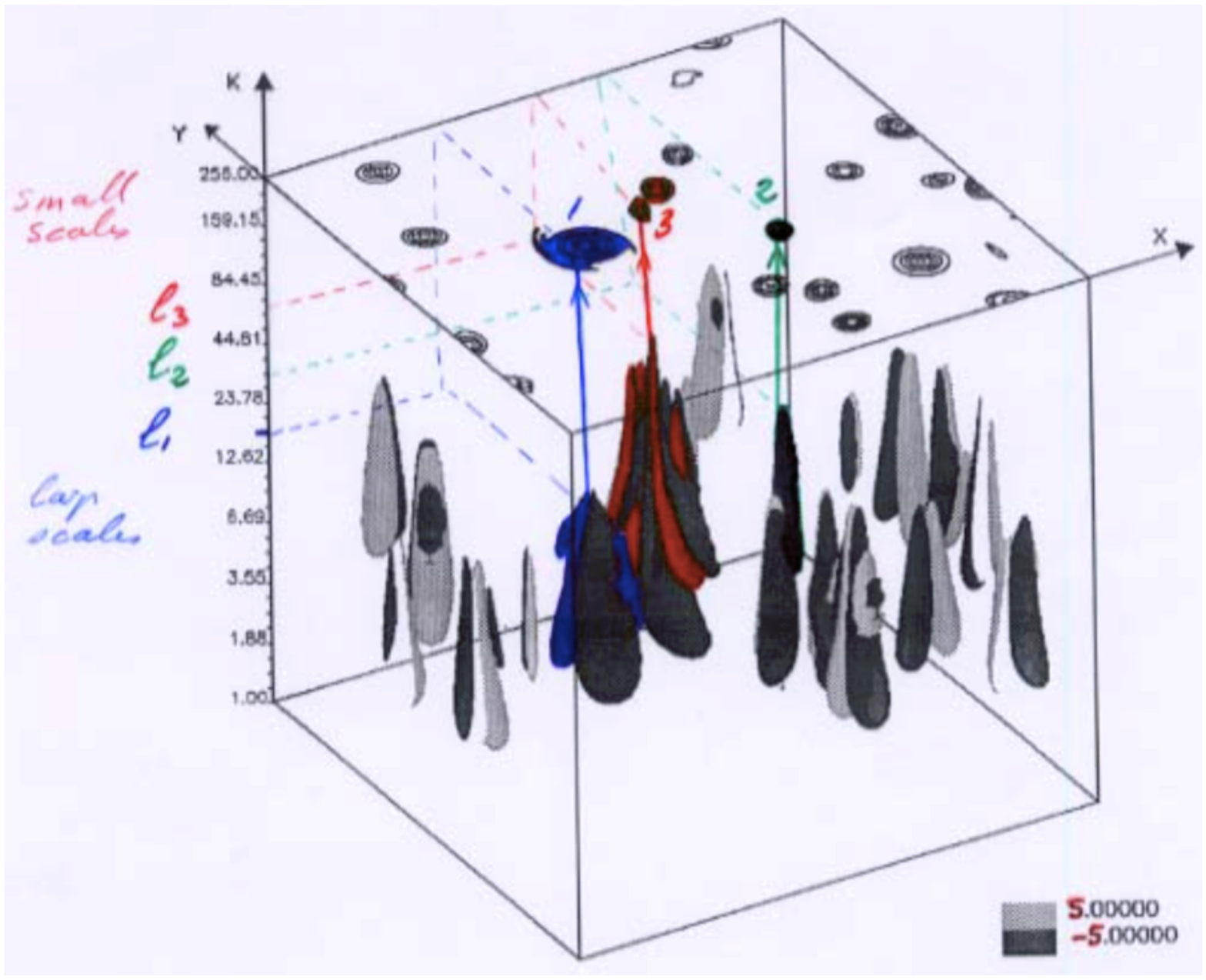}\\
  \caption{Energy transfers of a two-dimensional turbulent flow studied by numerical experiment seen via the continuous wavelet representation \cite{Farge et al. 1993b}.}
  \label{Figure15}
\end{figure}

We also computed the continuous wavelet transform of the non-linear term of the Navier--Stokes equation using a Morlet complex-valued wavelet to obtain the energy transfers, as shown in {\bf [Figure \ref{Figure15}]}. We observe that where vortices non-linearly interact, {\it e.g.}, during the merging of opposite-sign vortices, energy is spread in scale towards small scales. This wavelet representation shows that the `turbulent cascade' is local in both space and scale, while the Fourier representation is blind to physical space.

\bigskip


\bigskip
\bigskip


\section{Continuous wavelets to analyse turbulent flows}
\label{sec:6}

In 1992 {\it Annual Review of Fluid Mechanics} invited me to write a review on {\it `Wavelet transforms and their applications to turbulence'}, where I introduced several diagnostics based on the continuous wavelet coefficients \cite{Farge 1992}. I will present three of them: the local energy spectrum, the space-scale Reynolds number and the intermittency measure. 

\subsection{Scalogram and local spectrum}
\label{subsec:6.1}

A `local spectrum' is a paradoxical notion since, due to the uncertainty principle, one cannot be both local in physical space and local in spectral space. In contrast, there is no longer a paradox if, instead of wavenumbers, we think in terms of scales and consider the `scalogram'. To this end, we compute the continuous wavelet coefficients of each velocity component and take the square of their modulus to get the energy density as a function of both space and scale. If we integrate in space and plot the logarithm of the energy versus the logarithm of the scale we obtain the energy scalogram. To convert it to a local energy spectrum we compute an equivalent wavenumber by renormalising the scale with the centroid wavenumber of the chosen wavelet ({\it i.e.}, the barycentre of the wavelet support in Fourier space). We thus obtain thus a `stabilised energy spectrogram', which is the Fourier spectrum weighted by the square of the Fourier transform of the analysing wavelet at each wavenumber \cite{Farge et al. 1996}. Therefore we no longer need to use an {\it ad hoc} method ({\it e.g.}, Welch or Bartlett methods) to stabilise the spectrogram by reducing the noise due to the finite number of data samples.

\bigskip

We can segment the velocity field into regions exhibiting different dynamic behaviour, using the Weiss criterion \cite{Weiss 1991}: 

\begin{itemize}
\item
the elliptical regions corresponding to the cores of the coherent structures where rotation exceeds strain, 
\item
the parabolic regions corresponding to the shear layers at the periphery of the coherent structures where rotation is in balance with strain and velocity is strong,
\item
the hyperbolic regions corresponding to the vorticity filaments of the incoherent background flow where strain exceeds rotation and velocity is weak.
\end{itemize}

\noindent
Then, instead of integrating all wavelet coefficients in space at once, we integrate them separately according to the flow regions to which they belong.

\bigskip

\begin{figure}[ht!]
\centering
\includegraphics[width=1.0\linewidth]{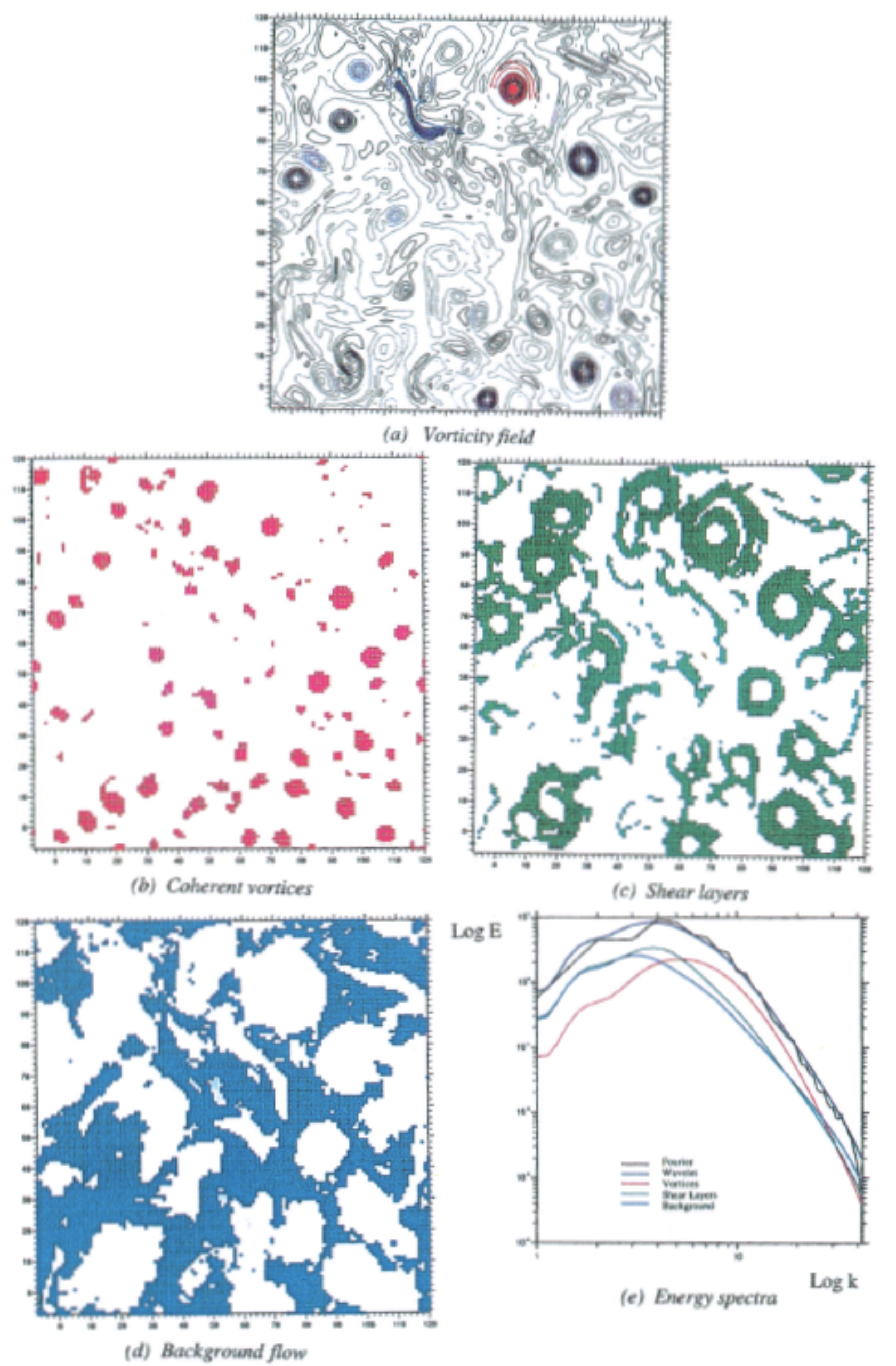}\\
  \caption{Conditional energy spectra. (a) Vorticity field of a turbulent flow, computed by direct numerical simulation of the two-dimensional Navier--Stokes equation in a periodic domain, using a pseudo-spectral method at resolution $N=128^2$ grid points. (b) Elliptic region where rotation dominates strain, due to vortices. (c) Parabolic region where rotation and strain are balanced, due to shear layers. (d) Hyperbolic region where strain dominates rotation, due to incoherent background. (e) The corresponding energy spectra: in black the Fourier spectrum (also called spectrogram) of the total flow which scales as $k^{-4.5}$, in brown the scalogram of the total flow which is a smoothed Fourier spectrum (stabilised spectrogram), in red the conditional spectrum of the vortical flow which scales as $k^{-5}$, in green the conditional spectrum of the sheared flow which scales as $k^{-4}$, and in blue the conditional spectrum of the background flow which scales as $k^{-3}$ \cite{Farge et al. 1996}.}
  \label{Figure16}
\end{figure}

{\bf [Figure \ref{Figure16}]} shows three local energy spectra computed by integrating in space the continuous wavelet coefficients of the velocity field of a two-dimensional turbulent flow. Using the Weiss criterion \cite{Weiss 1991} it has been segmented into regions characterised by three different dynamics, namely elliptical, parabolic and hyperbolic.We observe that the coherent flow (corresponding to elliptical regions) scales like $k^{-6}$, the sheared flow (corresponding to the parabolic regions) scales like  $k^{-4}$, while the incoherent flow (corresponding to hyperbolic regions) scales like $k^{-3}$. Therefore, the more coherent and bursty the flow, the steeper its spectrum, while the more incoherent and homogeneous the flow, the flatter its spectrum. 

\clearpage


\clearpage

\subsection{Intermittency measure}
\label{subsec:6.2}

Intermittency is defined as localised bursts of high-frequency activity, and therefore intermittent phenomena are localised in both physical and spectral space. Hence, a suitable basis for representing intermittency should reflect this dual localisation. The Fourier basis is well localised in spectral space but delocalised in physical space, therefore when a signal or a field is filtered using a high-pass Fourier filter and then reconstructed in physical space, its spatial information is lost. Moreover, Fourier filtering smooths strong gradients and produces spurious oscillations, because the phases of the discarded Fourier modes are lost, which impairs the measure of intermittency. When for instance a scalar field $F(\vec x)$ ({\it e.g.}, a velocity derivative) is intermittent it contains rare but strong events, {\it i.e.}, bursts of intense activity, which correspond to large deviations reflected in the heavy tails of its probability distribution function. Second-order statistics ({\it e.g.}, energy spectrum, second-order structure function, flatness) are relatively insensitive to such rare events because their time or space support is very small and does not dominate the integrals that compute them. In contrast, rare events become increasingly important for higher-order statistics, where they ultimately become dominant. Therefore the higher their order, the more sensitive statistics are to intermittency.

\bigskip

In \cite{Farge et al. 1990a} I proposed to define the intermittency measure $I(\vec x,l)$ of a field $F$ as the square of its wavelet coefficients modulus $|\widetilde{F}(\vec x,l)|^2$, renormalised by its spatial average at each scale. This provides information on the spatial variance of its energy at a given scale and therefore:

\begin{itemize}
\item 
if $I(\vec x,l)=1$, the field is homogeneous and non-intermittent, because it does not exhibit spatial variance of its energy at this scale,
\item
if $I(\vec x,l) > 1$, the field is intermittent, because there is some spatial variance of its energy at a given scale which comes from some excited regions of limited spatial support, 
\item
if $I(\vec x,l) \gg 1$, the field is very intermittent, because all spatial variance of its energy at a given scale comes from highly excited regions of small spatial support.
\end{itemize}

\begin{figure}[ht!]
\centering
\hspace{-2.3cm}
\includegraphics[width=1.2\linewidth]{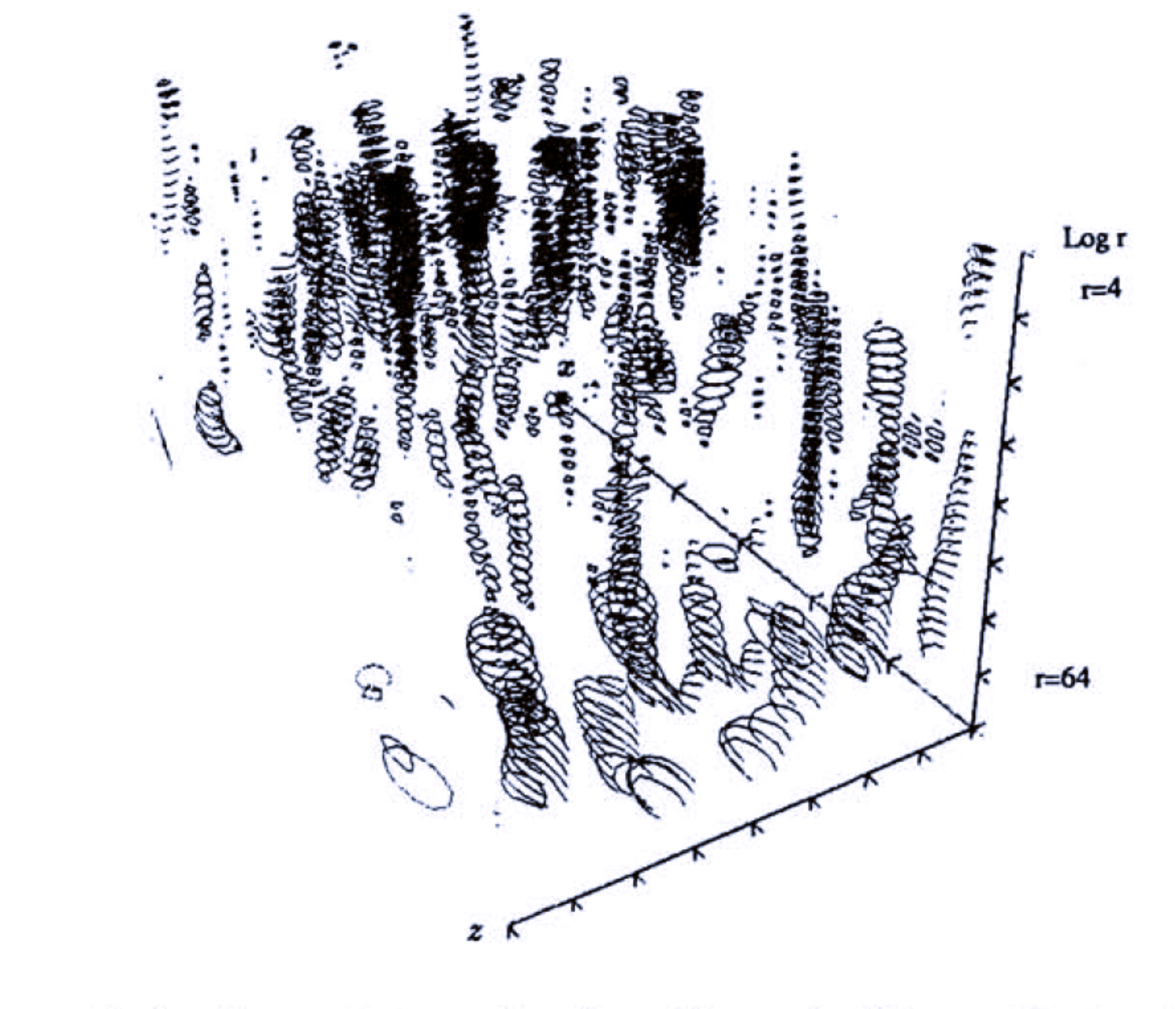}\\
 \caption{Intermittency measure of the vertical vorticity component of a turbulent three-dimensional mixing layer, computed by direct numerical simulation using a Galerkin spectral method with $N= 64 \times 128 \times 64$ Fourier-Jacobi modes \cite{Rogers and Moser 1992, Farge 1990}. The horizontal plane represents the spanwise direction $x$ and the streamwise direction $z$, while the vertical logarithmic axis $r$ goes from large scale $r=64$ to small scale $r=4$. The strongest intermittency value, $I (\vec x,l)=145$ (in black), is reached at small scales. It is localised in regions where the streamwise vorticity tubes are stretched by the spanwise vortices, which have developed in the central region due to Kelvin-Helmholtz instability \cite{Farge et al. 1990a}.} 
  \label{Figure17}
\end{figure}

\noindent
During the Summer Program 1990 of CTR (Center for Turbulence Research of Stanford University and NASA-Ames), in collaboration with Charles Meneveau who was doing his postdoc there, we used this new diagnostic to analyse a turbulent three-dimensional mixing layer computed by Moser and Rogers (NASA-Ames Research Center, California) using a Galerkin spectral method with $N= 64 \times 128 \times 64$ Fourier-Jacobi modes \cite{Rogers and Moser 1992}. {\bf [Figure \ref{Figure17}]} shows the intermittency measure of the vertical vorticity component. We observe a strong intermittency that increases with scale up to $I (\vec x,l)=145$ and is localised in the central region where the streamwise vorticity tubes, called `ribs', are engulfed and stretched by a few large spanwise vortices formed after the mixing transition, due to the Kelvin-Helmholtz instability. In {\bf [Figure \ref{Figure17}]} we also notice the return to isotropy of the turbulent flow at small scales, because intermittency becomes increasingly isotropic from large to small scales:  at large scales it is elongated in the streamwise direction, while at small scales it becomes isotropic.


\clearpage

\subsection{Space-scale Reynolds number}
\label{subsec:6.3}

\begin{figure}[ht!]
\centering
\hspace{-2.3cm}
\includegraphics[width=1.2\linewidth]{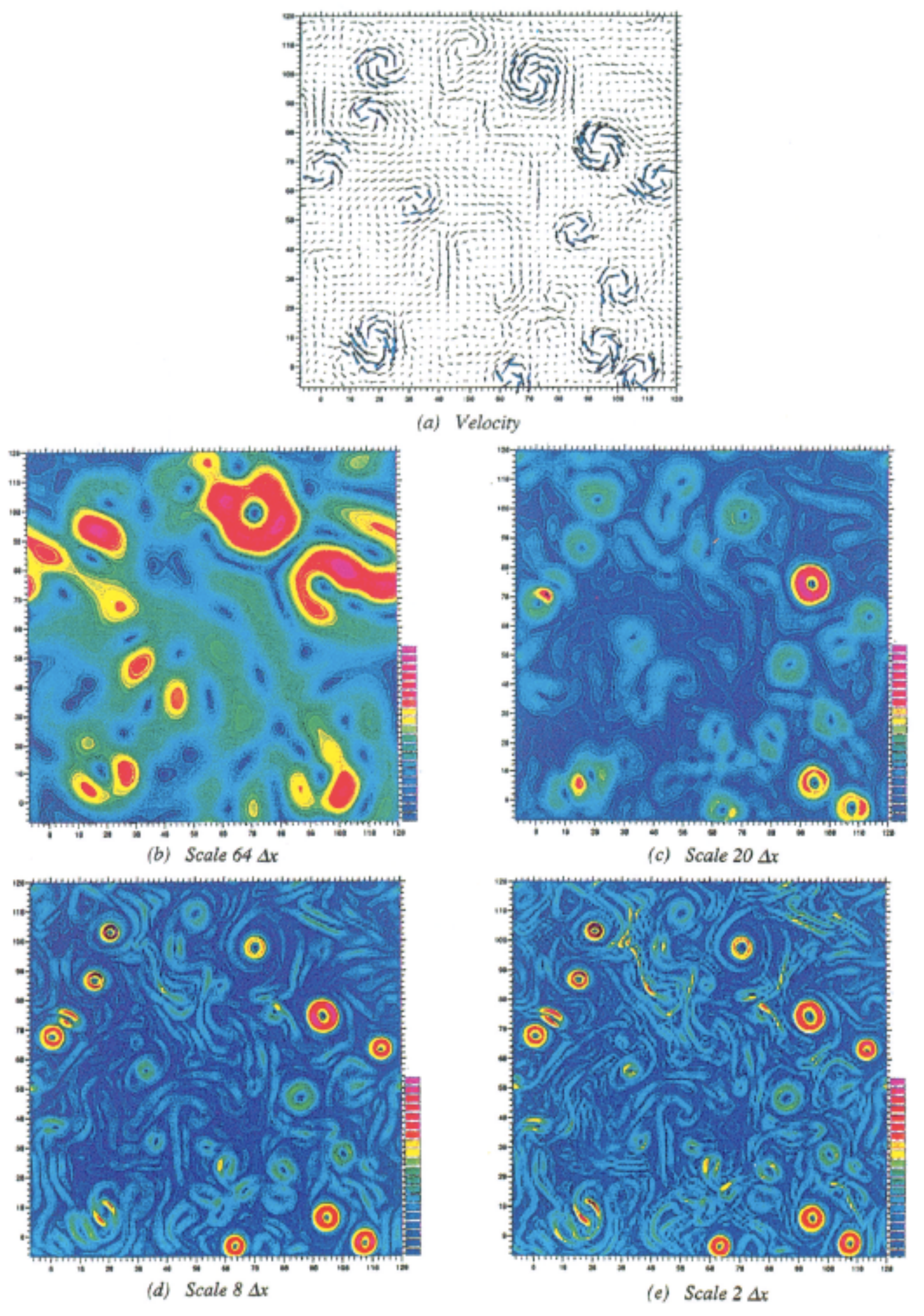}\\
 \caption{Space-scale Reynolds number of a two-dimensional turbulent flow computed by direct numerical simulation of Navier--Stokes equation, with the mesh size $\Delta x$ chosen as unit length. (a) Velocity field computed with $N=128^2$ grid points. (b) Reynolds number at scale $64 \Delta x$, which fluctuates between $148$ and $2700$, with a mean value of $1713$. (c) Reynolds number at scale $20 \Delta x$, which fluctuates between $31$ and $578$, with a mean value of $365$. (d) Reynolds number at scale $2 \Delta x$, which fluctuates between $0$ and $3$, with a mean value of $2$ \cite{Farge et al. 1996}.}
  \label{Figure18}
\end{figure}

In 1990 I introduced a space-scale Reynolds number $Re({\vec x},l)$ defined as the velocity modulus at location $\vec x$ and scale $l$, multiplied by that scale, and divided by the kinematic viscosity of the fluid \cite{Farge et al. 1990b}, \cite{Farge 1992}. Its value varies from $1$ at the Kolmogorov dissipative scale, where dissipation due to the fluid viscosity damps the non-linear instabilities of the turbulent flow, up to the value of the usual Reynolds number $Re$, which is based on velocity modulus at the integral scale ({\it i.e.}, the scale where energy is maximal). {\bf [Figure \ref{Figure18}]} shows the space-scale Reynolds number of a two-dimensional turbulent flow generated by direct numerical simulation of the two-dimensional Navier--Stokes equation in a periodic domain, computed using a pseudo-spectral method with resolution $N= 128^2$.

\begin{figure}[ht!]
\centering
\includegraphics[width=1.2\linewidth]{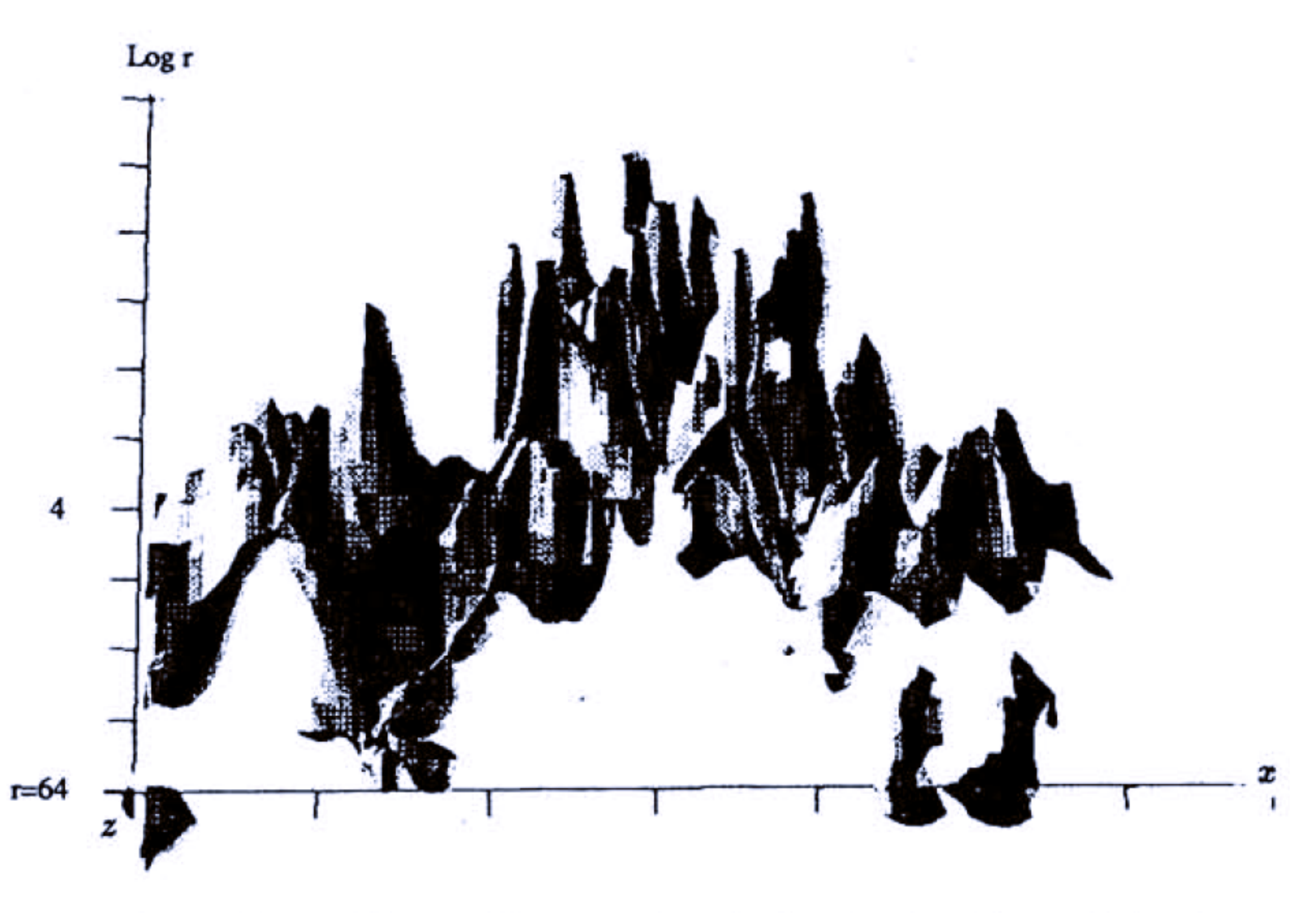}\\
 \caption{Isosurface on which the space-scale Reynolds is equal to one plotted as a function of the streamwise direction $x$ on the horizontal axis and the scale $r$ on the vertical logarithmic axis \cite{Farge et al. 1990a}.}
  \label{Figure19}
\end{figure}

To analyse the space-scale Reynolds number it is also recommended to plot the iso-surface $Re({\vec x},l)=1$, because:

\begin{itemize}
\item
if the iso-surface $Re({\vec x},l)=1$ is flat, there is only one value of the Kolmogorov dissipative scale everywhere and the turbulent flow is non-intermittent, 
\item
if the the iso-surface $Re({\vec x},l)=1$ is not flat, the value of the Kolmogorov scale varies in space from  $\eta_{min}$ to $\eta_{max}$. In this case the turbulent flow is intermittent and the ratio $\frac{\eta_{max}}{\eta_{min}}$ measures the level of intermittency,
\item
if at the smallest scale, {\it i.e.}, the grid size chosen for the computation, there are regions where  $Re({\vec x},l)$ remains larger than one, this means that locally the Kolmogorov dissipative scale is not sufficiently resolved. Therefore in some regions of the flow there are aliasing errors, because the resolved scales are not small enough to insure that the dissipative linear term of Navier--Stokes equation is able to control the non-linear term. In collaboration with Charles M\'eneveau, we found in 1990 such a problem in the results of a direct numerical simulation of a three-dimensional mixing layer \cite{Farge et al. 1990a}. {\bf [Figure \ref{Figure19}]} shows the iso-surface $Re({\vec x},l)=1$ plotted with on the horizontal axis the streamwise direction $x$ and on the vertical logarithmic axis the scale $r$. This diagnostic confirms that this turbulent flow is very intermittent and that the Kolmogorov scale $Re({\vec x},l)=1$ strongly varies in space: the larger the non-linear activity, the smaller the scale where the linear dissipation is able to control and damp any non-linear instabilities \cite{Farge et al. 1990a}.
\end{itemize}

\clearpage


\clearpage


\bigskip
\bigskip


\section{Continuous wavelets to filter turbulent flows}
\label{sec:7}

\subsection{Algorithm to extract vortices}
\label{subsec:7.1}

Because the continuous wavelet transform is invertible, it is always possible to select a subset of the wavelet coefficients and reconstruct a filtered version of a one-dimensional function, or of a $n$-dimensional field, from their wavelet coefficients. In 1991, in collaboration with Eric Goirand and Thierry Phillipovitch who were doing their PhD with me, we developed two methods based on the continuous wavelet transform to extract vortices out of turbulent flows. The first method extracts them individually, one after the other, and the second extracts all of them at once. 
 
 \bigskip
 
To illustrate these methods, we chose a vorticity field from a turbulent flow computed by direct numerical simulation of the two-dimensional Navier--Stokes equation with periodic boundary conditions, using a pseudo-spectral method with resolution $512^2$ (see {\bf [Figure \ref{Figure1}]}, {\bf [Figure \ref{Figure2}]} and {\bf [Figure \ref{Figure3}]}). For both methods, we first computed the continuous wavelet coefficients of the vorticity field $\omega(\vec x)=\omega(x,y)$, with $\vec x \in R^2$, using a Morlet wavelet, and obtained the complex-valued four-dimensional field $\widetilde{\omega}(\vec x, l, \theta)$. We then took its modulus and integrated it over all angles $\theta$ in order to reduce it to a real-valued three-dimensional field  $|\widetilde{\omega}(\vec x, l)|$, which is still quite challenging to visualise. To this end we produced 
{\bf [Movie \href{http://wavelets.ens.fr/TURBULENCE/3_MOVIES/Movie_3_Vorticity_evolution_of_a_2D_turbulent_flow_in_a_circular_domain_with_no_slip_boundary_conditions_computed_by_DNS_and_visualised_with_a_cartographic_view.mpg}{3}]}, with one snapshot in {\bf [Figure \ref{Figure12}]}) where we superposed: 

\begin{itemize} 
\item 
a cavalier projection \cite{Farge 1987} of the vorticity field in physical space $\omega(\vec x)$,
\item
the modulus of the vorticity wavelet coefficients  $|\widetilde{\omega}(x, y, l)|$, 
\item
a moving plane $(x,y)$ where $y$ evolved with time. 
\end{itemize}

\subsection{Extraction of one vortex}
\label{subsec:7.2}

The method for extracting vortices one by one involves the following steps:

\begin{itemize}
\item
sort the maxima of the vorticity $\omega(\vec x)$ by order of magnitude, select the vortex one wishes to extract and find its spatial position $\vec x_v$,
\item
retain the continuous wavelet coefficients $\widetilde{\omega}(\vec x_v, l, \theta)$ which are inside the wavelet influence cone pointing to $\vec x_v$ and cancel those outside the cone,
\item
invert the continuous wavelet transform of the retained continuous wavelet coefficients and reconstruct in physical space the vorticity field of the extracted vortex,
\item
retain the continuous wavelet coefficients which are outside the influence cone corresponding to the extracted vortex and cancel those which are inside the cone,
\item
invert the continuous wavelet transform of the retained continuous wavelet coefficients and reconstruct in physical space the vorticity field without the extracted vortex.
\end{itemize}

\begin{figure}[ht!]
\centering
\hspace{-3.5cm}
\includegraphics[width=1.3\linewidth]{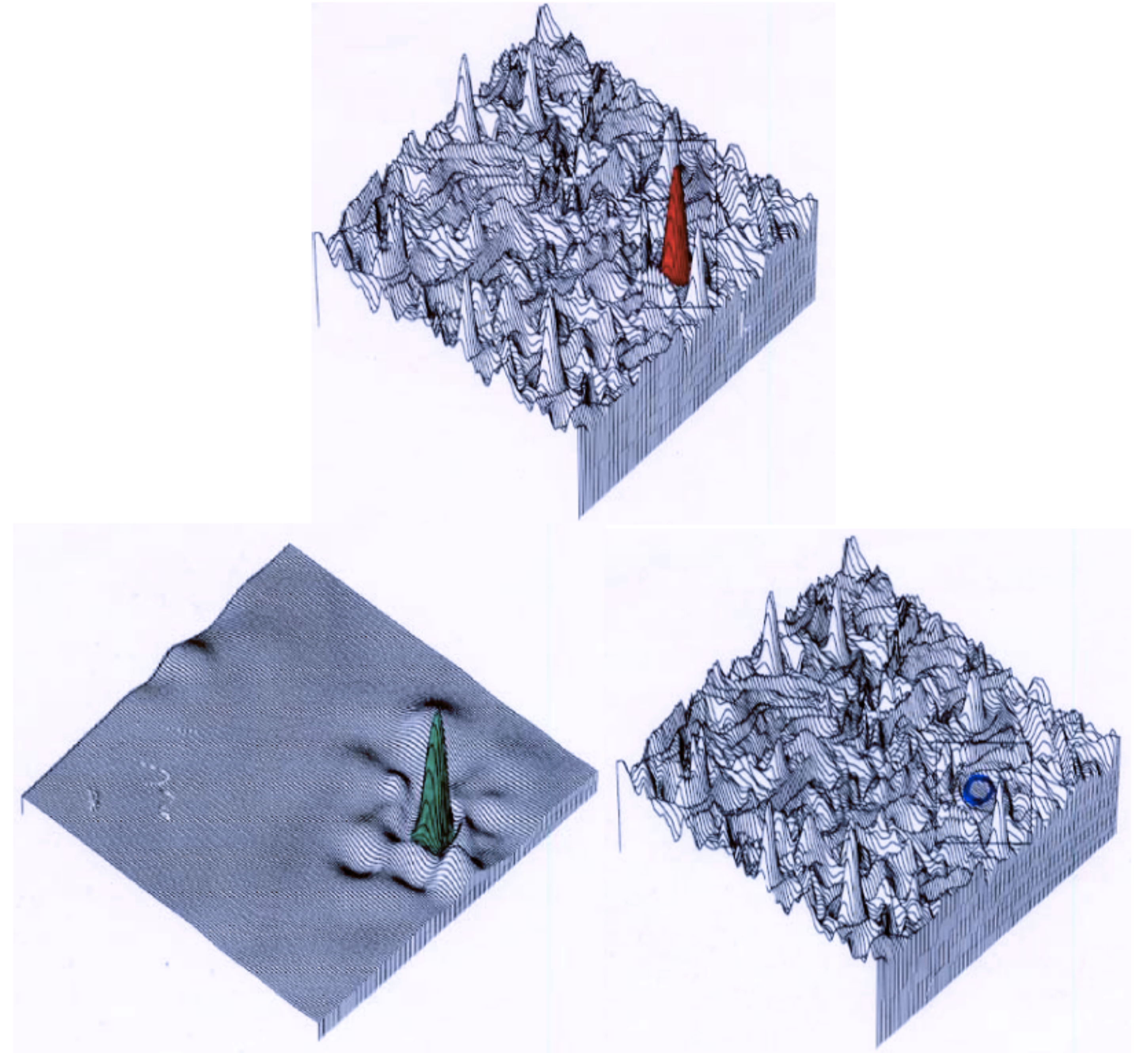}\\
  \caption{Extraction of one vortex out of a two-dimensional vorticity using the continuous wavelet representation. Top: vorticity field with the vortex to extract in red. Bottom left: the filtered vorticity field with only the extracted vortex in green. Bottom right: the filtered vorticity field with the remaining vortices and the location of the missing vortex highlighted in blue \cite{Farge et al. 1993b}.}
  \label{Figure20}
\end{figure}

\noindent
We applied this method to  a vorticity field from a turbulent flow computed by direct numerical simulation of the two-dimensional Navier--Stokes equation with periodic boundary conditions \cite{Farge et al. 1993b}. {\bf [Figure \ref{Figure20}]} shows the vorticity field with the vortex to extract (in red), the extracted vortex (in green) and the remaining vorticity field with the location of the missing vortex (in blue). 


\clearpage

\subsection{Extraction of all vortices}
\label{subsec:7.3}

The method to extract all vortices at once involves the following steps:

\begin{itemize}
\item
Fix a threshold value $T$ of the modulus of the continuous wavelet coefficients of the vorticity field, either chosen {\it a priori}, or found by trial and error,
\item
retain only the strongest wavelet coefficients, whose modulus is greater than or equal to the threshold value, and cancel the weaker coefficients,
\item
invert the continuous wavelet transform of the strongest wavelet coefficients to reconstruct in physical space the vorticity field with the extracted vortices,
\item
retain only the weaker wavelet coefficients whose modulus is below the threshold value and cancel the strongest coefficients,
\item
invert the continuous wavelet transform of the weaker wavelet coefficients to reconstruct in physical space the background vorticity field without the extracted vortices.
\end{itemize}

\begin{figure}[ht!]
\centering
\includegraphics[width=1.2\linewidth]{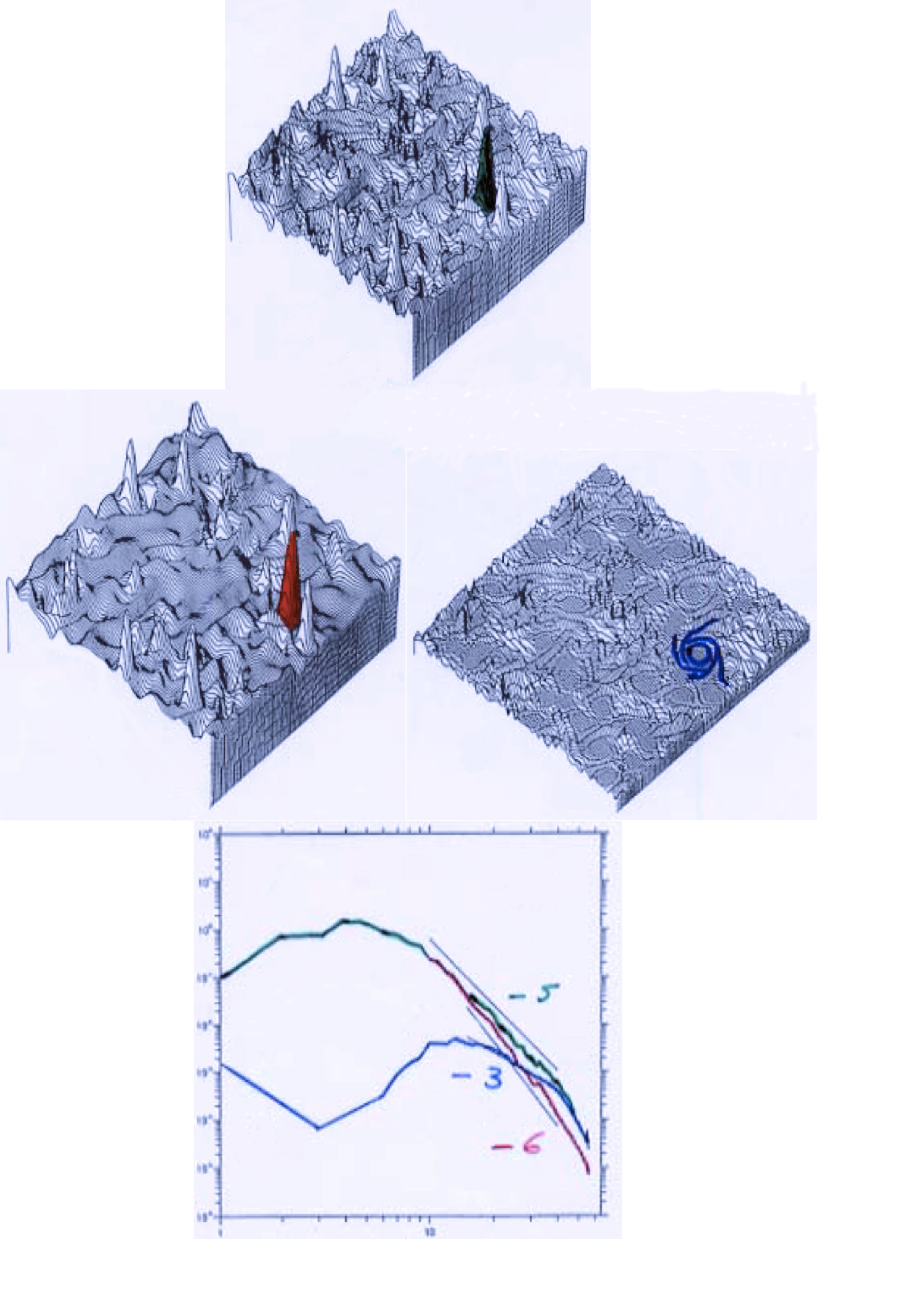}\\
  \caption{Extraction of all vortices out of a two-dimensional vorticity field using the continuous wavelet representation. Top: original vorticity field with one typical vortex highlighted in green. Middle left: the resulting coherent vorticity field, with all extracted vortices, the extracted typical vortex highlighted in red. Middle right: the incoherent vorticity field, with the region where the typical vortex has been removed highlighted in blue. Bottom: the energy spectrum (with the wavenumber along the horizontal axis and the logarithm of the energy along the vertical axis) for the three vorticity fields, highlighted in green for the original field, in red for the coherent vorticity field, and in blue for the incoherent vorticity field \cite{Farge et al. 1993b}.}
  \label{Figure21}
\end{figure}

\noindent
We applied this method to extract all coherent vortices out of  a vorticity field from a turbulent flow computed by direct numerical simulation of the two-dimensional Navier--Stokes equation with periodic boundary conditions. {\bf [Figure \ref{Figure21}]} shows the result \cite{Farge et al. 1993b} where the original vorticity field has been decomposed into an intermittent and coherent vorticity field retaining all vortices, and a non-intermittent and incoherent vorticity field made of vorticity filaments without any vortices. Using this decomposition we can then plot the energy spectrum of each of those three fields. We observe that the energy spectrum of the total vorticity field scales as $k^{-5}$, the energy spectrum of the coherent vorticity scales as $k^{-6}$ and the energy spectrum of the incoherent vorticity scales as $k^{-3}$.


\clearpage

In {\bf [Figure \ref{Figure12}]}, which is a snapshot of 
{\bf [Movie \href{http://wavelets.ens.fr/TURBULENCE/3_MOVIES/Movie_3_Vorticity_evolution_of_a_2D_turbulent_flow_in_a_circular_domain_with_no_slip_boundary_conditions_computed_by_DNS_and_visualised_with_a_cartographic_view.mpg}{3}]}, one can see the continuous wavelet representation of the vorticity modulus $|\widetilde{\omega}(\vec x, l)|$, where the threshold $T$ corresponds to a space-scale manifold which separates the retained from the discarded wavelet coefficients. In 
{\bf [Movie \href{http://wavelets.ens.fr/TURBULENCE/3_MOVIES/Movie_3_Vorticity_evolution_of_a_2D_turbulent_flow_in_a_circular_domain_with_no_slip_boundary_conditions_computed_by_DNS_and_visualised_with_a_cartographic_view.mpg}{3}]} and {\bf [Figure \ref{Figure12}]} I have highlighted as a red isoline the location, in space and scale, where the interface $|\widetilde{\omega}(\vec x, l)|=T$ intersects the moving plane. It is interesting to study how it evolves in space and scale because, the more wrinkled and spiky this interface, the more intermittent the flow. We also produced another 
{\bf [Movie \href{http://wavelets.ens.fr/TURBULENCE/3_MOVIES/Movie_4_Modulus_of_the_continuous_wavelet_coefficients_of_the_vorticity_of_a_2D_turbulent_flow_computed_by_DNS.mpg}{4}]} (see one snapshot in {\bf [Figure \ref{Figure13}]}) to show the time evolution of the interface, which corresponds to the iso-surface $|\widetilde{\omega}(\vec x, l)|=T$, visualised with a cavalier view \cite{Farge 1987}. It is interesting to compare 
{\bf [Movie \href{http://wavelets.ens.fr/TURBULENCE/3_MOVIES/Movie_2_Vorticity_evolution_of_a_2D_turbulent_flow_with_periodic_boundary_conditions_computed_by_DNS_and_visualised_with_a_cartographic_view.mpg}{2}]} and 
{\bf [Movie \href{http://wavelets.ens.fr/TURBULENCE/3_MOVIES/Movie_4_Modulus_of_the_continuous_wavelet_coefficients_of_the_vorticity_of_a_2D_turbulent_flow_computed_by_DNS.mpg}{4}]} with a snapshot in {\bf [Figure \ref{Figure12}] and [Figure \ref{Figure13}] }), which present the evolution of the same two-dimensional vorticity field, in physical space for 
and in wavelet space for 
{\bf [Movie \href{http://wavelets.ens.fr/TURBULENCE/3_MOVIES/Movie_2_Vorticity_evolution_of_a_2D_turbulent_flow_with_periodic_boundary_conditions_computed_by_DNS_and_visualised_with_a_cartographic_view.mpg}{2}]}
and in wavelet space for
{\bf [Movie \href{http://wavelets.ens.fr/TURBULENCE/3_MOVIES/
Movie_4_Modulus_of_the_continuous_wavelet_coefficients_of_the_vorticity_of_a_2D_turbulent_flow_computed_by_DNS.mpg}{4}]}. In both physical space and wavelet space we observe a similar transport of vortices and the merging of same-sign vortices. All of those movies are in open access and can be freely downloaded from {\it http://wavelets.ens.fr/CONTINUOUS\_WAVELETS}.


\bigskip
\bigskip
\bigskip
\bigskip
\bigskip

\begin{tcolorbox}[width=\linewidth]
In conclusion, the complex-valued continuous wavelet representation is, in my opinion, the appropriate representation to study turbulence, because it preserves the invariance by translation and dilation which orthogonal wavelets lose. Unfortunately, the study of three-dimensional flows in the strong turbulence regime using continuous wavelets remains beyond the reach of current supercomputers. However, complex-valued continuous wavelets could be useful for detecting finite-time singularities in solutions of the Navier--Stokes equations, as we did in 1990 \cite{Farge et al. 1990a}. More recently, we have detected such a tendency for two-dimensional turbulence by numerically studying a vortex dipole crashing onto a wall \cite{Nguyen et al. 2018}. The fourth millennium problem \cite{Fefferman 2006} addresses the question of finite-time singularities in the solutions of the three-dimensional Navier--Stokes equations in unbounded space. This is probably easier to prove mathematically than in the presence of solid walls, but less realistic physically. This reminds me of a remark made by Einstein on January 27 1921 in his Berlin lecture on {\it `Geometrie und Erfahrung'}: {\it `To the extent that the laws of mathematics refer to reality, they are not certain; and to the extent that they are certain, they do not refer to reality'} \cite{Einstein 1921}. 
\end{tcolorbox}

\clearpage

\bigskip
\bigskip
\bigskip
\bigskip
\bigskip
\bigskip
\bigskip

\noindent
{\bf Acknowledgements}

\bigskip

\noindent
I am very grateful to my friends and colleagues, Laurette Tuckerman, Patrick Chassaing, Greg Eyinck, Jean-Pierre Hansen, Yves Meyer, Thierry Paul, Kai Schneider, for their help in improving this text. I would also like to thank the `Biblioth\`eque Nationale de France', the caf\'e `Le Bucheron' and the family of Professor Tony Maxworthy for provide me pleasant conditions in Paris and in Morro Bay (California), respectively, when I was writing this article.

\bigskip
\bigskip

\noindent
{\bf Copyright and licence}

\bigskip

\noindent
This text is based on several articles and courses I have written over the past thirty years. In accordance with the policy of my employer, the `Centre National de la Recherche Scientifique', the European Commission and `cOAlition S' ({\it \url{https://www.coalition-s.org}}), I retain my copyright and provide a Creative Commons CC-BY license ({\it \url{https://creativecommons.org}}), so that this article is in open access  for everyone to freely disseminate it. 

\bigskip
\bigskip

\noindent
{\bf Documents} 

\bigskip

\noindent
Articles, figures and movies related to this article, as well as photos and interviews of Alex Grossmann and me, are freely available in open access on my website:

\medskip
\noindent
{\it \url{http://wavelets.ens.fr/CONTINUOUS\_WAVELETS}}

\bigskip

\noindent
The video and slides of a talk entitled {\it `Production of dissipative vortices by solid bodies in incompressible fluid flows: comparison between Prandtl, Navier--Stokes and Euler solutions'}, that I gave on $1^{st}$ October $2014$ at IPAM (Institute of Pure and Applied Mathematics), UCLA, Los Angeles, are freely available on IPAM website:

\medskip
\noindent
{\it \url{http://www.ipam.ucla.edu/abstract/?tid=12194\&pcode=MTWS1}}

\bigskip

\noindent
The video and slides of a talk entitled {\it `Wavelet-based definition of turbulent dissipation'}, that I gave on $13^{th}$ January $2017$ at IPAM (Institute of Pure and Applied Mathematics), UCLA, Los Angeles, are freely available on IPAM website:

\medskip
\noindent
{\it \url{https://www.ipam.ucla.edu/programs/workshops/turbulent-dissipation-mixing-and-predictability/?tab=schedule}}

\clearpage


\end{document}